\documentclass[%
 print,
superscriptaddress,
 amsmath,amssymb,
 aps,
]{revtex4-2}

\usepackage{xcolor}
\usepackage{graphicx}
\usepackage{dcolumn}
\usepackage{bm}
\usepackage{hyperref}

\begin{document}

\preprint{APS/123-QED}

\title{The impact of harmonic inflow variations on the size and dynamics of the separated flow over a bump}

\author{Himpu Marbona}
\email{himpu.marbona@upm.es}
\affiliation{Universidad Polit\'{e}cnica de Madrid, Plaza Cardenal Cisneros 3, E-28040 Madrid, Spain 
}
\affiliation{Center for Computational Simulation, Universidad Polit\'{e}cnica de Madrid, Campus de Montegancedo, Boadilla del Monte, 28660 Madrid, Spain
}
\author{Daniel Rodr\'iguez}%
\affiliation{Universidad Polit\'{e}cnica de Madrid, Plaza Cardenal Cisneros 3, E-28040 Madrid, Spain 
}
\author{Alejandro Mart\'inez-Cava}
\affiliation{Universidad Polit\'{e}cnica de Madrid, Plaza Cardenal Cisneros 3, E-28040 Madrid, Spain 
}
\affiliation{Instituto Universitario ``Ignacio Da Riva'' (IDR/UPM), Universidad Polit\'{e}cnica de Madrid, Plaza Cardenal Cisneros 3, E-28040 Madrid, Spain
}
 \author{Eusebio Valero}
\affiliation{Universidad Polit\'{e}cnica de Madrid, Plaza Cardenal Cisneros 3, E-28040 Madrid, Spain 
}
\affiliation{Center for Computational Simulation, Universidad Polit\'{e}cnica de Madrid, Campus de Montegancedo, Boadilla del Monte, 28660 Madrid, Spain
}


\date{\today}

\begin{abstract}
The separated flow over a wall-mounted bump geometry under harmonic oscillations of the inflow stream is investigated by direct numerical simulations. The bump geometry gives rise to streamwise pressure gradients similar to those encountered on the suction side of low-pressure turbine (LPT) blades. Under steady inflow conditions, the separated-flow laminar-to-turbulent transition is initiated by self-sustained vortex shedding due to Kelvin-Helmholtz (KH) instability. In LPTs, the dynamics are further complicated by the passage of the wakes shed by the previous stage of blades. The wake-passing effect is modeled here by introducing a harmonic variation of the inflow conditions. Three inflow oscillation frequencies and three amplitudes are considered. The frequencies are comparable to the wake-passing frequencies in practical LPTs.The amplitudes range from 1\% to 10\% of the inflow total pressure. The dynamics of the separated flow are studied by isolating the flow components that are respectively coherent with and uncorrelated to the inflow oscillation. Three scenarios are identified. The first one is analogous to the steady inflow case. In the second one, the KH vortex shedding is replaced during a part of the inflow period by the formation and release of a large vortex cluster. The third scenario consists solely of the periodic formation and release of the vortex cluster; it leads to a consistent reduction of the separated flow length over all the period compared to the steady inflow case and would be the most desirable flow condition in a practical application.
\end{abstract}

\maketitle

\section{\label{sec:Introduction}Introduction\protect }

Laminar boundary layer separation is a ubiquitous phenomenon present in several aeronautical applications, such as low-pressure turbines (LPT) at high-altitude flight \citep{Mayle01,Bons02,CoullJFM2011} and Unmanned Aerial Vehicle (UAV) or Micro Aerial Vehicle (MAV) wings at steady and pitching conditions \citep{Ol10}. Separated flow is associated with detrimental effects on aerodynamics and performance; flow control strategies that lead to reductions in the size of the separated flow region are thus a way forward towards improving their efficiency \citep{GreenblattPAS2000,Volino11}. 

 The laminar-to-turbulent transition process has a dominant role in the reattachment of the separated flow, thus determining the size and dynamics of separation bubbles \citep{Gaster1967,Mayle01,Dovgal94,McAuliffeJTurb2010} and also the structural loading of LPT blades (e.g. \cite{Curtis97}). The details of the transition process in separated flows where the incoming flow and the aerodynamic surface are under steady conditions and low environmental disturbance levels are relatively well known. In scenarios representative of leading edge separation on airfoils and LPT blades with a moderate load, the transition is initiated by Kelvin-Helmholtz (KH) instability of the separated shear layer, that amplifies small-amplitude disturbances existing in the pre-separated boundary layer \citep{Dovgal94,DiwanRamesh2009}, quickly leading to the formation of spanwise vortices and triggering transition via nonlinear interactions of these vortices \citep{Marxen:JFM13}. Alternative or complementary transition scenarios have been proposed for stronger adverse pressure gradients, that involve the onset of absolute inflectional instability \citep{HammondRedekopp1998,RistMaucher2002,Avanci19}, spanwise modulation of the recirculation region due to a self-excited global mode \citep{THD,gallaire_marquillie_ehrenstein_2007,Rodriguez:JFM10,Rodriguez:JFM13}, or a combination of them \citep{Passaggia12,Rodriguez:AeroJ19,Rodriguez21}. 
The transition process can be altered significantly in the presence of moderate to elevated levels of free-stream disturbances. The incoming turbulence generates streaky structures in the attached boundary layer upstream of separation, referred to as Klebanoff modes \citep{JacobsDurbin:JFM2001}, which prevent the formation of spanwise-homogeneous KH vortices. The interaction between the streaks and the inflectional instability leads to the formation of short-span KH structures which enhance momentum transfer in the wall-normal direction and result in a faster reattachment compared to cases with low free-stream disturbances \citep{McAuliffeJTurb2010,CoullJFM2011,Simoni:EF2017,Hosseinverdi:JFM19}. 

The flow dynamics on the suction side of LPT blades, and especially the transition process, are complicated further by the inherent unsteadiness of the multi-stage machines, as the wakes shed by one blade stage convect through the downstream passages periodically disturbing the flow conditions of the next stage. These disturbances excite the formation of intense vortical structures that shed and pull fluid from the recirculation region downstream, thus temporally reducing the separated flow extent. Subsequently, the separation bubble regenerates, increasing in size until it reaches a stationary value or the influence of the next wake impacts the flow. The transition to turbulence is thus multi-modal, promoted by periodic fluctuations of the incoming free-stream flow and the own instabilities of the separated flow \citep{Hodson:ARFM2005}. Wake-induced transition is very sensitive to the combined effects of the adverse pressure gradient and intensity, temporal dependence and frequency of the inflow free-stream fluctuations. Several investigations have addressed the impact of the wake-passing period on the length of the separated flow region, modeling the periodic passage of wakes either as an inlet velocity deficit localized in space and time \citep{CoullJFM2011,Volino:JTurb2012,Gungor:JTurb2012,Karaca16} or as a harmonic change of the cross-sectional inlet conditions \citep{Wissink:FTurbCom2003,Wissink:AMM2006}. While the two approaches present some differences, their results agree qualitatively on the impact of the wake-passing frequency and intensity on the flow. A dimensionless frequency $F$, often referred to as reduced frequency, is defined based on the characteristic free-stream velocity and a streamwise length representative of the extent of the adverse pressure gradient region in the absence of wakes. The reduced frequency is thus the ratio of the convective time scale and the period of the imposed free-stream variation. Ambiguity exists in the practical characterization of $F$: first, the length of the steady-flow separated flow region is problem dependent and only known \textit{a posteriori}. Second, under unsteady inflow conditions, both the free-stream velocity and the streamwise pressure gradients change continuously and can differ significantly from the values obtained under the steady inflow conditions.

Wake-passing frequencies $F$ close to but slightly lower than 1 are representative of aero-engine LPTs \citep{CoullJFM2011}. For low $F$, the wake passing period is long compared with the characteristic times of both the KH-related vortex shedding and the regeneration of the separation bubble; the impact on the time-average separated flow and aerodynamic performance is expected to be comparatively weak. For $F$ above 1, successive wakes pass by the separated flow region before the separation bubble has time to fully regenerate, which leads to significantly shorter time-averaged bubbles. These observations agree with investigations on active flow control of separated flows by means of periodic excitation using wall suction and blowing or geometries with moving parts, reviewed by \citet{GreenblattPAS2000}, which conclude that the forcing frequency that minimises the size of the separated flow is $F \sim 1$. Interestingly, this frequency is typically lower than that of the KH instability and scales with the global length of the separated flow rather than with the local properties of the separated boundary layer. 

\begin{figure}
\centering{\includegraphics[width=.7\textwidth,trim={0in 0in 0in 0.95in},clip] {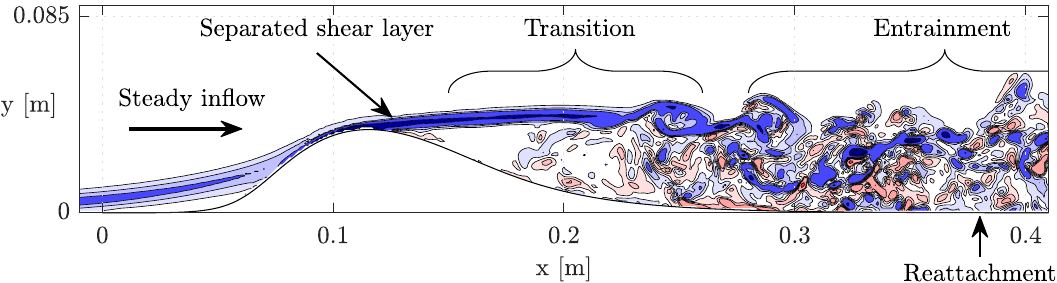}}
 \caption{Spanwise vorticity field around a wall-mounted bump under steady inflow conditions.}
\label{fig:sketch_bump}
\end{figure}

The present work studies the separated flow formed over a wall-mounted bump geometry that reproduces some characteristics of the suction side of a LPT blade under periodic fluctuations of the inflow stream. Similar geometries have been employed in the past both experimentally and numerically to study the fundamental physics of separated flow and possible means for flow control \citep{Seifert:AIAAJ2002,Bernard:JFM2003,Marquillie03,Passaggia12,Pescini:ETFS2017,Saavedra18,Saavedra21}. 
Direct numerical simulations are carried out for ten different inflow conditions. The first one is a steady inflow prescribed by a constant pressure difference between the inflow and outflow that results in the transition scenario shown in figure \ref{fig:sketch_bump}. Under steady inflow conditions, the laminar boundary layer separates just upstream of the bump summit. The separated shear layer sustains a self-excited shedding of spanwise-dominant KH vortices, followed by their breakdown in smaller three-dimensional structures and an abrupt transition to turbulence. The entrainment provided by turbulence leads to the mean flow reattachment. 
The other nine cases study different scenarios of wake-induced transition. The large-scale action of passing wakes is modeled as a harmonic fluctuation of the inflow conditions (total pressure and bulk velocity) in a manner analogous to  \citet{Wissink:FTurbCom2003,Wissink:AMM2006}. 
 Three different frequencies, each one with three fluctuation amplitudes, are prescribed, resulting in distinct scenarios for the dynamics of the separated flow and relevant modifications of the separated flow length. The scenario most favorable is characterized by the periodic formation and release of a large vortex cluster, considerably larger in size than the KH vortices, and that is phase-locked to the oscillations of the inflow conditions.

 The rest of the paper is organized as follows. Section \ref{sec:Numerics} describes the geometry, computational domain, boundary conditions and numerical methods used in the simulations. Section \ref{sec:Results} presents and discusses the results. A qualitative description of the flowfield evolution is given in Section \ref{sec:instantaneous}. The triple decomposition proposed by \citet{Hussain70} is applied to separate flow components that are coherent (in-phase) with the inflow oscillation from those occurring randomly. Section \ref{sec:Phase-averaged} describes the phase-averaged fields for three representative cases. The phase-averaged data provides information on the impact of the inflow oscillation over the length of the separated flow and how it evolves over the inflow period, which is discussed in Section \ref{sec:length}. Section \ref{sec:Vortex_dynamics} presents the flow component that is incoherent (i.e. uncorrelated) with the inflow oscillation. Monitoring this component sheds light on the vortex dynamics that ultimately govern the behaviour of the separated flow. Section \ref{sec:spectra} presents frequency spectra at different probe locations. In combination, these results show that the separated flow subject to inflow oscillations can present three different scenarios regarding the flow dynamics and their impact on the length of the separated flow region. These scenarios are thus fully characterized in Section \ref{sec:Results} and discussed in Section \ref{sec:Conclusion}, along with their connection with active flow control strategies. To verify that present findings are general and not exclusive to the particular geometry of the wall-mounted bump used, results of an analogous study considering the related yet different NASA hump geometry \cite{Seifert:AIAAJ2002,GreenblattAIAAJ2006} are presented in Appendix \ref{sec:appendix_NASA_hump}.

\section{\label{sec:Numerics}Numerical Approach}
\subsection{Geometry and domain}
\begin{figure}
\centering{\includegraphics[width=.8\textwidth,trim={0.1in 2.55in 0in 0in},clip] {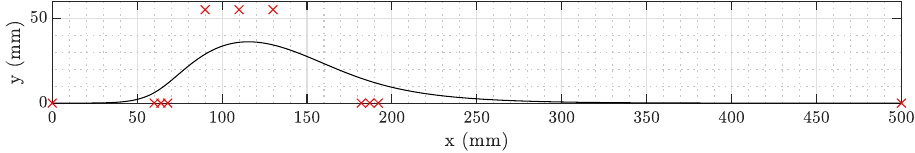}}
\caption{Definition of the wall-mounted bump geometry. The symbols show the location of the control points defining the Bezier curves.}
\label{fig:geometry}
\end{figure}

The geometry of interest is shown in Fig. \ref{fig:geometry}. \textcolor{black}{This geometry is based on the experimental set-up used by \citet{Saavedra18, Saavedra21}, consisting on a plane channel with a bump protruding from one of the walls. The bump geometry is defined by Bezier curves using 11 control points, to ensure continuity of the surface up to the second derivatives. The bump maximum height (summit) and the channel height are 0.036 m and 0.17 m, respectively, resulting in a throat width of 0.134 m and a blockage ratio of 21.2 \%. This relatively large blockage ratio impacts the pressure gradient over the bump and especially near the bump summit. However, for the related geometry of the NASA hump, \citet{GreenblattAIAAJ2006} concluded that reducing the blockage ratio indeed modifies the wall pressure distribution but does not affect the separation and reattachment locations nor the root-mean-squared pressure distribution. }

\begin{figure}
\includegraphics[width=0.9\textwidth,trim={0in 1.3in 0.1in 2.75in},clip] {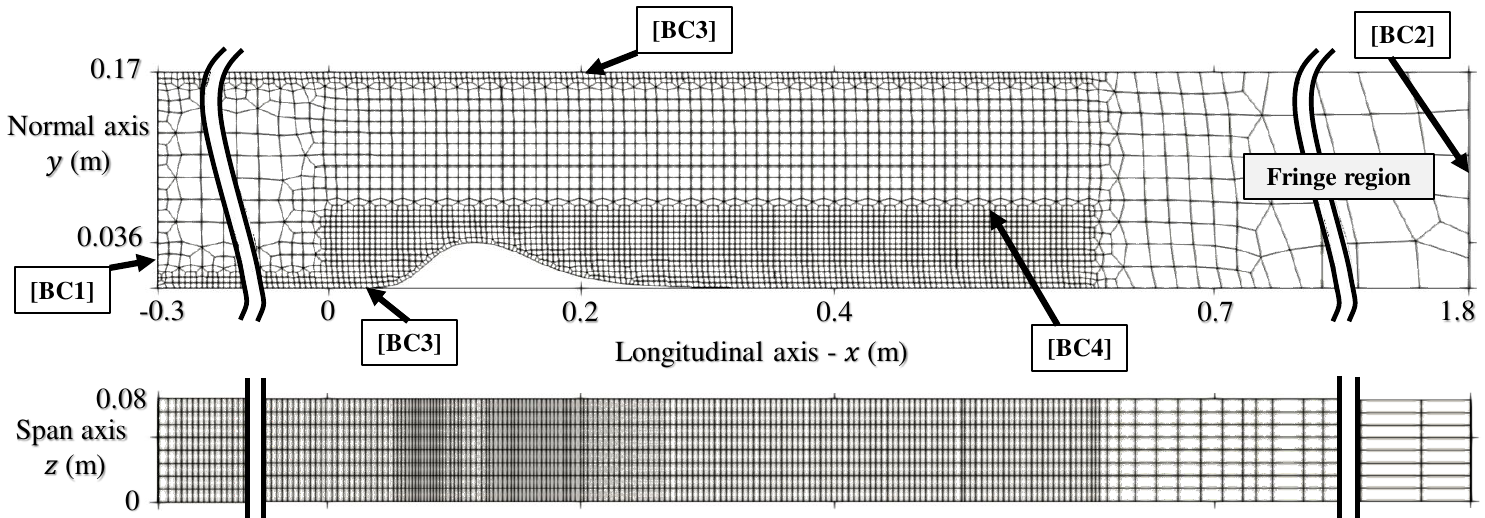}
\caption{Computational domain and representative mesh for spectral element computations.}
\label{fig:mesh}
\end{figure}

The computational domain and a representative mesh used are shown in Fig. \ref{fig:mesh}. The channel is considered to be homogeneous in the spanwise direction. The spanwise size of the computational domain is set as 0.08 m (more than twice the bump height), and periodic boundary conditions are imposed on the lateral walls. The computational domain is extended in the streamwise direction with respect to the reference experiments and simulations to minimise the impact of the boundary conditions and to allow for the introduction of a Fringe region before the downstream boundary. The inlet section (denoted by BC1) is located 0.3 m upstream of the beginning of the bump, i.e., upstream of the first control point in the Bezier curves as shown in Fig. \ref{fig:geometry}. This relatively long distance between the inlet section and the bump is intended to allow for the development of the incoming boundary layer.  

A high-order mesh generation tool called High Order Hex-Quad Mesher (HOHQ) \citep{HOHQMesh} is used to generate a mesh suitable for spectral element computations. The curvature of the elements is represented using a 5th-order polynomial.
 Local mesh refinement is done in the regions adjacent to the wall and around and downstream of the bump, where the strongest velocity gradients are expected to appear.  Section \ref{sec:Numerics_convergence} summarises the mesh refinement studies. The mesh finally used consists of 31 640 rectangular elements with 8 elements in the spanwise, $z-$direction, and a 3rd-order polynomial in all the elements and directions. 

\subsection{Simulation setup and boundary conditions}
\label{sec:set-up}

Table \ref{tab:BCs} describes the boundary conditions. The flow is driven by a pressure gradient imposed through the inflow and outflow boundary conditions. The static pressure $p_{Ref} = $ 101325 Pa is imposed at the outlet and used as reference pressure. Uniform values of total pressure $p_t$ and temperature $T_t$ are prescribed at the inlet. 
A reference case is defined in which the inlet total pressure is constant and equal to $p_t = $ 105319 Pa and the inlet total temperature is $T_t = $ 291.2 K; this case is referred to as \emph{steady inflow} case. Under these conditions, the mean velocity at the inlet section is $U_{Ref} = 68.158$ m/s, the static temperature at inflow (which is taken as reference temperature) is $T_{Ref} = $ 288.9 K, and the resulting flow has (unit length) Reynolds number $Re = \rho_{Ref} U_{Ref} L_{Ref} / \mu_{Ref} \approx 100 \ 000$ and Mach number $M = U_{Ref}/c_{Ref} = 0.2$, where $\rho_{Ref}$, $\mu_{Ref}$ and $c_{Ref}$ are respectively the density, dynamic viscosity and speed of sound based on the reference temperature and pressure ($T_{Ref}$ and $p_{Ref}$), and $L_{Ref}$ is an arbitrary reference length which is chosen as $L_{Ref} = $ 1 m.
\begin{table*}
\caption{Boundary Conditions}
\label{tab:BCs}
\def~{\hphantom{0}}
\begin{ruledtabular}
\begin{center}
\begin{tabular}{lcl} 
{[}BC1] Inlet   & \mbox{   } & Total pressure $p_t$ as in Eq. \ref{eq:Ptotal} and total temperature $T_t = 291.2$ K \\
{[}BC2] Outlet  & & Static pressure: $p = 101 \ 325$ Pa  \\
{[}BC3] Wall    & & No-slip, adiabatic   \\
{[}BC4] Sides   & & Periodicity  \\
\end{tabular}
\end{center}
\end{ruledtabular}
\end{table*}

In the other cases, referred to as \emph{harmonic inflow} cases, a periodic variation of the total pressure at the inlet is imposed, while the total temperature remains constant as in the steady inflow case. The total pressure at the inlet is defined as 

\begin{equation}
{p_t\left(t^*\right)} = p_{t,steady}\ (1 +A_{in} \ \sin(2\pi f_{in}^* (t^*-t^*_0)))
\label{eq:Ptotal}
\end{equation}

\noindent where $p_{t,steady}$ is the total pressure of the steady inflow case, $A_{in}$ is the amplitude of the harmonic oscillation, $f_{in}^*$ is a dimensionless frequency, $t^*$ is a dimensionless time and $t^*_0$ is a reference instant. Dimensionless velocity, time, and frequency are defined using $U_{Ref}$ and the unit length ($L_{Ref}$),

{\color{black}
\begin{equation}
u^*= \ \frac{ u }{U_{Ref}}, \quad t^*= t  \ \frac{ U_{Ref} }{L_{Ref} }, \quad f^* = f  \ \frac{L_{Ref} \ }{U_{Ref}} . 
\end{equation}
}

\noindent Note that the definition of $f^*$ is not the same as the reduced frequency $F$ discussed in Section \ref{sec:Introduction}. The definition of $F$ is based on the representative length of the separated flow region, which is not known \textit{a priori}. Also, the specification of $f^*$ based on the unit length simplifies the data temporal sampling and subsequent analyses.  Nine cases with harmonic inflow variation are considered, comprising three amplitudes ($A_{in}=0.01,0.05$ and 0.1) and three different frequencies ($f_{in}^*=0.5,1$ and 2).  

Early simulations showed the presence of non-physical disturbance reflections from the outlet boundary which impacted notably the dynamics of the separated flow. To minimize their upstream influence, a Fringe region is applied as in \citet{Spalart88,Spalart_Watmuff_93, Nordstrom99}. This approach imposes a forcing term into a bounded spatial region extending upstream from the outlet, to drive the flowfield towards prescribed values and effectively dampen the flow fluctuations before they reach the outlet boundary. The forcing term takes the form

\begin{equation}
{\bf{s}}= - \bar{\lambda} \ F \left( \frac{x-x_{start}}{\Delta_{rise}} \right)  \left( {\bf q} - {\bf q}_{target} \right), 
 \label{eq:FringeForcing}
 \end{equation}

\noindent where $\bar{\lambda}$ is a constant controlling the strength of the forcing, $F(r)$ is a smooth function describing the spatial structure of the forcing term, $x_{start}$ is the coordinate where the Fringe region starts, $\Delta_{rise}$ is the length of the Fringe region, $\bf q$ is the vector of fluid variables, written in conservative variables (as described in section \ref{sec:horses}), and ${\bf q}_{target}$ is the prescribed flow to be recovered at the outlet. Following \citet{Spalart_Watmuff_93}, the function $F(r)$ is defined as: 

\begin{equation}
F \left(r\right) = \Bigg \{ \begin{array}{l rcl}
     0, & & r & \leq 0,  \\
     1/ \left[1+\exp\left(\frac{1}{r-1}+\frac{1}{r}\right)\right],    & 0<&r&<1, \\
     1, & 1 \leq & r & . 
\end{array}
 \label{eq:Fringe}
\end{equation}

 \textcolor{black}{The reference values of the density $\rho_{Ref}$  and temperature $T_{Ref}$, and the streamwise velocity of the steady inflow are imposed for the target values ${\bf{q}}_{target}$, together with the vanishing of the wall-normal and transversal velocity components. The Fringe parameters used in the simulations have been determined from different tests and ensure that the flowfield does not fluctuate at the outlet and has negligible upstream effects. The Fringe forcing is very effective in dampening vortical disturbances and short Fringe regions are typically used for incompressible flow. Conversely, present computations consider compressible flow at relatively low Mach number and consequently additional disturbances of acoustic nature are introduced.  
Previous tests simulating plane channels under the same inflow conditions as the wall-mounted bump showed that Fringe regions comprising at least one acoustic wavelength were required to effectively damp the acoustic reflections and prevent acoustic feedback with the current numerical set-up.
The Fringe region finally used starts a sufficient distance from the bump summit, i.e. at $x_{start}=0.7$ m and extends to the outlet located at $x_{outlet} = 1.7$ m. The parameter $\Delta_{rise}$ is set equal to $x_{outlet} - x_{start}$ in order to provide the lowest gradient possible, and $\bar{\lambda}=400$. 
 }

\subsection{Computational methods}
\label{sec:horses}
Direct numerical simulations are performed using the in-house discontinuous Galerkin spectral element code HORSES3D \citep{Ferrer22}. 
The flow variables are made dimensionless using  $L_{Ref}$, $T_{Ref}$, $p_{Ref}$ and the reference velocity $U_{ref}$ for the steady inflow case, resulting in a unit-length Reynolds number $Re = $100 000 and Mach number 0.2, as described in Section \ref{sec:set-up}.  The dimensionless compressible Navier-Stokes equations in conservative form take the form

\begin{equation}
\frac{\partial \bf{q}}{\partial t} +  \boldsymbol\nabla \cdot \left( \bf{f}^{a} -\bf{f}^{v} \right) = \bf{s},
 \label{eq:NS Equation}
\end{equation}

\noindent where ${\bf{q}}=[\rho, \ \rho u, \ \rho v, \ \rho w, \ \rho E ]^{T}$ are the conservative variables, $E$ is the specific internal energy and $ \bf{s}$ is a source or volumetric forcing term. Advective ($\bf{f}^{a}$) and viscous ($\bf{f}^{v}$) fluxes are expressed in primitive variables as equations (\ref{eq: Advection Fluxes}) and (\ref{eq: Viscous Fluxes}), respectively. 

\begin{equation}
{\bf{f}}^{a}_{1}=\begin{bmatrix}
\rho u \\ p + \rho u^2 \\ \rho u v \\ \rho u w \\ u(\rho E +p)   
\end{bmatrix}, \ \
{\bf{f}}^{a}_{2}=\begin{bmatrix}
\rho v \\ \rho u v \\ p + \rho v^2 \\ \rho v w \\ v(\rho E +p)    
\end{bmatrix}, \ \
{\bf{f}}^{a}_{3}=\begin{bmatrix}
\rho w \\ \rho u w \\ \rho v w \\ p + \rho w^2 \\ w(\rho E +p)    
\end{bmatrix},
\label{eq: Advection Fluxes}
\end{equation}

\begin{equation}
{\bf{f}}^{v}_{1}=\frac{1}{Re}\begin{bmatrix}
0 \\ \tau_{xx} \\ \tau_{xy} \\ \tau_{xz} \\ v_{i} \tau_{1i} + \kappa \partial_{x} T   
\end{bmatrix}, \ \
{\bf{f}}^{v}_{2}=\frac{1}{Re}\begin{bmatrix}
0 \\ \tau_{yx} \\ \tau_{yy} \\ \tau_{yz} \\ v_{i} \tau_{2i} + \kappa \partial_{y} T   
\end{bmatrix} , \ \ 
 {\bf{f}}^{v}_{3}=\frac{1}{Re}\begin{bmatrix}
0 \\ \tau_{zx} \\ \tau_{zy} \\ \tau_{zz} \\ v_{i} \tau_{3i} + \kappa \partial_{z} T   
\end{bmatrix},  
\label{eq: Viscous Fluxes}
\end{equation}

\noindent The equation of state for ideal gas takes the form
\begin{equation}
p=\left( \gamma -1 \right) \rho \left[ E-\frac{u^2 + v^2 + w^2}{2} \right],
\label{eq: pressure equation}
\end{equation}

\noindent and Sutherland's law is used for the dynamic viscosity
\begin{equation}\label{eq: Sutherland Law}
\mu =  \frac{1+T_{suth}/T_{Ref}}{T+T_{suth}/T_{Ref}} T^{\frac{3}{2}},
\end{equation}
\\
where $T_{suth}= 110.4$ K. The dimensionless thermal conductivity is expressed as 
\begin{equation}
\kappa = \frac{\mu}{\left( \gamma -1 \right) Pr  \ M},
\end{equation}

\noindent where $Pr$ is the Prandtl number, assumed to be constant and equal to 0.72. The stress tensor,  using Stokes hypothesis, is defined as
\begin{equation}
 {\boldsymbol \tau}=\mu \left( \left(  \nabla {\boldsymbol v} \right)^{T} +  \nabla {\boldsymbol v} \right) - \frac{2}{3} \mu \left( \boldsymbol \nabla \cdot {\boldsymbol v}\right) \ {\boldsymbol I} 
\label{eq: stress tensor}
\end{equation}

The simulations are performed using 3rd-order polynomials with Gauss nodes. Time integration is performed with an explicit 3rd-order Runge-Kutta scheme. A standard discontinuous Galerkin discretization of the inviscid fluxes is done using Roe's method for the Riemann problem and the Bassi-Rebay 1 scheme is used for the discretization of the viscous fluxes. Further details on the numerical implementation can be found in \citet{Ferrer22} and references therein.

\subsection{Convergence study} \label{sec:Numerics_convergence}
\begin{table*}
\caption{Convergence study of spatial and temporal resolutions.}
\label{tab:spatial_resolution}
\begin{ruledtabular}
\begin{center}
\def~{\hphantom{0}}
\begin{tabular}{cccrcccccrr} 
Discretisation & $h/h_{ref}$ & $p$ & \hspace{10pt} DOF \hspace{10pt} & $\Delta t^*$ &  $L_s$ [m] & Relative cost & $N_{s}$ \hspace{4pt} &  $N_{ss}$\hspace{4pt}  \\  [3pt]  \hline
Baseline & 1   & 3 & 2 024 960 & $3\times10^{-5}$ &  0.2782 & 1      & 2 410 000 & 33 332 \\
Fine     & 0.8 & 3 & 5 012 480 & $5\times10^{-6}$ &  0.2818 & 13.82  & 2 400 000 & 200 000 \\
Coarse   & 1.2 & 3 & 1 343 616 & $3\times10^{-5}$ &  0.2836 & 0.60   & 1 350 000 & 33 332 \\
Baseline + low $p$ & 1   & 2 & 854 280   & $3\times10^{-5}$  & 0.2931 & 0.68   & 1 000 000 & 33 332 \\
Baseline + high $p$ & 1   & 4 & 3 955 000 & $2\times10^{-5}$  &  0.2858 & 2.31   & 550 000   & 50 000\\
\end{tabular}
\end{center}
\end{ruledtabular}
\end{table*}

\begin{table}
\caption{Coordinates of the reference point and sampling probes.}
\label{tab:Ref_Sampling_Probes}
\begin{ruledtabular}
\begin{center}
\def~{\hphantom{0}}
\begin{tabular*}{.6\textwidth}{@{\extracolsep{\fill}}clll} 
Probe & x [m] & y [m] & $\xi$ [m]  \\ [3pt] \hline
Reference & 0.0   & 0.15 & - \\
1 & 0.165   & 0.0417 & 0.06 \\
2 & 0.225   & 0.0425 & 0.12 \\
3 & 0.304   & 0.0348 & 0.2  \\
4 & 0.2   & 0.05 & -
\end{tabular*}
\end{center}
\end{ruledtabular}
\end{table}

\begin{figure}[t]
\centering{
\includegraphics[width=.45\textwidth,trim={0in 0in 0in 1.4in},clip] {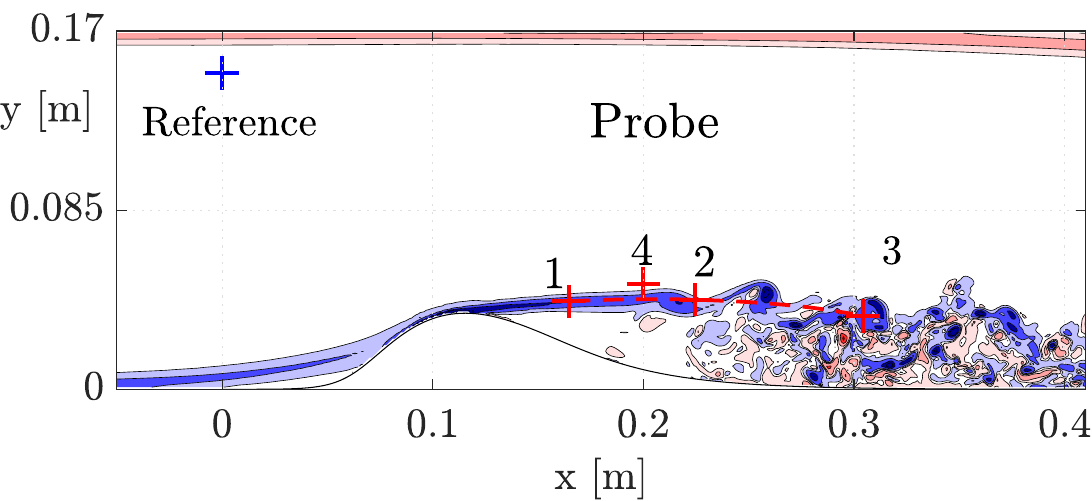} \\
}
\caption{Location of the reference point and sampling probes. }
\label{fig:probes}
\end{figure}

\begin{figure}
\centering{
\includegraphics[width=1\textwidth,trim={0in 0in 0in 0in},clip] {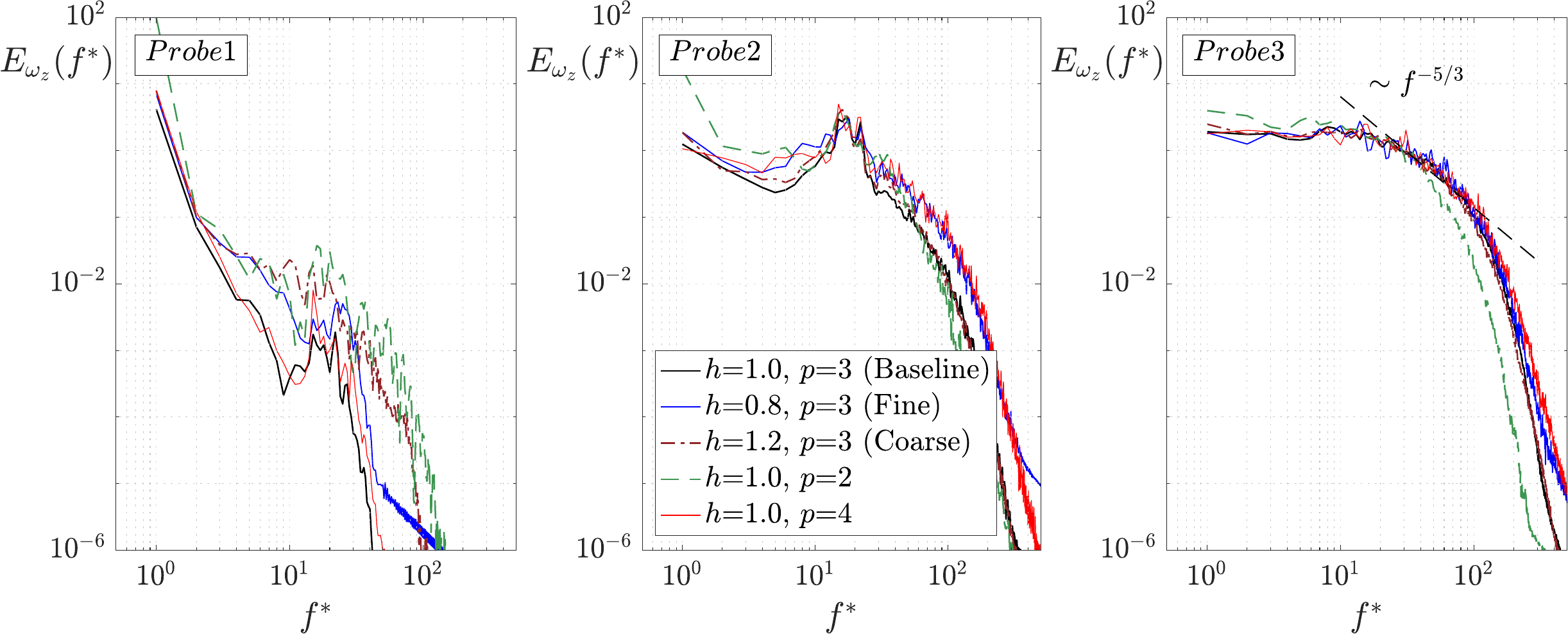} \\
}
\caption{PSD of spanwise vorticity at selected probe locations for different spatial and temporal discretizations. }
\label{fig:esd}
\end{figure}

Preliminary simulations are performed to study the robustness of the results with respect to the spatial discretization and numerical parameters. The case with steady inflow is considered. Because the nominal Reynolds number is the same for all cases, the spatial resolution requirements are considered to be analogous. Similarly, the period of the inlet fluctuations is much longer than the characteristic times associated with the KH vortices (see Appendix \ref{sec:appendix_KH}) and the series of increasingly smaller vortices typical of turbulent flow and thus the temporal resolution requirements are the same for all cases.

Spectral element methods allow the spatial discretization to be refined either by reducing the size of the elements ($h$-refinement) or by increasing the order of the polynomial used in the evaluation of derivatives within each element ($p$-refinement). A baseline mesh is defined based on a regular element size and different levels of zonal refinement, as shown in Fig. \ref{fig:mesh}. To study the $p$-refinement, 2nd, 3rd and 4th order polynomials are used together with the baseline mesh. To study the $h$-refinement, two additional meshes were built with the element size respectively increased and decreased a 20$\%$ with respect to the baseline mesh. The parameters of the 5 discretizations tested are given in Table \ref{tab:spatial_resolution}, together with the number of spatial degrees of freedom (DOF) and the time step $\Delta t^*$ used.
 The mean length of the separated flow region $L_s$ is compared for the different discretizations. The $h$-refinement study shows that $L_s$ changes are below $2\%$, while the relative computational cost increases an order of magnitude for the finer mesh. The effect of the polynomial order $p$ is found to be stronger. However, increasing the order from $p=3$ to $p=4$ results in a change of $L_s$ below $3\%$. 

The power spectra density (PSD) of the spanwise vorticity at different probe locations are computed and shown in figure \ref{fig:esd}. Power spectral densities are estimated using Welch's method \citep{Welch:1967}. For each spatial discretization, different time-step sizes $\Delta t^*$ are used and varying total numbers of snapshots $N_s$ are stored. The PSD is computed by averaging the periodograms of segments of $N_{ss}$ snapshots each with 50$\%$ overlap and a Hamming window. The segment length $N_{ss}$ is chosen separately for each simulation to ensure that the frequency bin $\Delta f^* = 1$. 

The location of the probes is illustrated in Fig. \ref{fig:probes} and given in Table \ref{tab:Ref_Sampling_Probes}, together with other locations that are used throughout the paper. 
Probes 1, 2 and 3 are located along the mean shear layer. The reference point and Probe 4 will be referred to in Section \ref{sec:Results}.
Probe 1 is located relatively close to the separation point, upstream of the region where the first KH vortices are observed. Probe 2 is in the region in which KH vortices are shed. Probe 3 is located downstream, where vortical structures of a broad range of scales are already present. 
 For all five spatial resolutions, the energy spectra are roughly the same except for the case with 2nd-order polynomial, coarser mesh, and at higher frequencies. These frequencies are associated with small scales and present very low energy levels, and the deviation of the spectra may indicate that the turbulent cascade is not fully resolved. However, the focus of the simulations is on the dynamics of the larger structures dominating the separated flow dynamics, which correspond to substantially lower frequencies and are robustly captured by the different spatial and temporal resolutions. This concludes that the baseline spatial discretization and time-step (cf. Table \ref{tab:spatial_resolution}) deliver robust results. More expensive simulations with increased spatial resolution do not result in significant differences in the physics of interest.


\section{Results}
\label{sec:Results}

\begin{table*}
\caption{Summary of the cases simulated, including the definition of the inlet pressure condition and the characterization of time-averaged and phase-dependent values of the reference streamwise velocity and the length of the recirculation region $L_s$.  }
\label{tab:simulations_details}
\begin{ruledtabular}
\begin{center}
\def~{\hphantom{0}}
\begin{tabular}{ccccccccc} 
$\Delta t^*$ & $CFL_{max}$ & $t^*_{data}$ & $A_{in}$ & $f_{in}^*$ & $\bar{u}^*_{@ Ref}$ & $\Delta {u}^*_{@ Ref}$ & $L_s$ [m]   & $\Delta{L_s}$ [m] \\ [1pt] \hline

 $3\times 10^{-5}$ & 0.9 & 72.3 & -  & -  & 1.11     & -       &  0.2782  &  - \\
 & & & & & & & & \\
 $2\times 10^{-5}$ & 0.7 & 50 & 0.01 & 0.5&  1.1025  & 0.0293  &  0.2782  &  0 \\
                   & 0.7 & 50 & 0.01 &  1 &  1.1078  & 0.0301  &  0.2780  &  -0.0002 \\
                   & 0.7 & 50 & 0.01 &  2 &  1.1072  & 0.0282  &  0.2625  &  -0.0157 \\
                   & & & & & & & & \\
                   & 0.7 & 50 & 0.05 & 0.5&  1.1030  & 0.1467  &  0.2727  &  -0.0055\\
                   & 0.7 & 50 & 0.05 &  1 &  1.1022  & 0.1467  &  0.2600  &  -0.0182\\
                   & 0.7 & 50 & 0.05 &  2 &  1.1008  & 0.1401  &  0.1966  &  -0.0816\\
                   & & & & & & & & \\
                   & 0.8 & 50 & 0.1 & 0.5 &  1.0872  & 0.2929  &  0.2575  &  -0.0207\\
                   & 0.8 & 50 & 0.1 &  1  &  1.0840  & 0.2961  &  0.2323  &  -0.0459\\
                   & 0.8 & 50 & 0.1 &  2  &  1.0871  & 0.2772  &  0.1536  &  -0.1246
\end{tabular}
\end{center}
\end{ruledtabular}
\end{table*}

\begin{figure}[t]
\centering{\includegraphics[width=.8\textwidth,trim={0in 0in 0in 0in},clip] {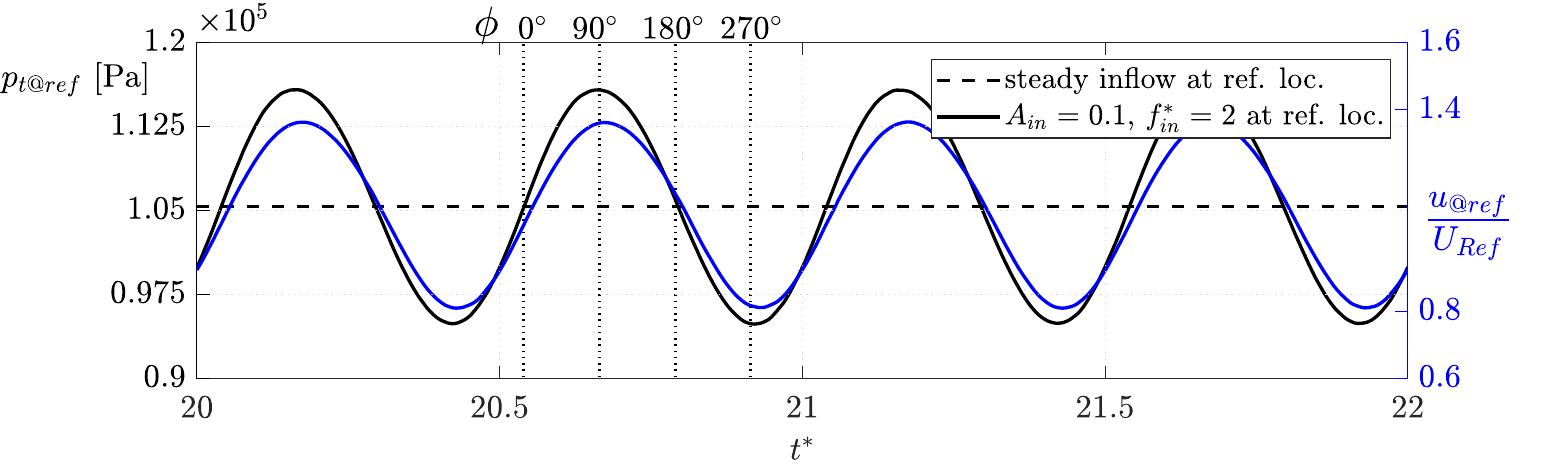}}
\caption{Evolution of total pressure and streamwise velocity at the reference point. }
\label{fig:Pt_Ref}
\end{figure}

Table \ref{tab:simulations_details} summarizes the different simulations performed in this work, comprising the reference case that features a steady inflow and nine cases in which the inflow has a harmonic component of different frequency $f_{in}^*$ and amplitude $A_{in}$. All the harmonic inflow cases use the same timestep $\Delta t^*=2\times 10^{-5}$. The initial transient, different for each case but comprising several flow-through times, was discarded and the following simulation sampling time $t^*_{data} = 50$ was collected for the subsequent analysis.

The steady inflow case is considered as the baseline. The length of the separated flow region, defined as the streamwise distance from the time-averaged separation to reattachment points, is $L_{s,steady} = 0.2782$ m. The Reynolds number based on the mean inflow velocity and $L_{s,steady}$ is $Re_{L_s} = 27 000$, which is comparable to the works by e.g. \citet{Embacher:JFM2014,Kurelek:PRF2019,Dellacasagrande:EF20}. The Reynolds number based on the maximum height of the recirculation region, approximated here as the height of the bump, is $Re_h = 3600$, which lies in the range of values reported by \citet{Gaster1967} and other experiments cited by \citet{DiwanChetanRamesh2006}. The Reynolds number based on the boundary layer momentum thickness at the separation point is $Re_{\theta,s} = 27$. This number is significantly smaller than most of the values reported in the literature for laminar separation bubbles. However, this discrepancy is explained by the use of a wall-mounted bump, as opposed to flat plates under an adverse pressure gradient or airfoils with small to moderate camber. The strong acceleration of the flow in the upstream portion of the bump leads to a substantial reduction of the boundary layer thickness. The rather low value of $Re_{\theta,s}$ is comparable to that in geometry-induced separation bubbles like those reported by \citet{Aniffa:JFM2023}.

 The harmonic change of the inlet total pressure leads to a periodic acceleration and deceleration of the bulk flow. Due to the relatively long upstream extension of the domain, the inflow changes reach the bump with a delay. A reference point located just upstream of the bump (see Fig. \ref{fig:probes} and Table \ref{tab:Ref_Sampling_Probes}) is used to characterize the bulk flow changes in the bump region.
 Figure \ref{fig:Pt_Ref} shows the total pressure and streamwise velocity evolution at the reference point for the case $A_{in}=0.1$, $f_{in}^*=2$.  The total pressure at this point is used to define the phase of oscillation $\phi$ in the analysis done in the rest of the paper. The phase $\phi = 0^\circ$ is chosen as the time in which the total pressure is at its mean value and has maximum positive derivative (i.e. maximum acceleration); $\phi = 90^\circ$ and $270^\circ$ correspond respectively to the maximum and minimum values of the total pressure.

 The streamwise velocity also exhibits a sinusoidal behaviour, with the mean value $u_{mean}$ remaining the same as for the steady inflow case. A small delay $T^*_d$ exists between the maxima and minima of total pressure and streamwise velocity, stemming from flow inertia. Table \ref{tab:simulations_details} shows the normalized amplitude of the streamwise velocity fluctuation, $\Delta {u}^* = (u_{max} - u_{min})/(2 U_{Ref})$. This value is linearly proportional to $A_{in}$ and the normalized streamwise velocity at the reference point can be approximated by

 \begin{equation}
 u^* (t) \approx \bar{u}^*  + \Delta u^* \sin (2 \pi f^*_{in} (t^* + T^*_d).
 \end{equation}

  Table \ref{tab:simulations_details} also shows the mean value of the streamwise length of the flow recirculation region $L_s$, and its relative change with respect to the steady inflow case, $\Delta L_s = L_s - L_{s,steady}$ for each one of the cases with harmonic oscillation inflow. As opposed to $\Delta u^*$, $\Delta L_s$ is not proportional to $A_{in}$, indicating that essentially non-linear dynamics govern the separated flow and its reattachment. 
The non-linearity is associated with the complex vortex dynamics originated in the separated flow region and the impact of the flow acceleration and deceleration upon them, which will be studied in the next sections.

\begin{figure}
\centering{\includegraphics[width=.8\textwidth,trim={0in 0.8in 0.4in 0in},clip] {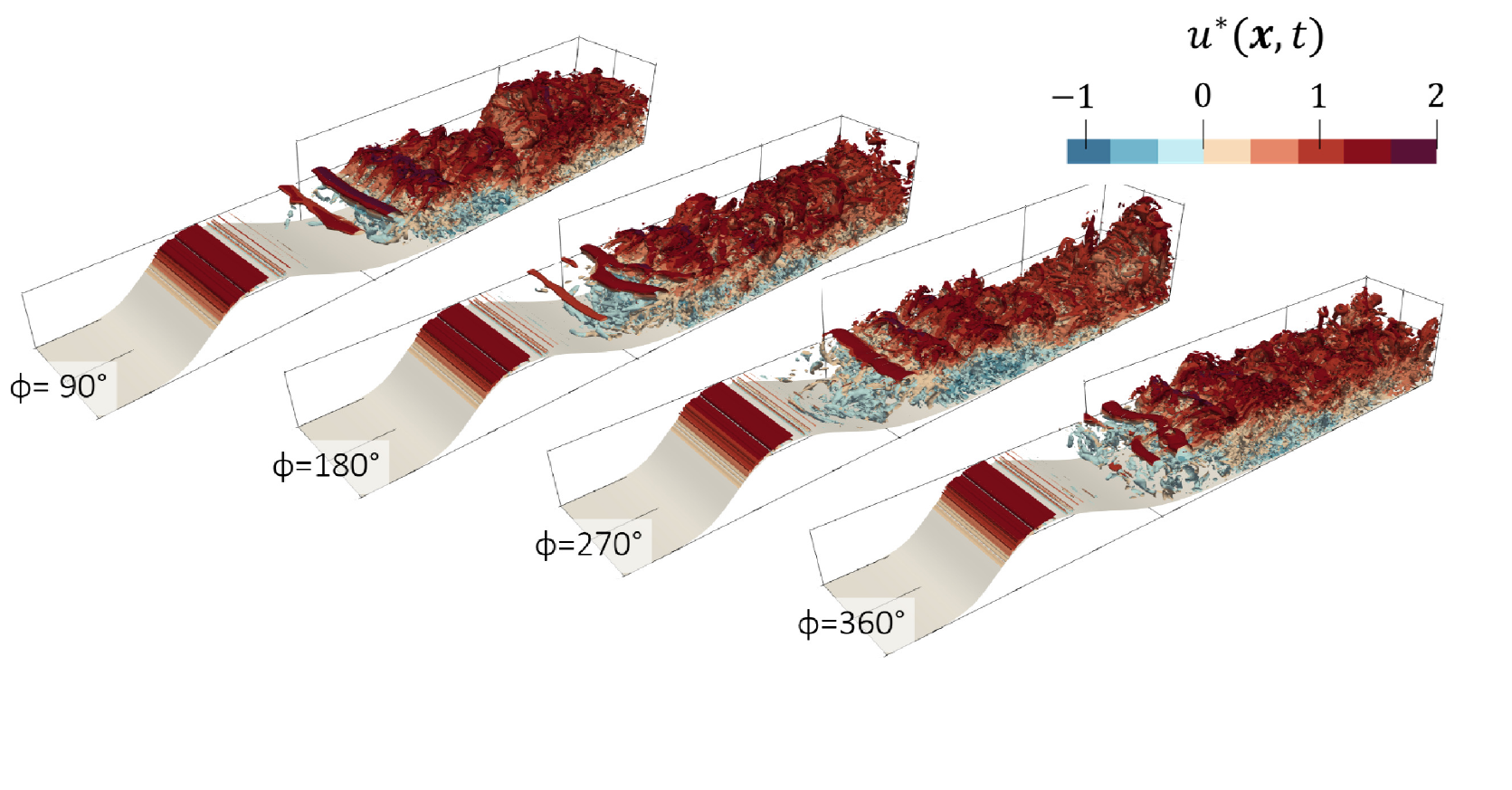}}
\caption{$Q(+)$ 
isosurface coloured by streamwise velocity. $A_{in}=0.01$ and $f_{in}^*=0.5$.  }
\label{fig:Qcriterion_A001_F05}
\end{figure}

\begin{figure}
\centering{\includegraphics[width=.8\textwidth,trim={0in 0.8in 0.4in 0in},clip] {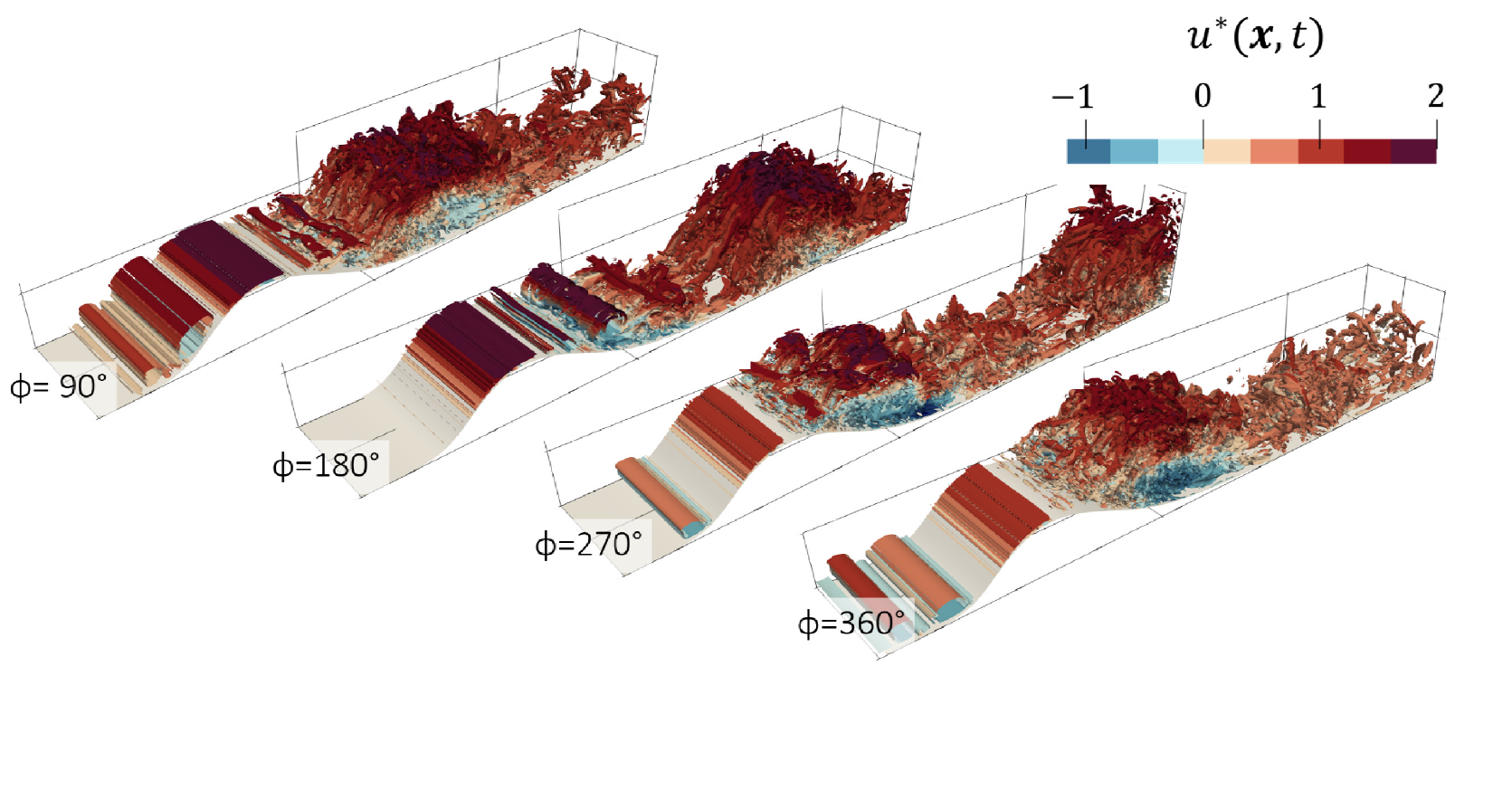}} 
\caption{$Q(+)$ isosurface coloured by streamwise velocity. $A_{in}=0.1$ and $f_{in}^*=2$.}
\label{fig:Qcriterion_A01_F2}
\end{figure}

\begin{figure}
\centering{
\includegraphics[width=.5\textwidth,trim={0.0in 2.3in 0in 0in},clip] {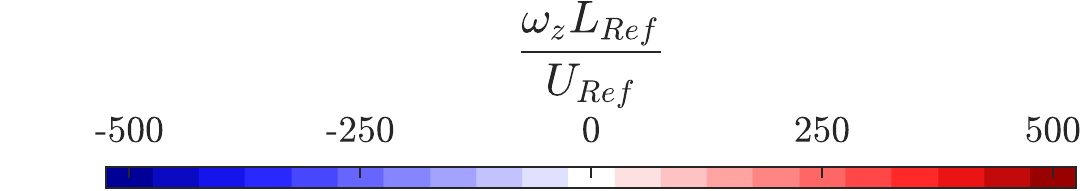}
\begin{tabular}{cc}
\includegraphics[width=.48\textwidth,trim={0.35in 0.88in 0.75in 1.5in},clip] {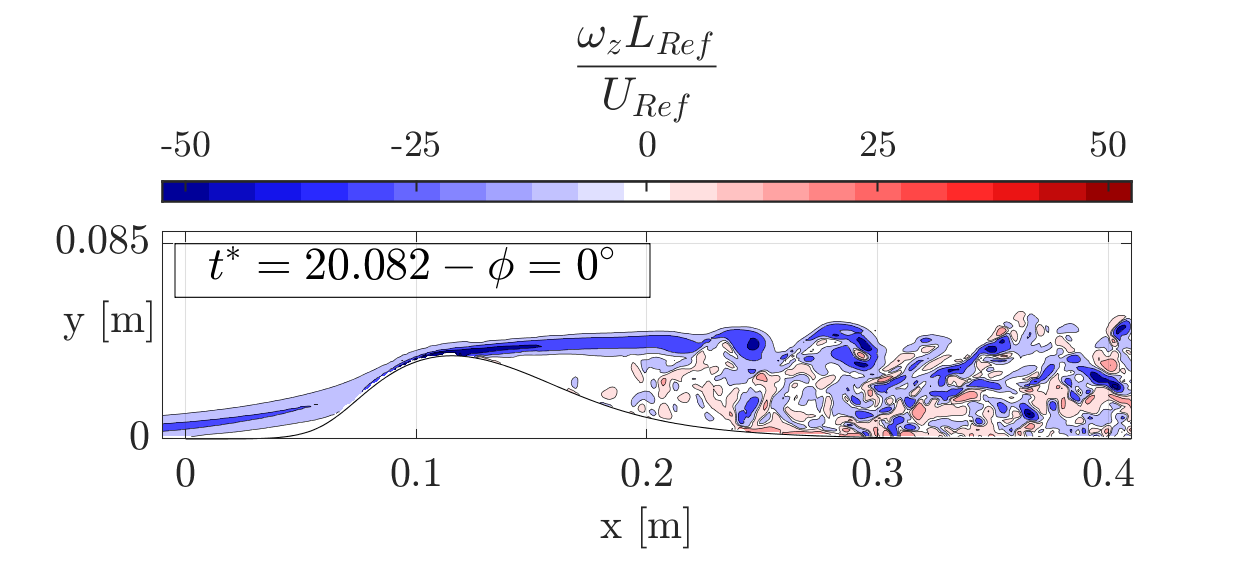} &
\includegraphics[width=.435\textwidth,trim={1.05in 0.88in 0.75in 1.5in},clip] {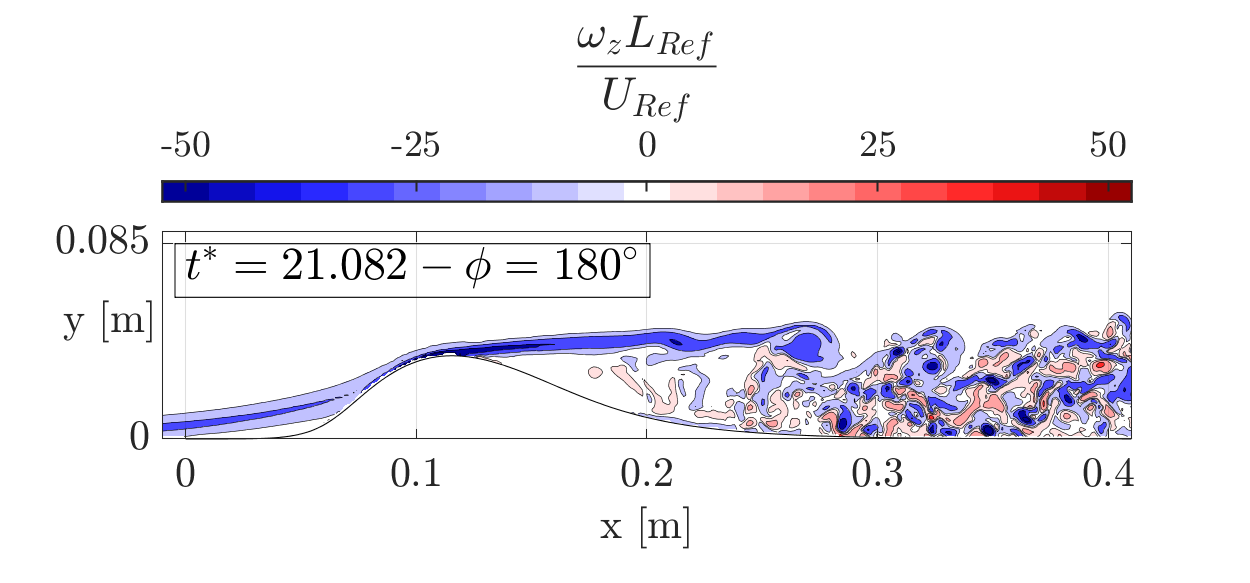} \\
\includegraphics[width=.48\textwidth,trim={0.35in 0.88in 0.75in 1.5in},clip] {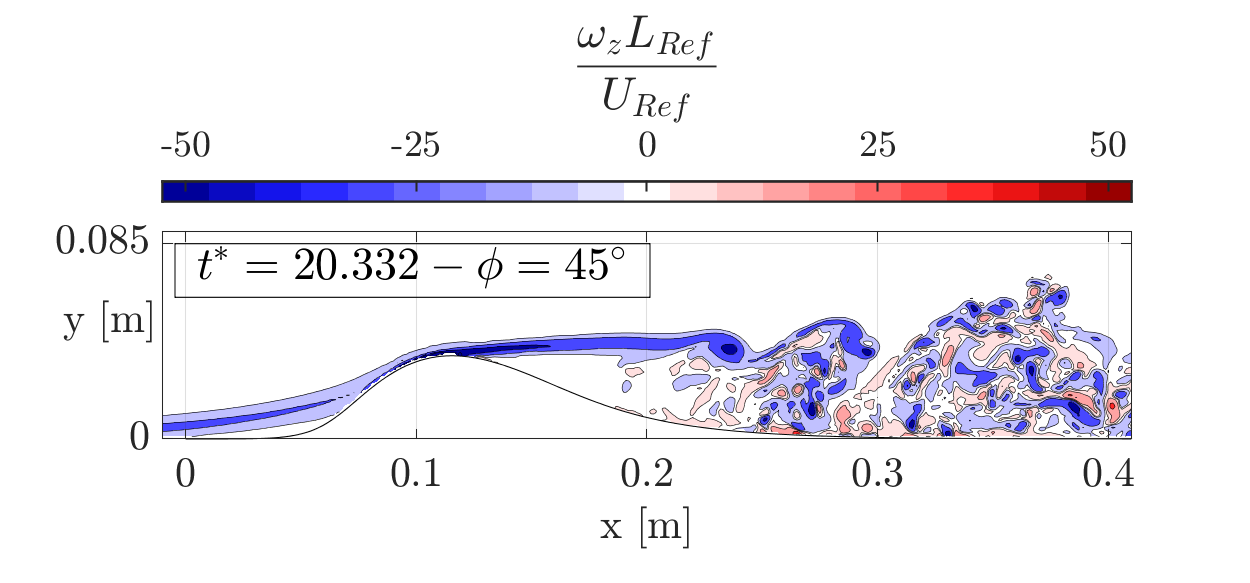} &
\includegraphics[width=.435\textwidth,trim={1.05in 0.88in 0.75in 1.5in},clip] {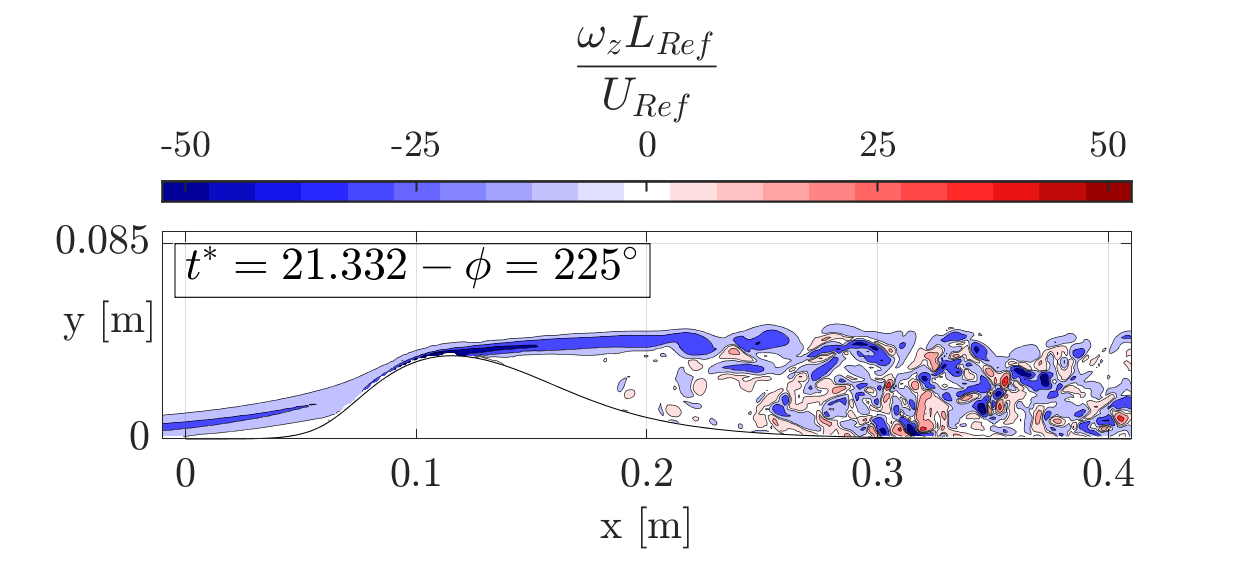} \\
\includegraphics[width=.48\textwidth,trim={0.35in 0.88in 0.75in 1.5in},clip] {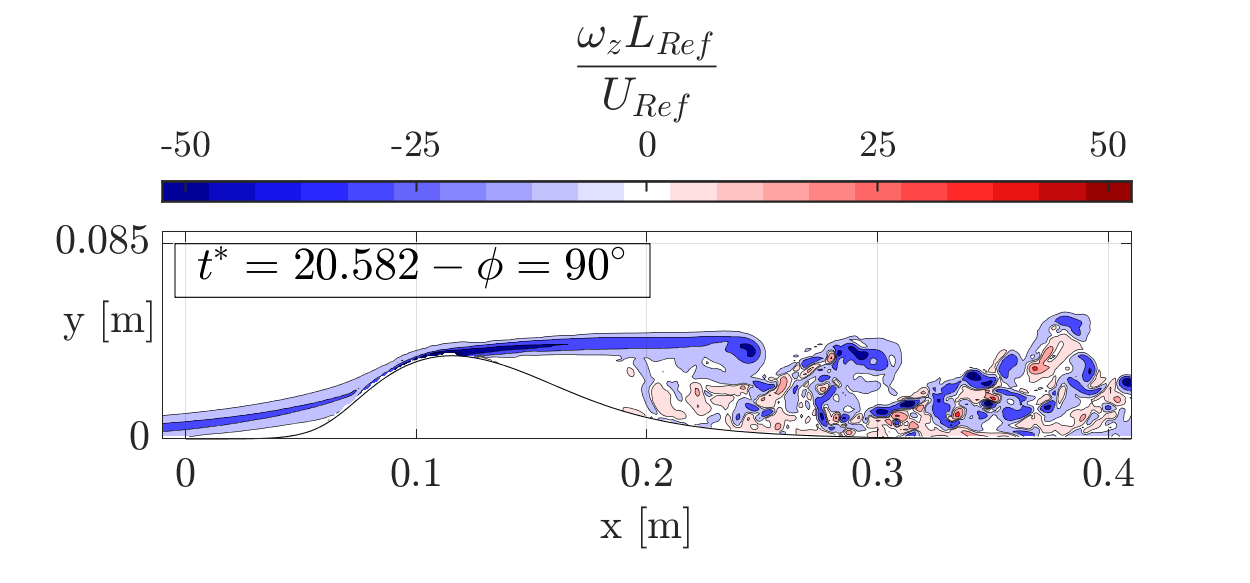} &
\includegraphics[width=.435\textwidth,trim={1.05in 0.88in 0.75in 1.5in},clip] {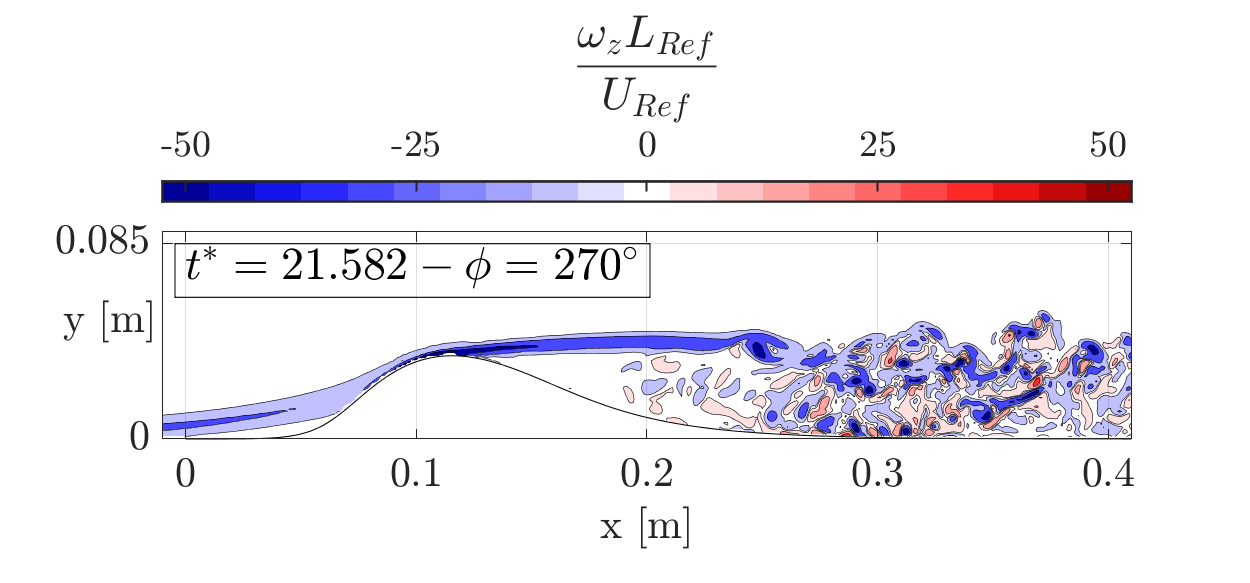} \\
\includegraphics[width=.48\textwidth,trim={0.35in 0.22in 0.75in 1.5in},clip] {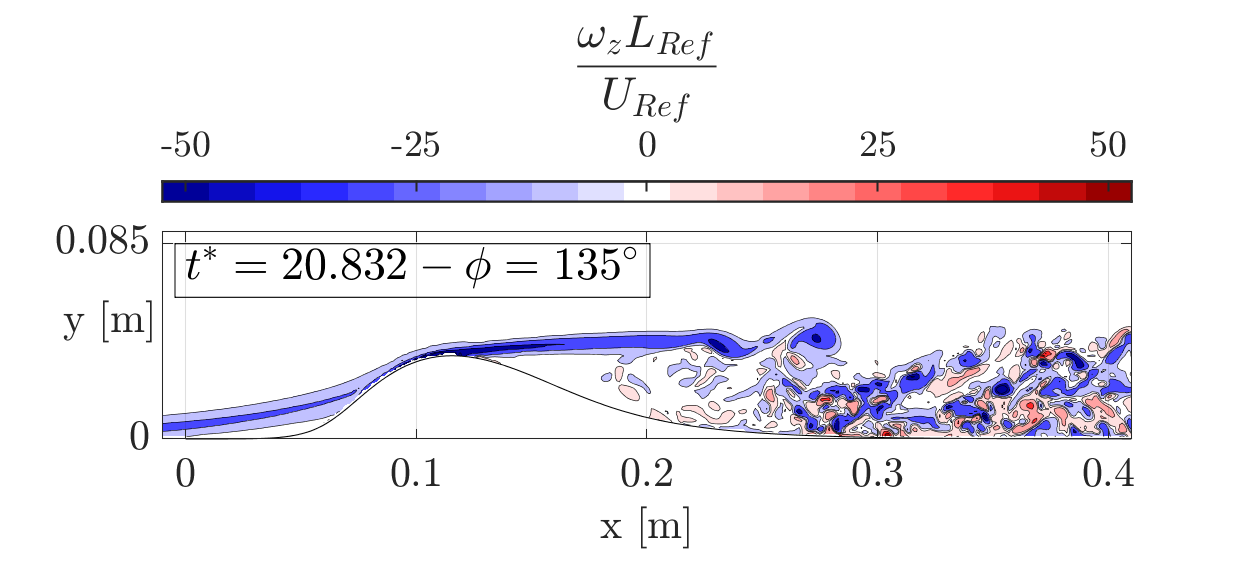} &
\includegraphics[width=.435\textwidth,trim={1.05in 0.22in 0.75in 1.5in},clip] {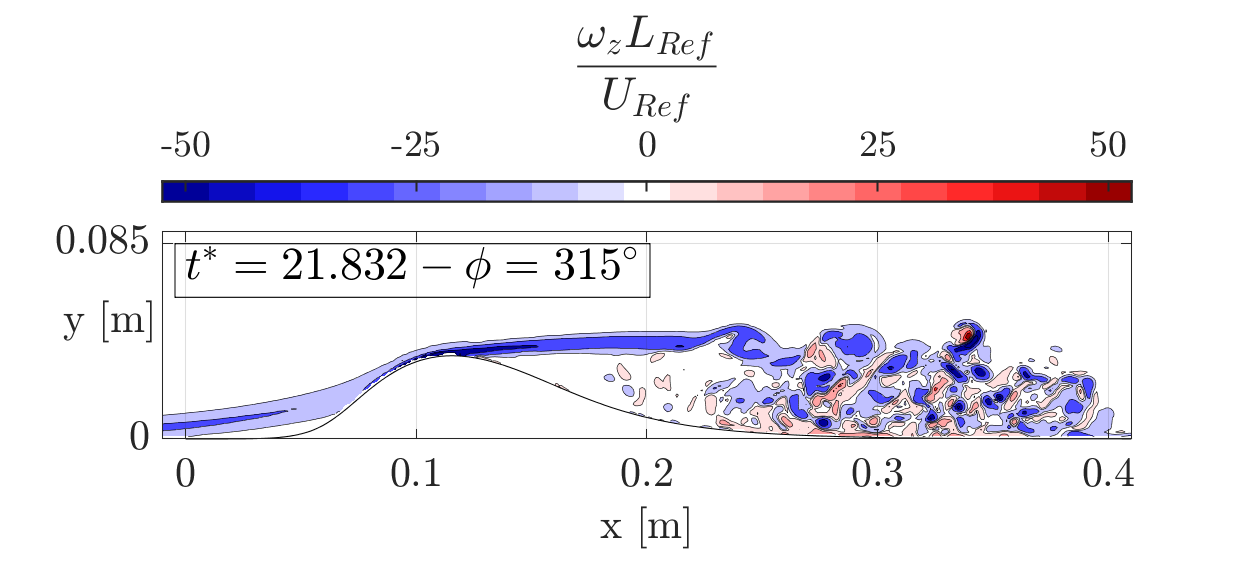} \\
\end{tabular}}
\caption{Instantaneous spanwise vorticity. $A_{in}=0.01$ and $f_{in}^{*}=0.5$. 
}
\label{fig:insta_vort_A001_f05}
\end{figure}

\begin{figure}
\centering{
\includegraphics[width=.5\textwidth,trim={0.0in 2.3in 0in 0in},clip] {instOmegaz_Scale10.eps}
\begin{tabular}{cc}

\includegraphics[width=.48\textwidth,trim={0.35in 0.88in 0.75in 1.5in},clip] {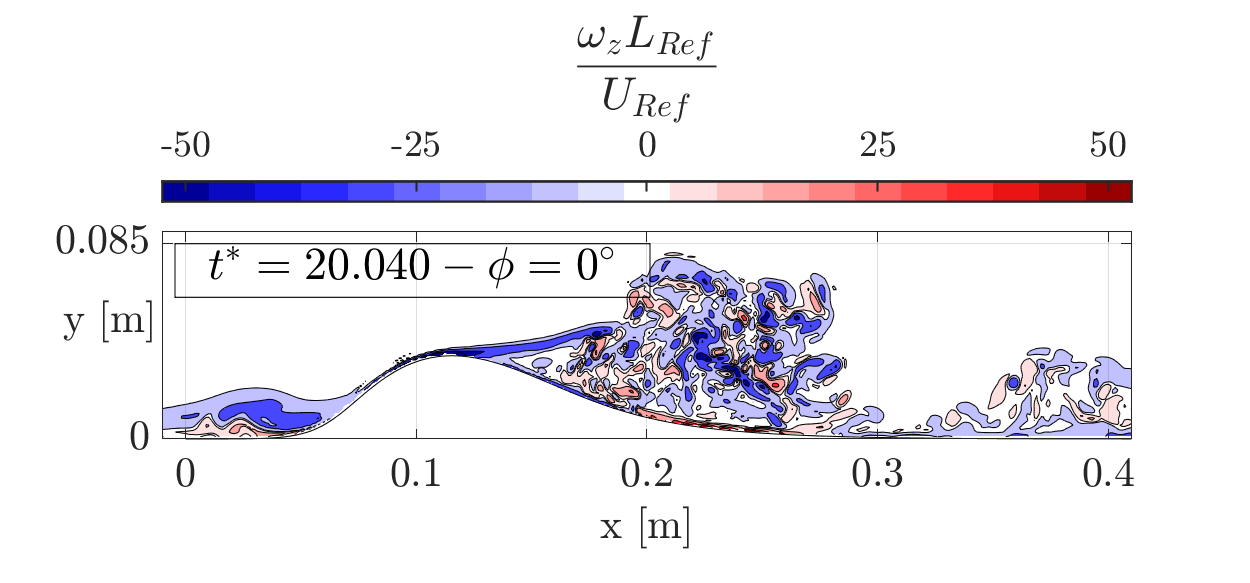} &
\includegraphics[width=.435\textwidth,trim={1.05in 0.88in 0.75in 1.5in},clip] {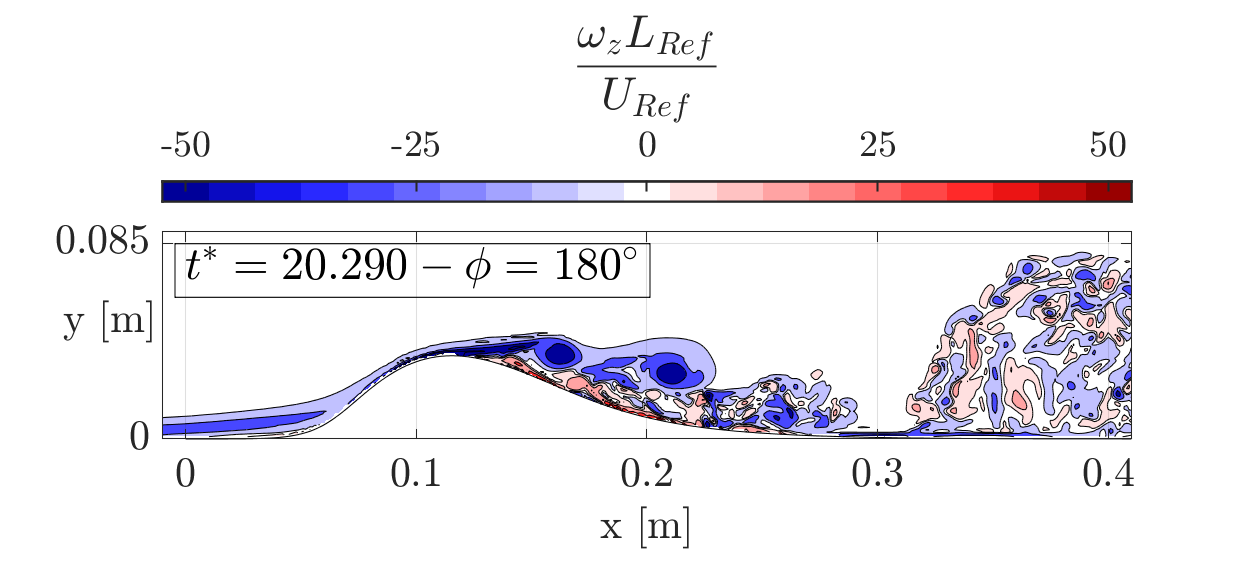} \\
\includegraphics[width=.48\textwidth,trim={0.35in 0.88in 0.75in 1.5in},clip] {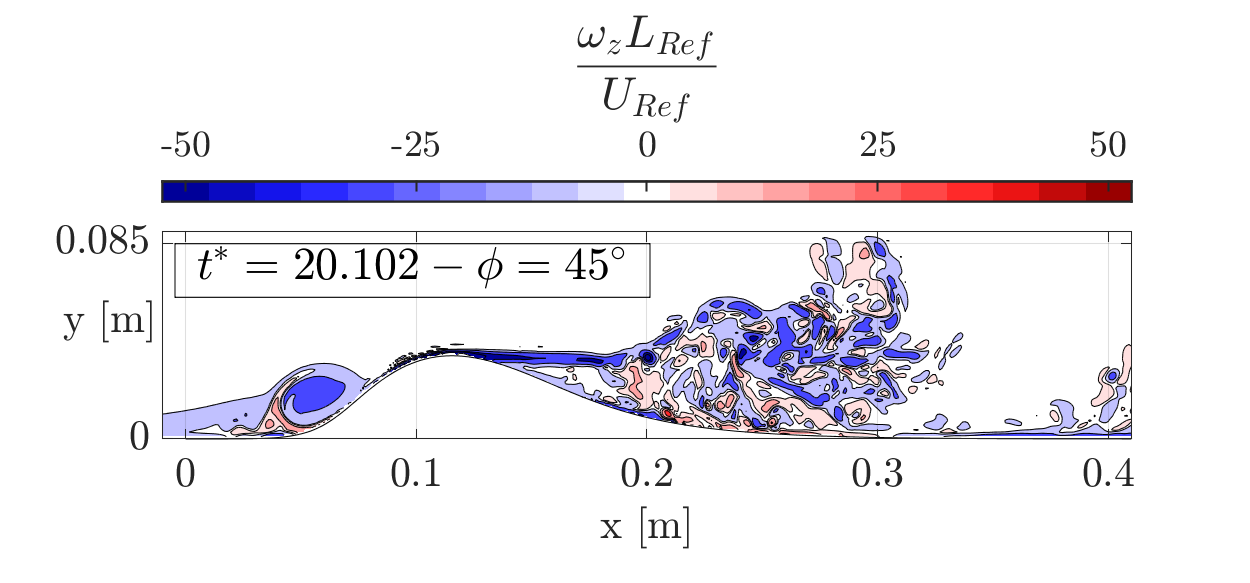} &
\includegraphics[width=.435\textwidth,trim={1.05in 0.88in 0.75in 1.5in},clip] {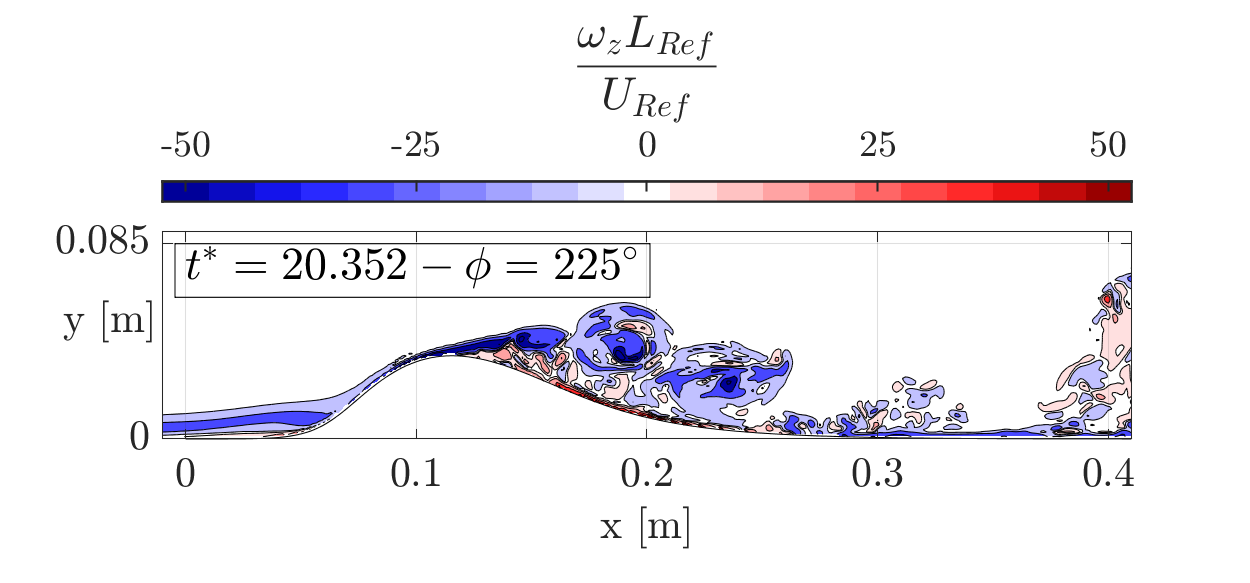} \\
\includegraphics[width=.48\textwidth,trim={0.35in 0.88in 0.75in 1.5in},clip] {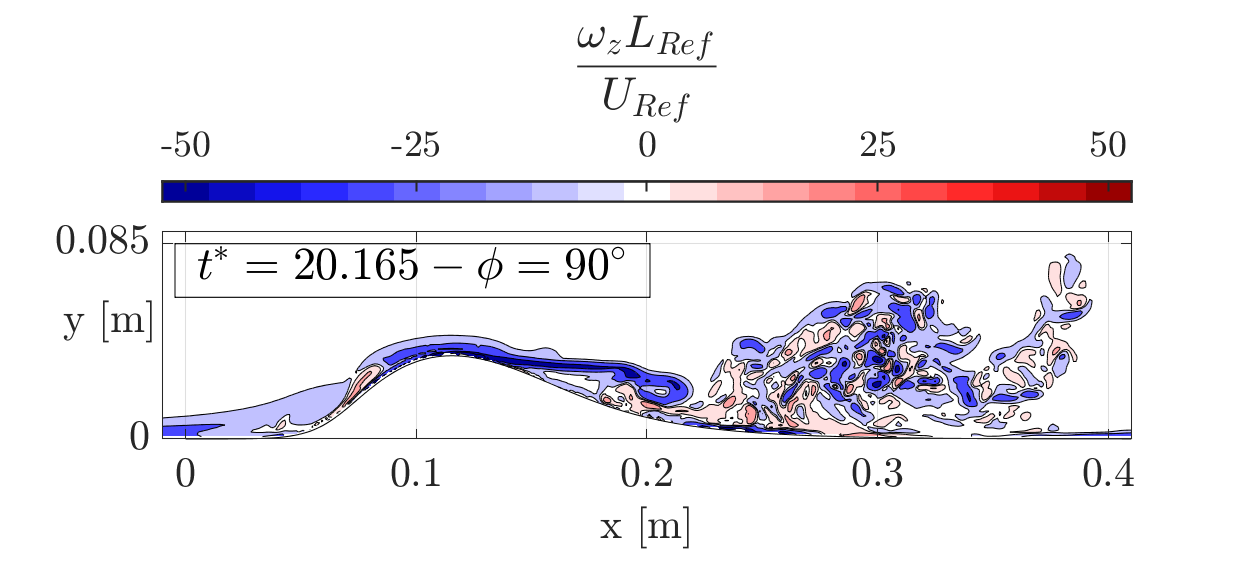} &
\includegraphics[width=.435\textwidth,trim={1.05in 0.88in 0.75in 1.5in},clip] {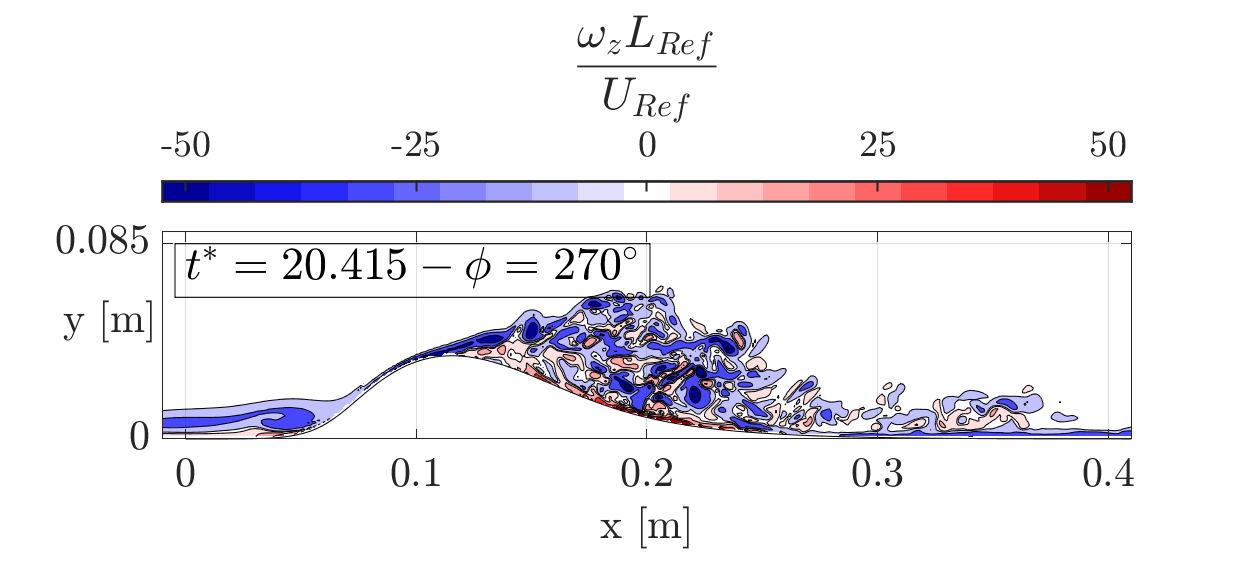} \\
\includegraphics[width=.48\textwidth,trim={0.35in 0.22in 0.75in 1.5in},clip] {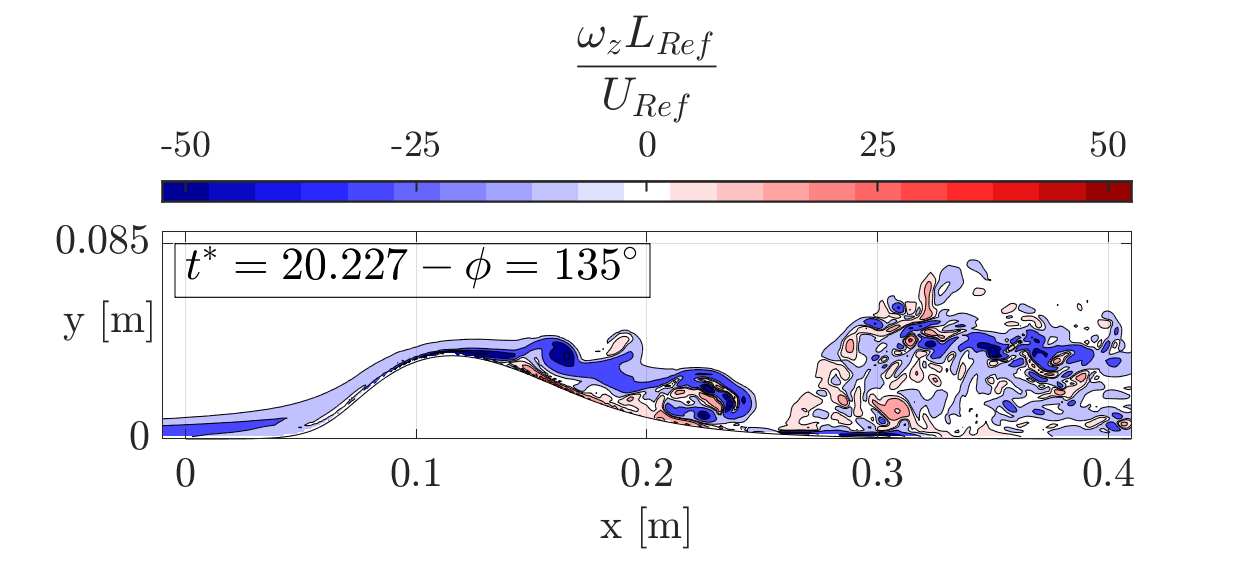} &
\includegraphics[width=.435\textwidth,trim={1.05in 0.22in 0.75in 1.5in},clip] {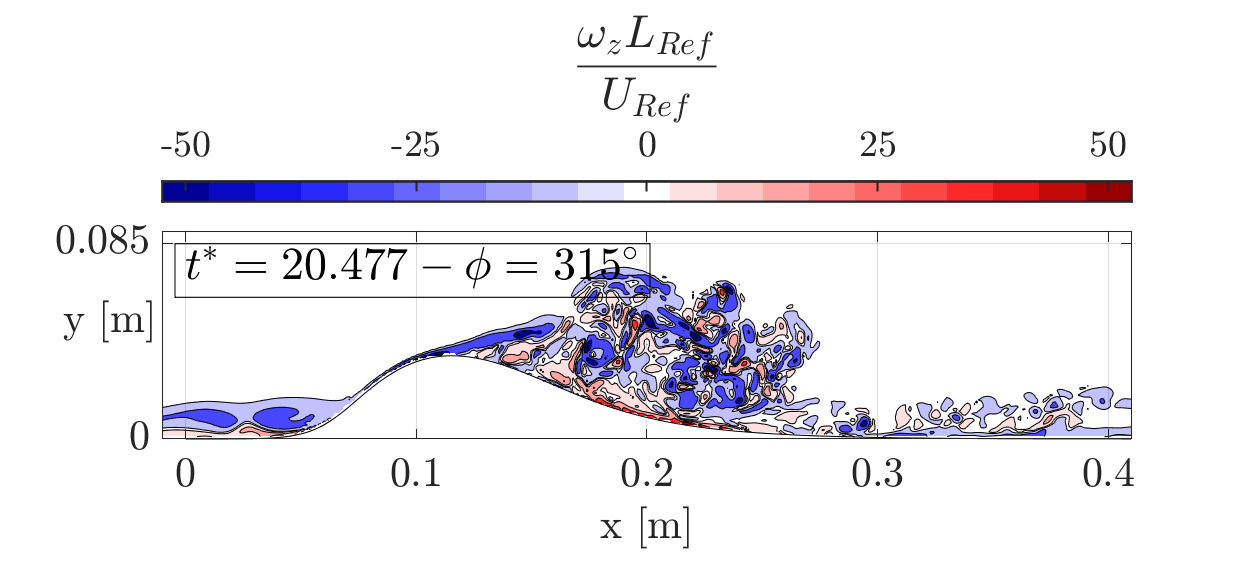} \\
\end{tabular}}
\caption{Instantaneous spanwise vorticity. $A_{in}=0.1$ and $f_{in}^{*}=2$.}
\label{fig:insta_vort_A01_f2}
\end{figure}

\subsection{Instantaneous flow fields}
\label{sec:instantaneous}

Figures \ref{fig:Qcriterion_A001_F05} and \ref{fig:Qcriterion_A01_F2} show the sequence of instantaneous three-dimensional visualizations of the vortical structures using the Q-criterion \citep{JeongHussain:JFM95} for two extreme cases, namely ($A_{in}=0.01$, $f_{in}^* = 0.5$) and ($A_{in}=0.1$, $f_{in}^* = 2$) (see supplementary movies 1 and 2). The isosurfaces, corresponding to the arbitrary value $Q=1000$, are colored using the streamwise velocity. Figures \ref{fig:insta_vort_A001_f05} and \ref{fig:insta_vort_A01_f2} show the respective spanwise vorticity fields at the mid-spanwise plane. In addition to the dimensionless time, each subfigure also indicates the corresponding phase $\phi$. 
The three-dimensional visualizations show that the flow is fully two-dimensional upstream of the bump summit, even in the presence of vortices in the upstream part of the bump. This rules out the presence of G\"ortler \citep{Marxen:JFM09} and centrifugal global instabilities \citep{THD,Rodriguez:JFM13} and ensures that the flow remains laminar and two-dimensional at separation in all cases.

The first case ($A_{in}=0.01$, $f_{in}^* = 0.5$) corresponds to the weakest inflow oscillations, i.e. lowest inflow frequency and amplitude, and results in a mean recirculation length virtually identical to that of the steady case. For this case, Fig. \ref{fig:insta_vort_A001_f05} shows a continuous shedding of KH vortices from the separated shear layer followed by a fast transition to turbulence and recirculation of vortical structures, qualitatively identical to the steady case (cf. Fig. \ref{fig:sketch_bump}). However, careful observation of the separated shear layer upstream of the vortices shows some differences that gradually become more relevant for cases with increasing $A_{in}$ and $f_{in}^*$. For $\phi = 0^\circ$ the reference total pressure starts to increase above its mean value, exerting an acceleration of the flow along the channel. Due to the increased mass and momentum fluxes, the separated shear layer is pushed towards the bump wall. As the flow is accelerated, the high vorticity region associated with the laminar-turbulent transition is also pushed towards the wall. The maximum inflow pressure occurs at $\phi = 90^\circ$, followed closely by the inflow bulk velocity. 
For $\phi$ between $90^\circ$ and $270^\circ$ the reference pressure is reduced up to its minimum value, imposing a gradual deceleration of the bulk flow. The separated shear layer moves away from the wall; the high vorticity region seems to detach from the wall and the apparent recirculation region extends farther in the streamwise direction.
Finally, when $\phi > 270^\circ$ the flow gradually re-accelerates closing the period.

The second case ($A_{in}=0.1$, $f_{in}^* = 2$) corresponds to the strongest inflow oscillation, i.e. the largest inflow oscillation frequency and amplitude, and results in a substantial reduction of the mean recirculation region ($\Delta L_s / L_{s,steady} \approx -44\%$). For this case, Fig. \ref{fig:insta_vort_A01_f2} does not show a periodic shedding of KH vortices from the separated shear layer, but rather the formation and release of a big vortex cluster following the harmonic change in the inflow conditions.

These observations suggest that two closely related but different physical mechanisms are at play when the inflow has a harmonic time dependence with relatively small $A_{in}$ and $f^*_{in}$. First, the periodic acceleration-deceleration of the bulk flow modifies the angle of the separated shear layer through changes in the intensity of the streamwise pressure gradient. 
The location of the separation point is nearly unaffected by this. 
Second, the periodic vertical motion of the shear layer can impact its hydrodynamic instability properties and the ensuing dynamics of the KH vortices and the laminar-turbulent transition. As will be quantified later, the frequency of the KH vortex shedding is distinctly separated from the frequency of the inflow changes. 
On the other hand, for relatively large $A_{im}$ and $f^*_{in}$, the organized shedding of KH vortices from the separated shear layer is replaced by a periodic formation and ejection of large vortex clusters that is driven by the bulk flow acceleration and deceleration. Intermediate cases are expected to show a transition from one behaviour to the other, as will be detailed in the following sections. 

With the aim of isolating the shedding of KH vortices from the bulk flow oscillations imposed by the inlet frequency, the triple decomposition proposed by \citet{Hussain70} is applied. This decomposition takes the form

\begin{equation}
{q}({\boldsymbol x},t) = {\bar{q}}({\boldsymbol x})+{\tilde{q}}({\boldsymbol x},t)+{q'}({\boldsymbol x},t),
\end{equation} 

\noindent where $\bar{q}$ stands for the mean (time-averaged) flow, $\tilde{q}$ is the oscillatory component coherent with the inflow oscillation and $q'$ is the incoherent component. The term ``coherent'' refers to flow fluctuations that occur in phase with the harmonic changes of the total pressure at the reference point. As such, the mean plus coherent components are gathered together in the phase-averaged flow

\begin{equation}
\langle{q}({\boldsymbol x},\phi)\rangle = \frac{1}{N} \sum_{n=0}^{N} {q} \left({\boldsymbol x},t_{\phi}+\frac{n}{f^*} \right),
\end{equation}

\noindent where $t_\phi$ is the time used as the phase reference, $f^*$ is the inlet frequency and $N$ is the number of periods used in the averaging. The incoherent part of the flow is computed as ${q}'({\boldsymbol x},t) = {q}({\boldsymbol x},t) - \langle {q}({\boldsymbol x},t)\rangle$. The same dimensionless time-lapse was used in the averaging for the three frequencies $f^*_{in}$, resulting in $N = 25$ for $f_{in}^* = 0.5$, $N = 50$ for $f_{in}^* = 1$ and $N= 100$ for $f_{in}^*= 2$.

\subsection{Phase-averaged flow fields}
\label{sec:Phase-averaged}

\begin{figure}
\centering{
\includegraphics[width=.5\textwidth,trim={0.0in 2.3in 0in 0in},clip] {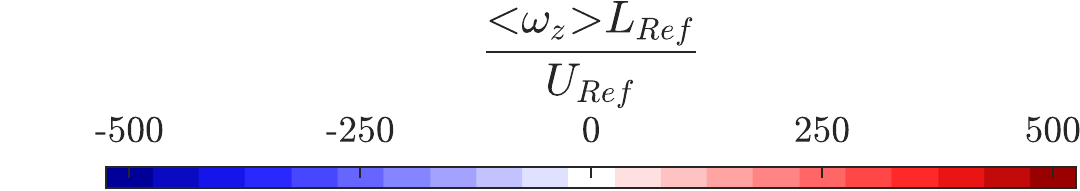} 
\begin{tabular}{cc}
\includegraphics[width=.48\textwidth,trim={0.35in 0.88in 0.75in 1.5in},clip] {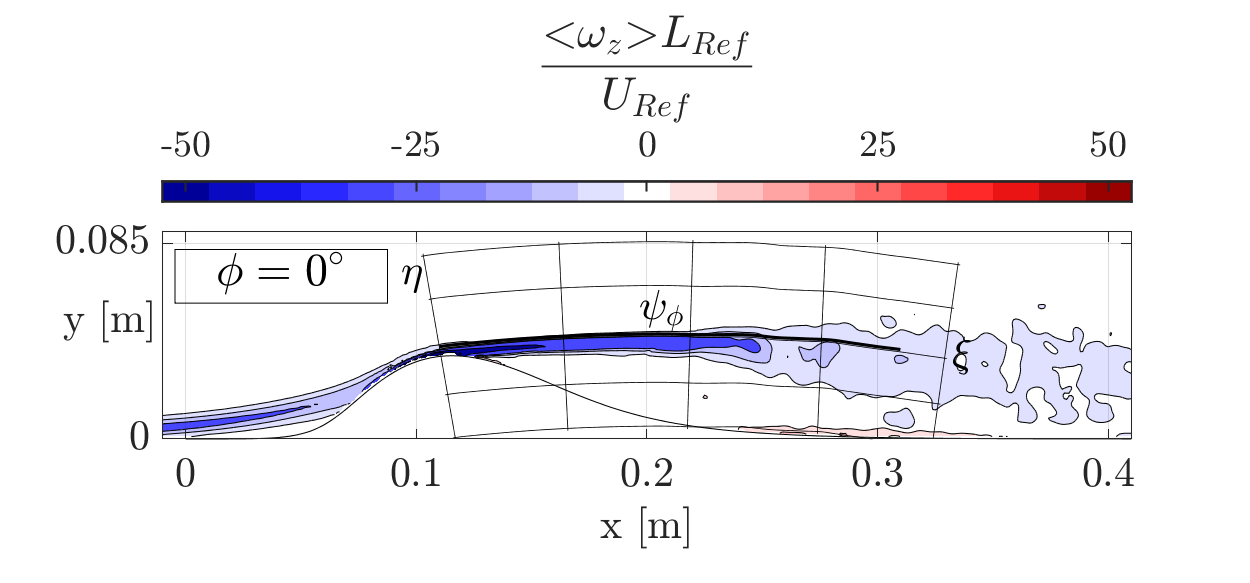} &
\includegraphics[width=.435\textwidth,trim={1.05in 0.88in 0.75in 1.5in},clip] {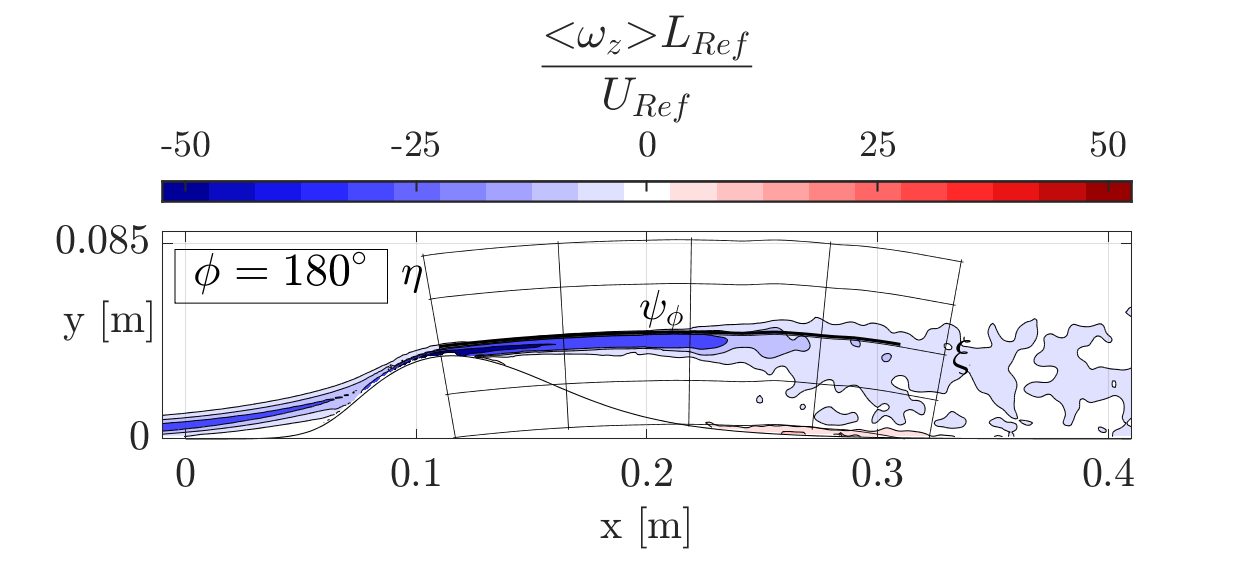} \\
\includegraphics[width=.48\textwidth,trim={0.35in 0.88in 0.75in 1.5in},clip] {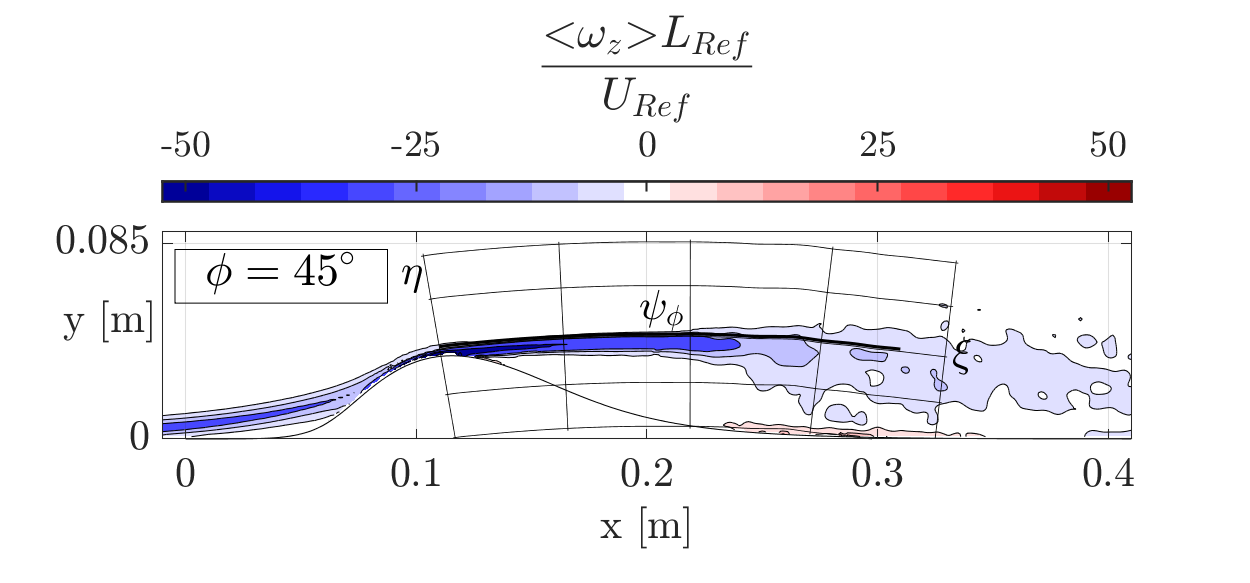} &
\includegraphics[width=.435\textwidth,trim={1.05in 0.88in 0.75in 1.5in},clip] {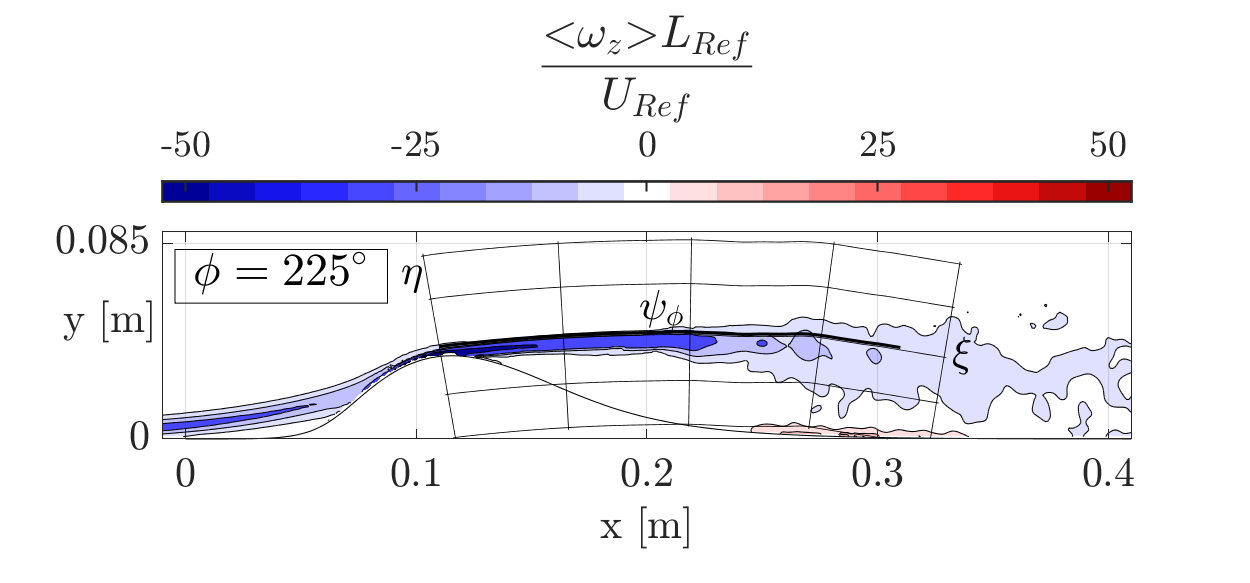} \\
\includegraphics[width=.48\textwidth,trim={0.35in 0.88in 0.75in 1.5in},clip] {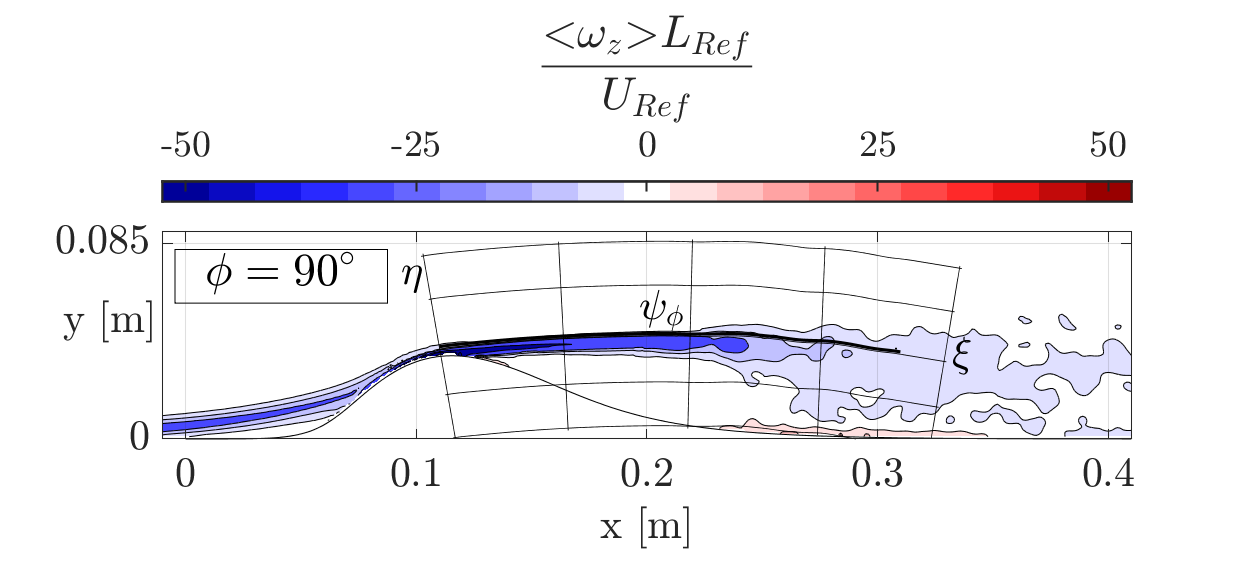} &
\includegraphics[width=.435\textwidth,trim={1.05in 0.88in 0.75in 1.5in},clip] {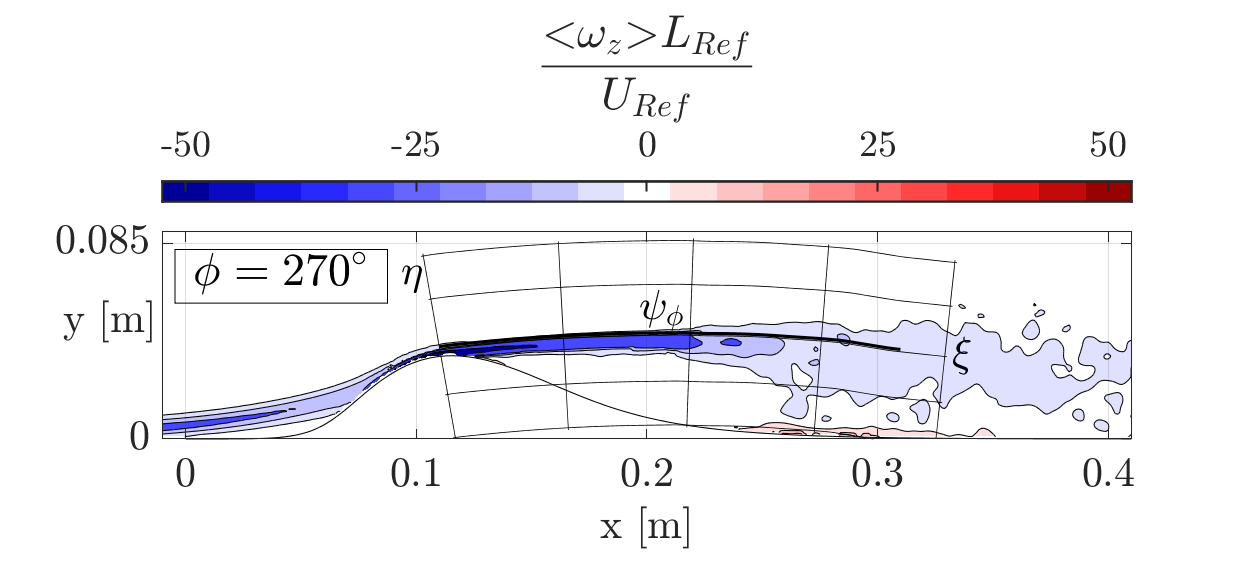} \\
\includegraphics[width=.48\textwidth,trim={0.35in 0.22in 0.75in 1.5in},clip] {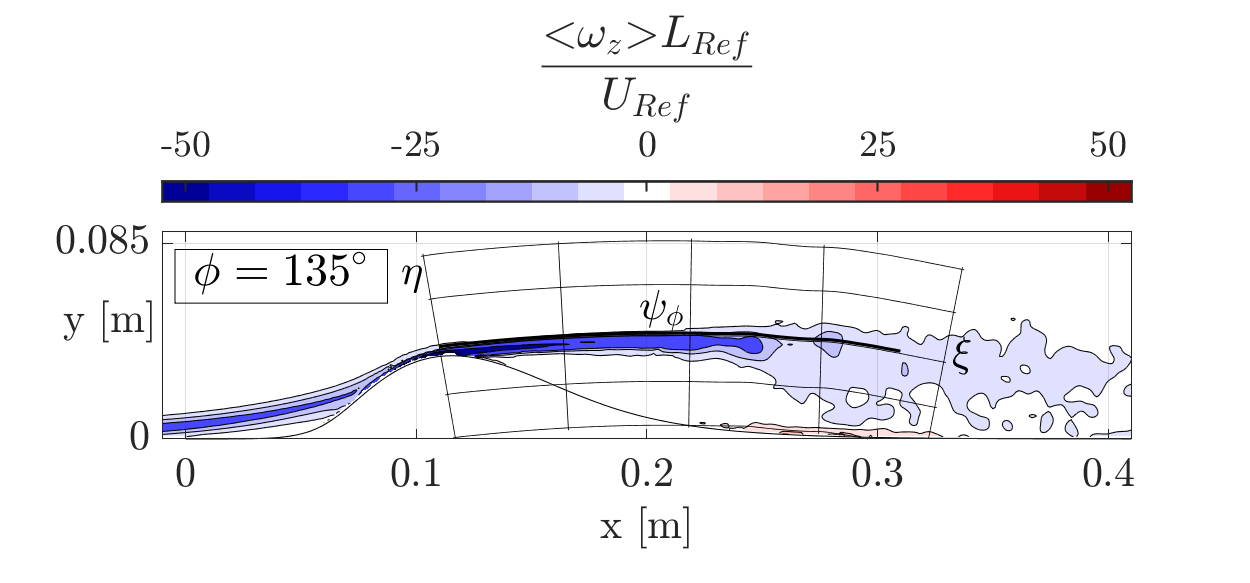} &
\includegraphics[width=.435\textwidth,trim={1.05in 0.22in 0.75in 1.5in},clip] {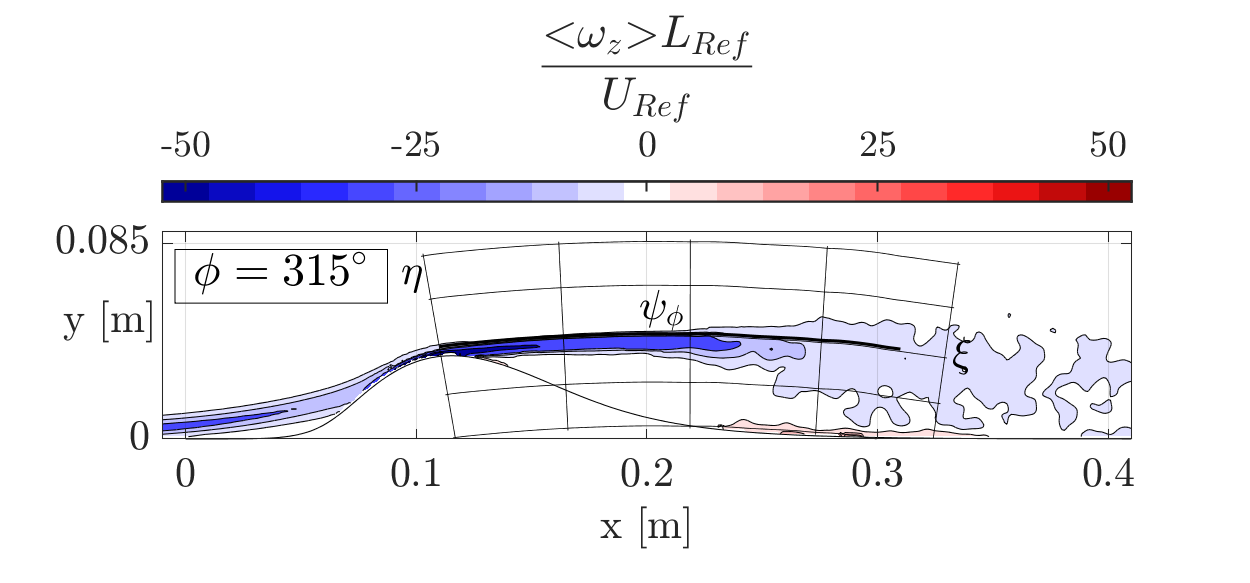} \\
\end{tabular}}
\caption{Phase-averaged spanwise vorticity. $A_{in}=0.01$ and $f_{in}^{*}=0.5$. }
\label{fig:phase_average_f05_A001}
\end{figure}

\begin{figure}
\centering{
\includegraphics[width=.5\textwidth,trim={0.0in 2.3in 0in 0in},clip] {phaseMeanPub_Scale10.eps} 
\begin{tabular}{cc}
\includegraphics[width=.48\textwidth,trim={0.35in 0.88in 0.75in 1.5in},clip] {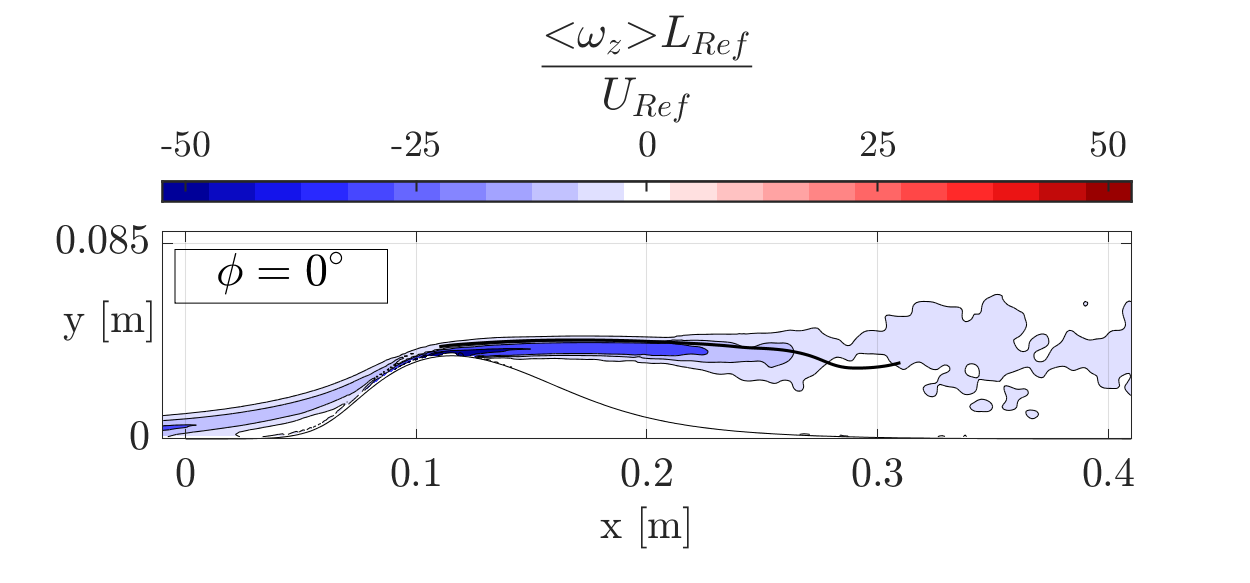} &
\includegraphics[width=.435\textwidth,trim={1.05in 0.88in 0.75in 1.5in},clip] {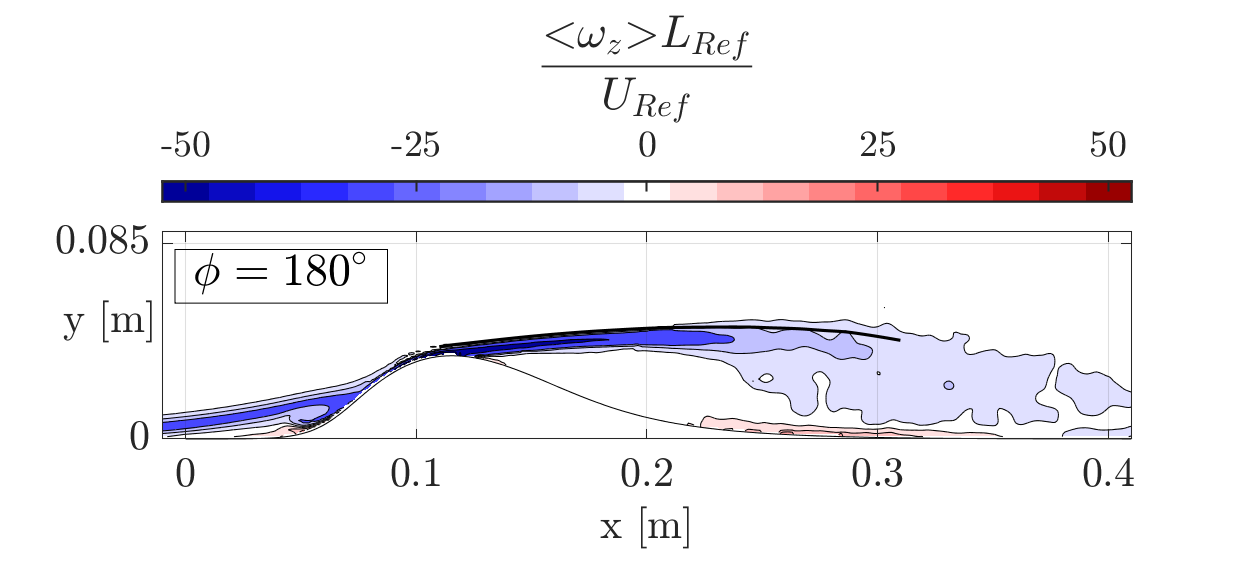} \\
\includegraphics[width=.48\textwidth,trim={0.35in 0.88in 0.75in 1.5in},clip] {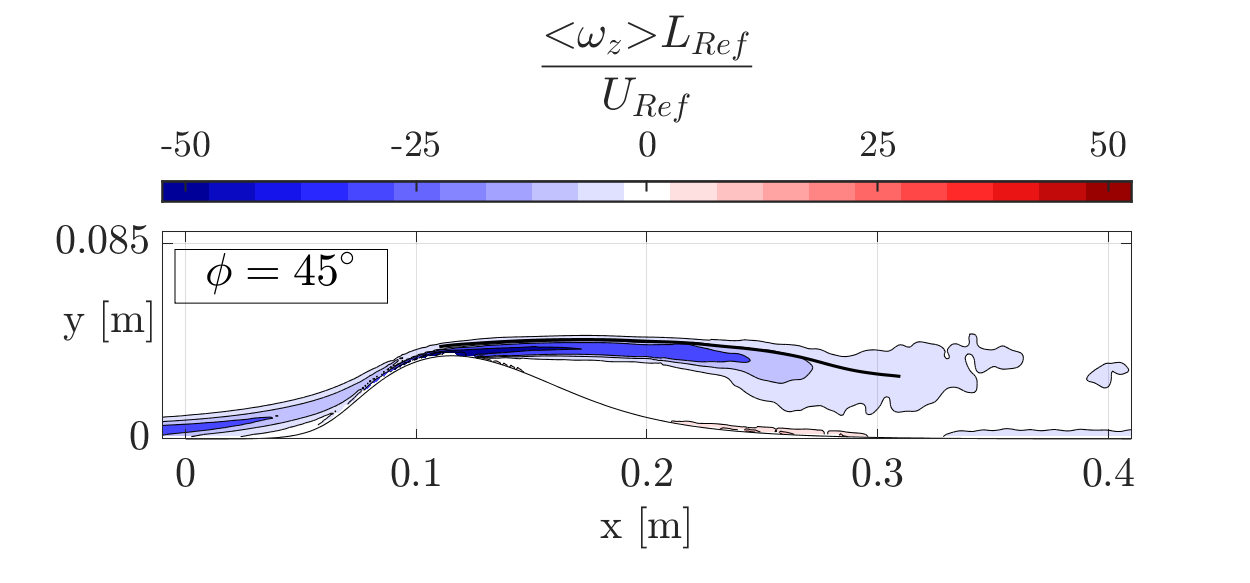} &
\includegraphics[width=.435\textwidth,trim={1.05in 0.88in 0.75in 1.5in},clip] {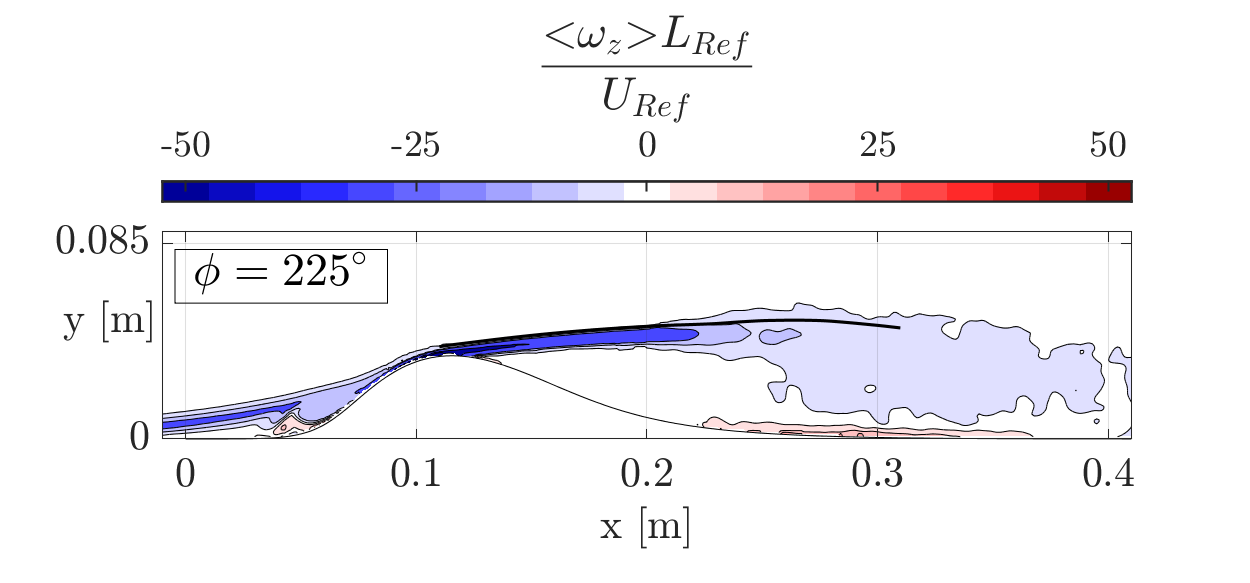} \\
\includegraphics[width=.48\textwidth,trim={0.35in 0.88in 0.75in 1.5in},clip] {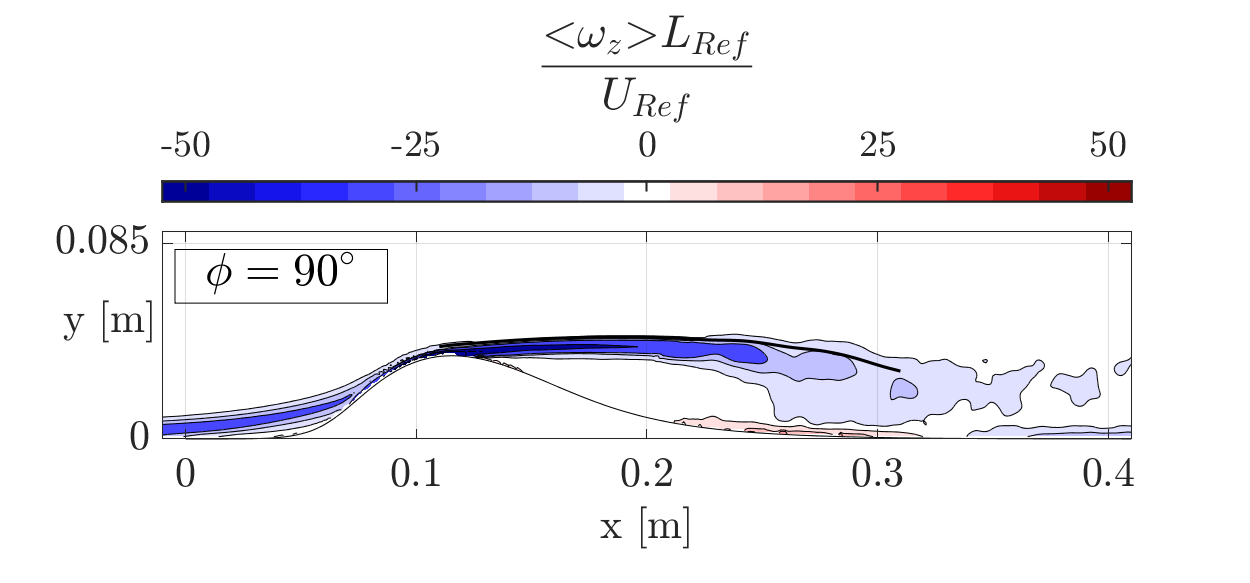} &
\includegraphics[width=.435\textwidth,trim={1.05in 0.88in 0.75in 1.5in},clip] {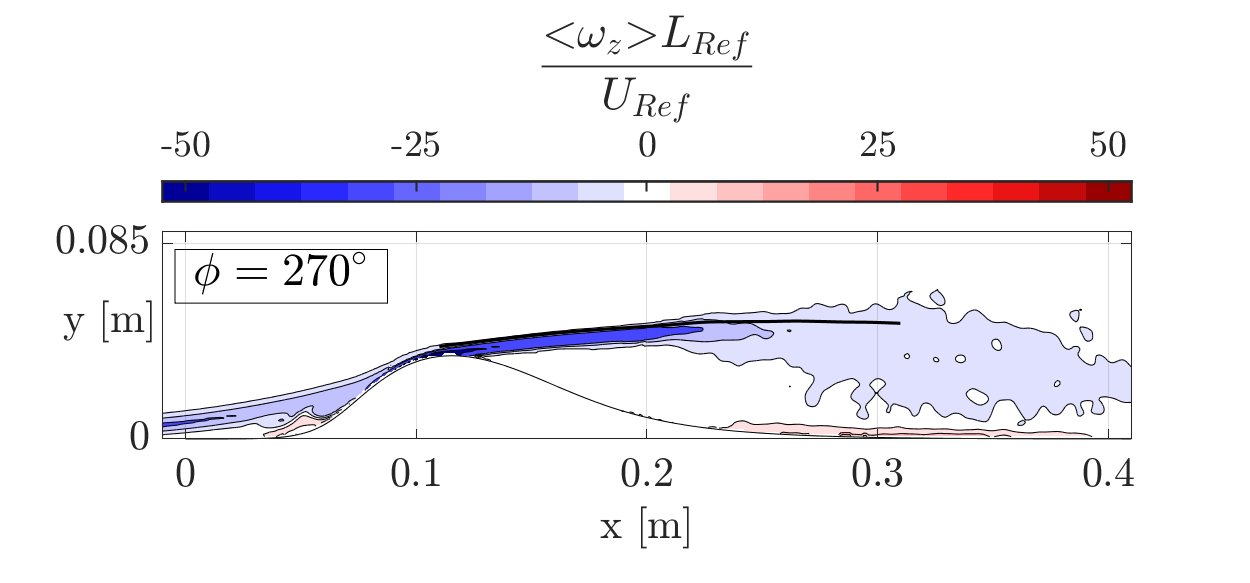} \\
\includegraphics[width=.48\textwidth,trim={0.35in 0.22in 0.75in 1.5in},clip] {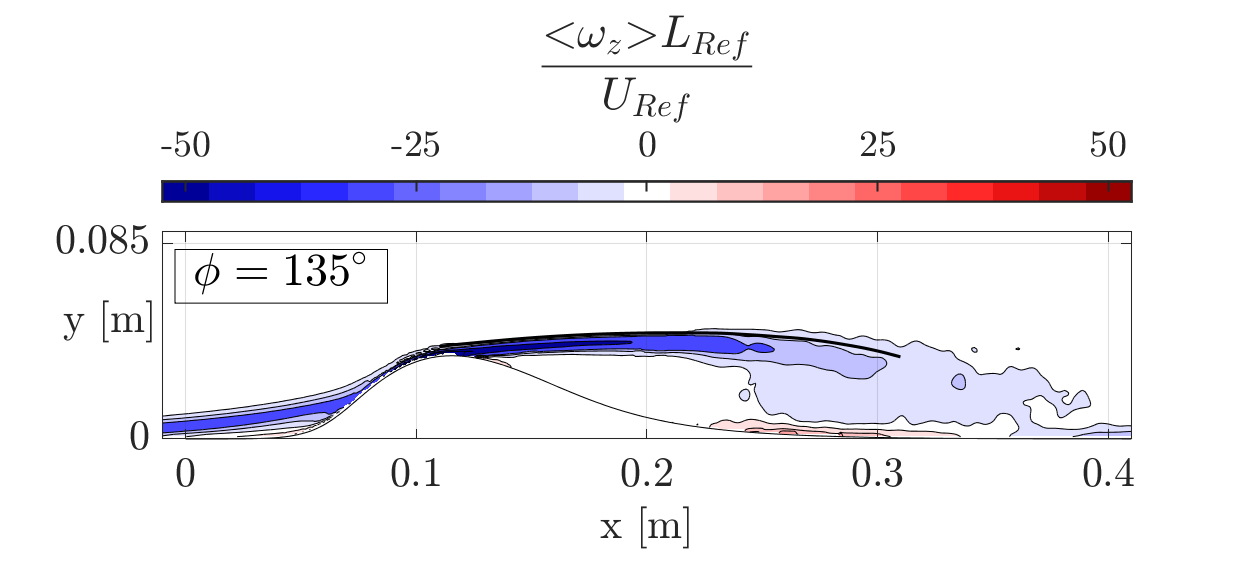} &
\includegraphics[width=.435\textwidth,trim={1.05in 0.22in 0.75in 1.5in},clip] {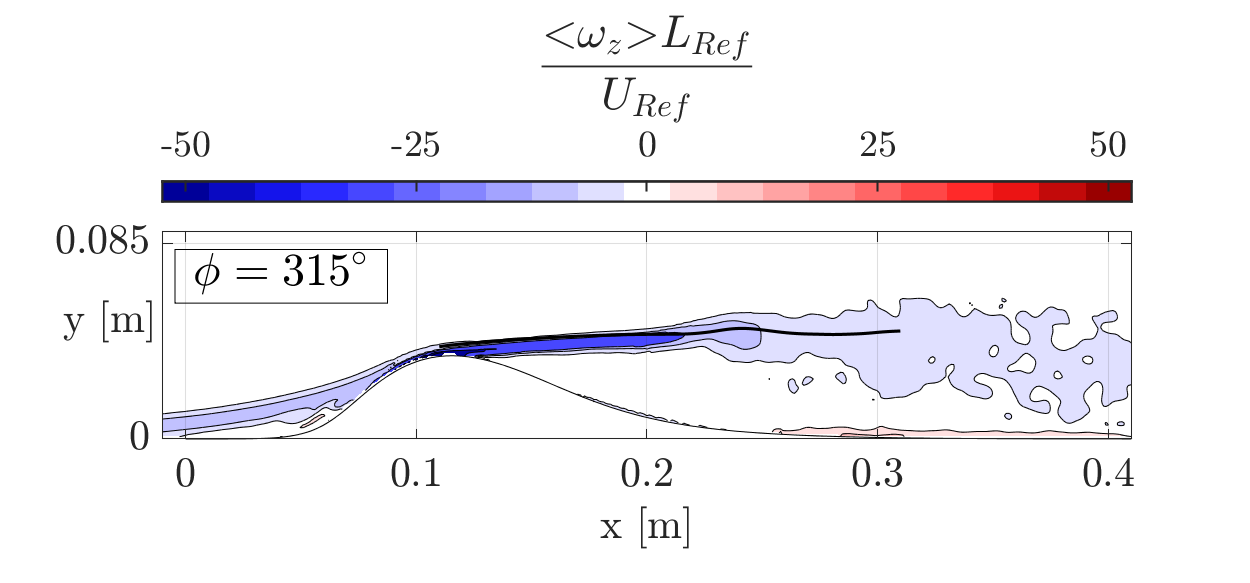} \\
\end{tabular}}
\caption{Phase-averaged spanwise vorticity.  $A_{in}=0.05$ and $f_{in}^{*}=1$.}
\label{fig:phase_average_f1_A005}
\end{figure}

\begin{figure}
\centering{
\includegraphics[width=.5\textwidth,trim={0.0in 2.3in 0in 0in},clip] {phaseMeanPub_Scale10.eps}
\begin{tabular}{cc}
\includegraphics[width=.48\textwidth,trim={0.35in 0.88in 0.75in 1.5in},clip] {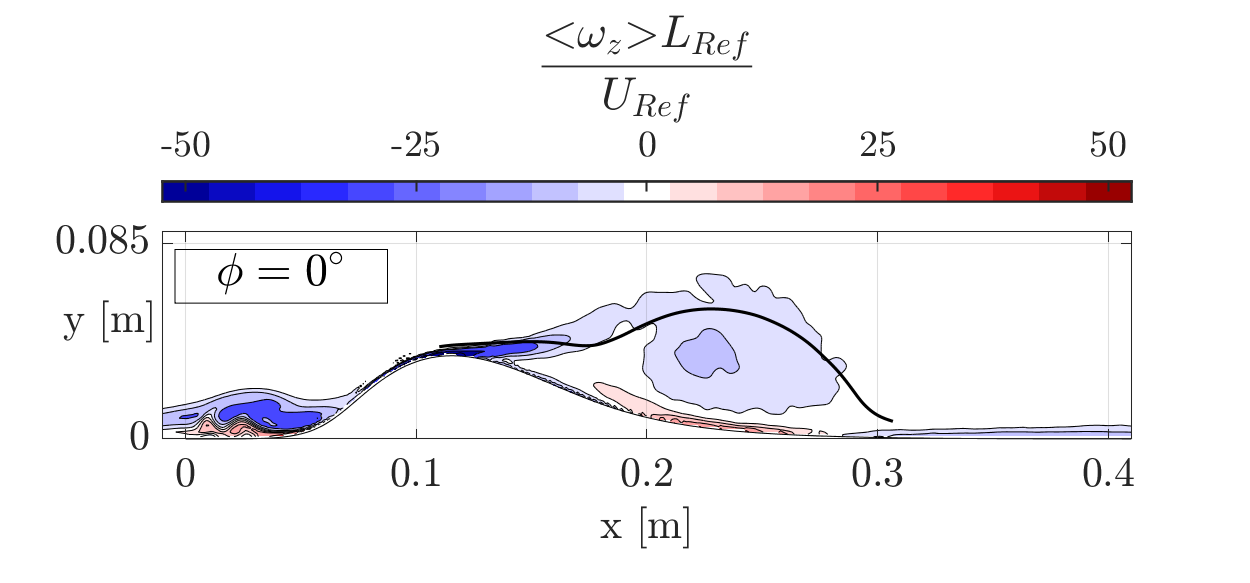} &
\includegraphics[width=.435\textwidth,trim={1.05in 0.88in 0.75in 1.5in},clip] {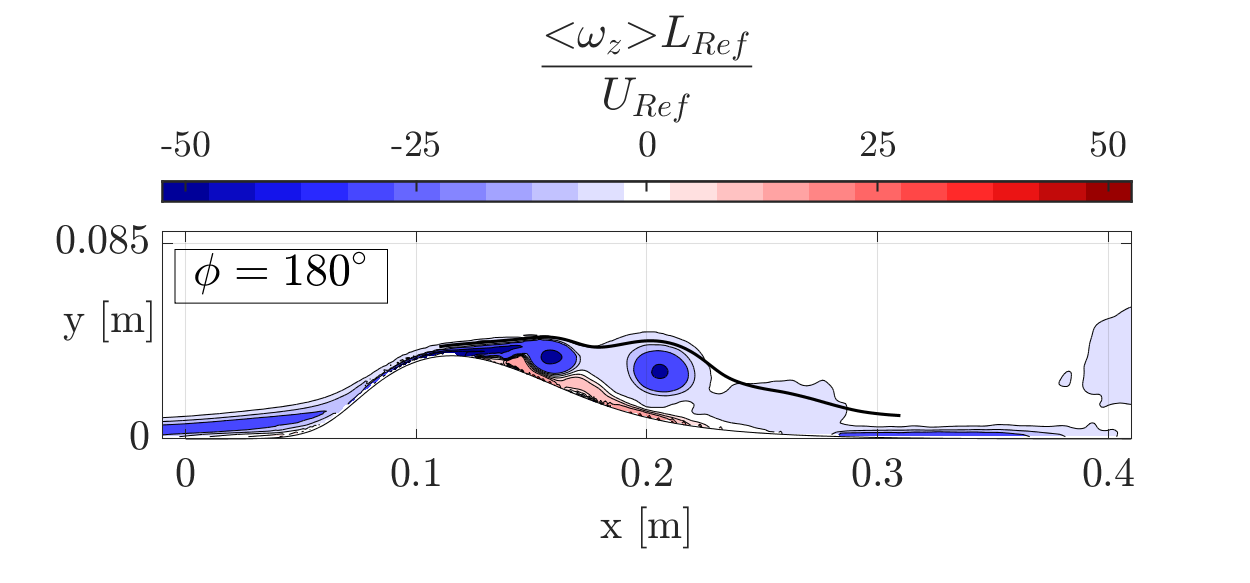} \\
\includegraphics[width=.48\textwidth,trim={0.35in 0.88in 0.75in 1.5in},clip] {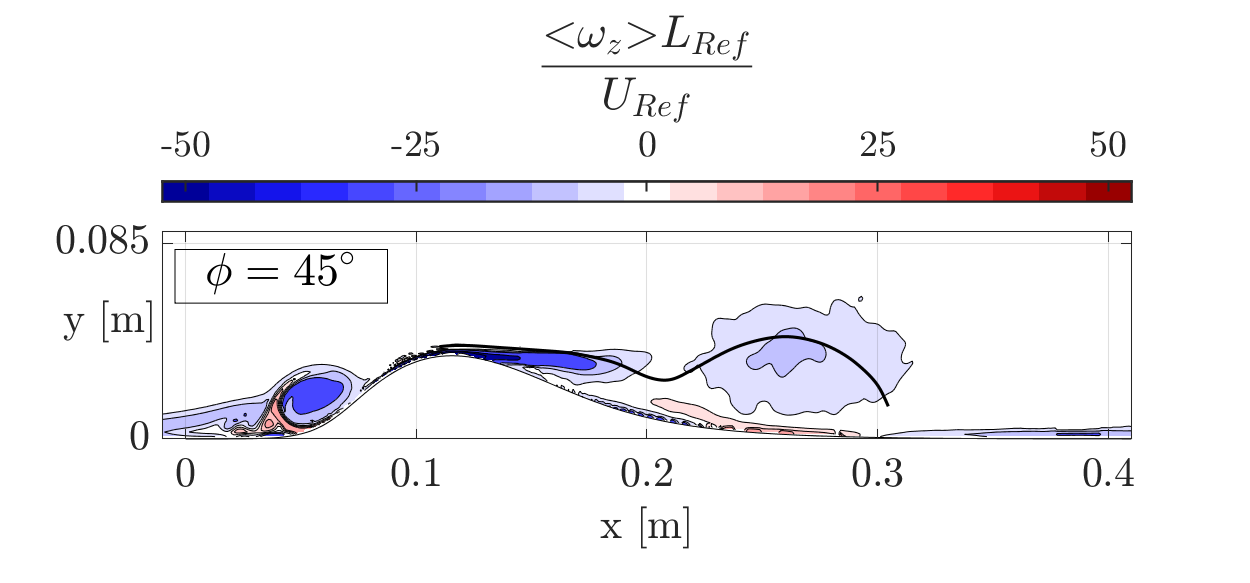} &
\includegraphics[width=.435\textwidth,trim={1.05in 0.88in 0.75in 1.5in},clip] {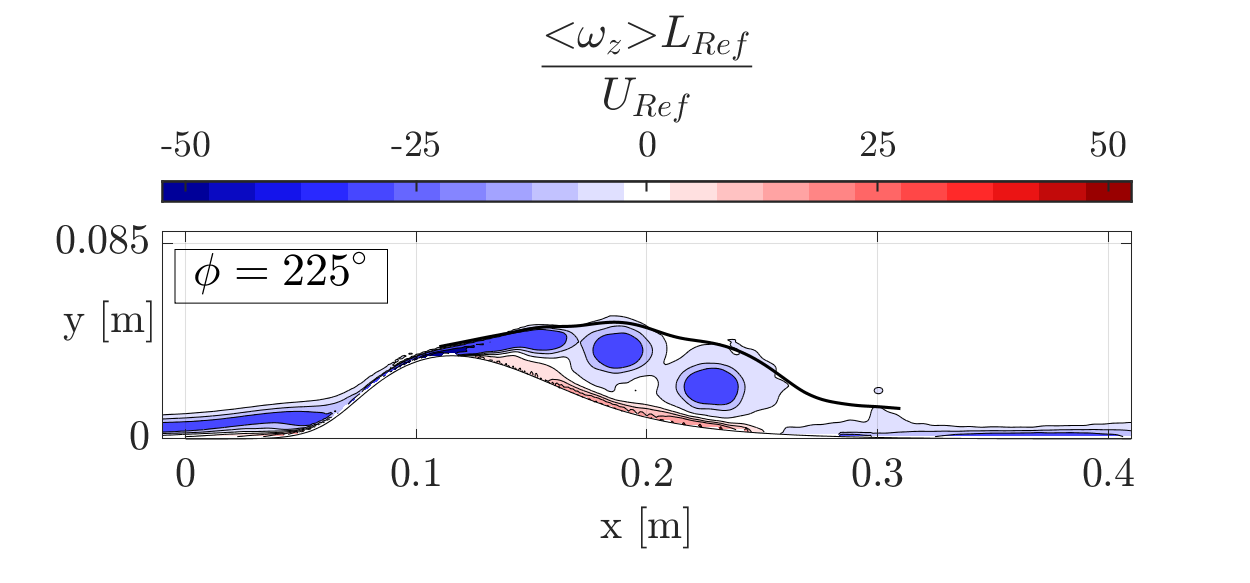} \\
\includegraphics[width=.48\textwidth,trim={0.35in 0.88in 0.75in 1.5in},clip] {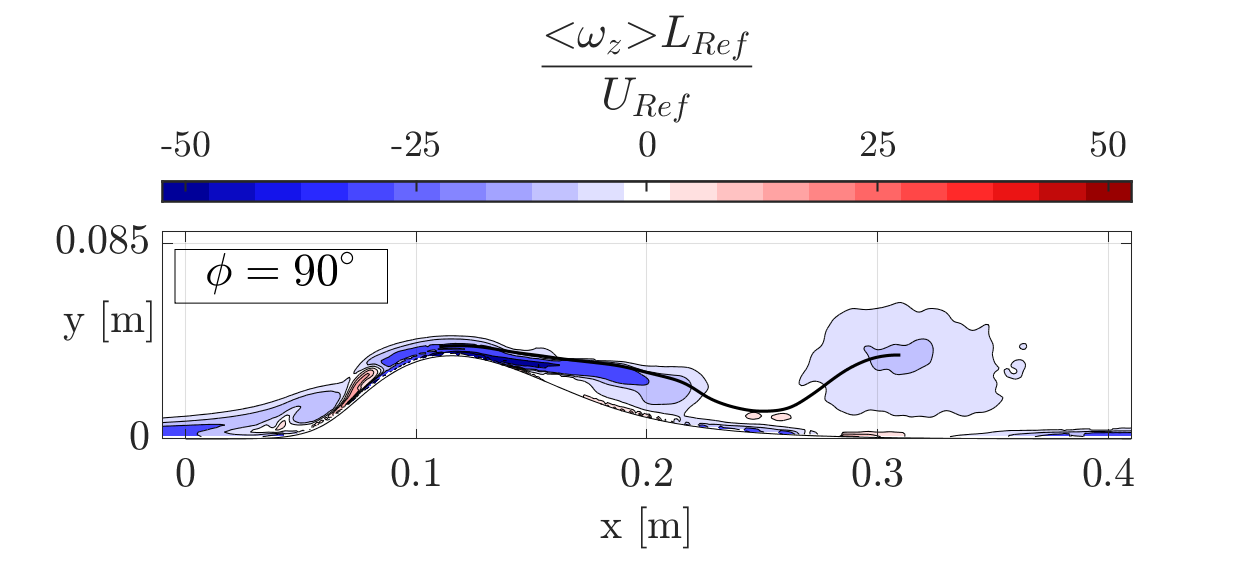} &
\includegraphics[width=.435\textwidth,trim={1.05in 0.88in 0.75in 1.5in},clip] {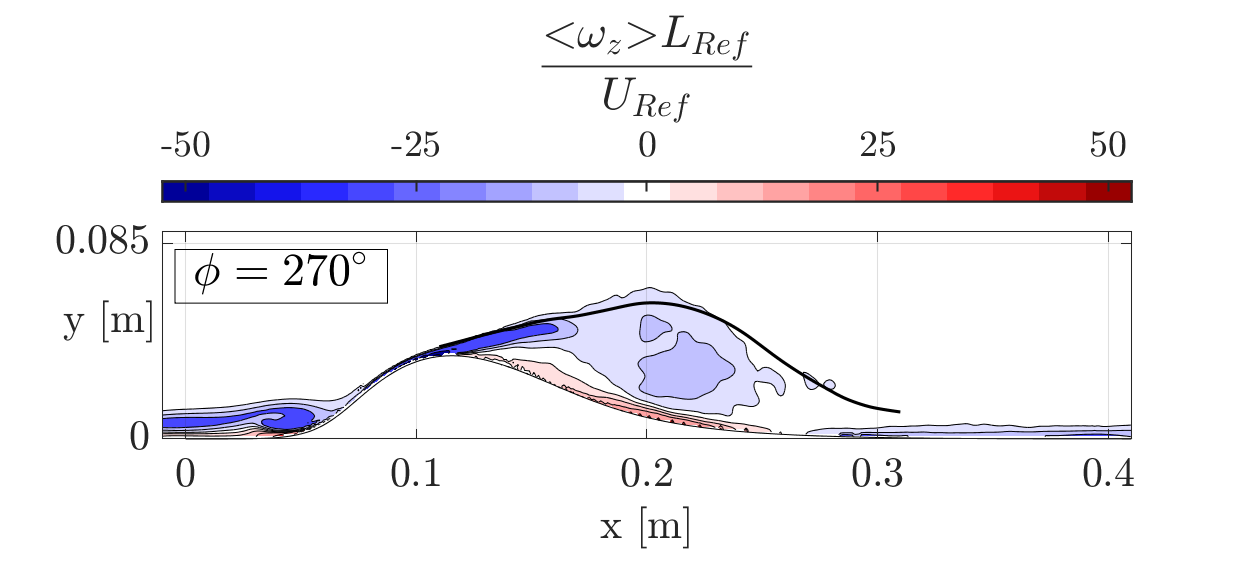} \\
\includegraphics[width=.48\textwidth,trim={0.35in 0.22in 0.75in 1.5in},clip] {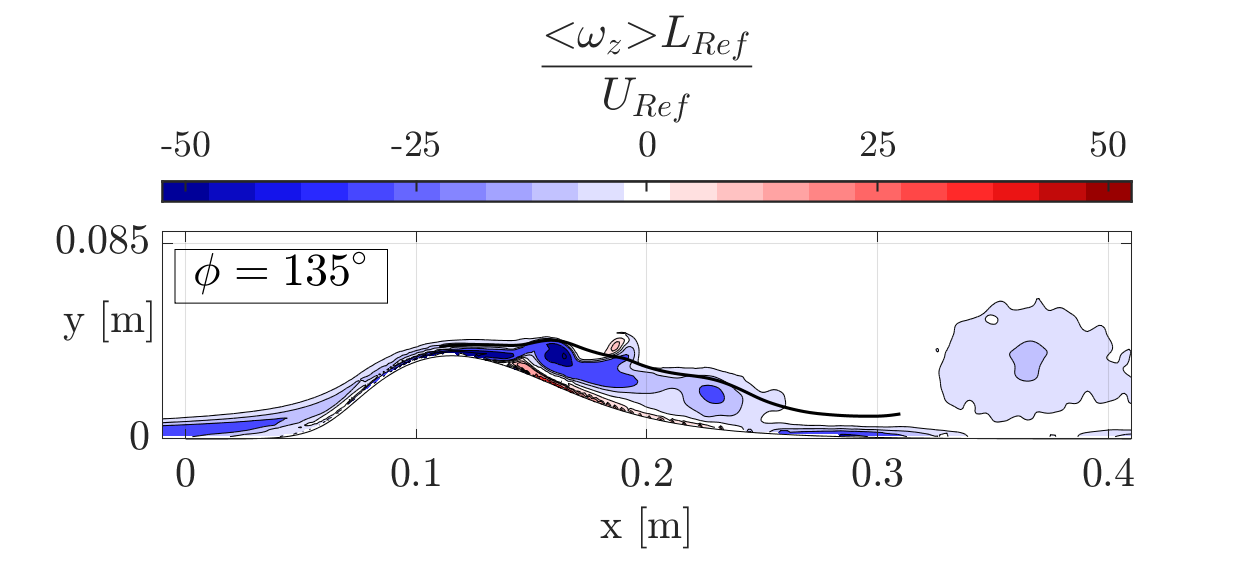} &
\includegraphics[width=.435\textwidth,trim={1.05in 0.22in 0.75in 1.5in},clip] {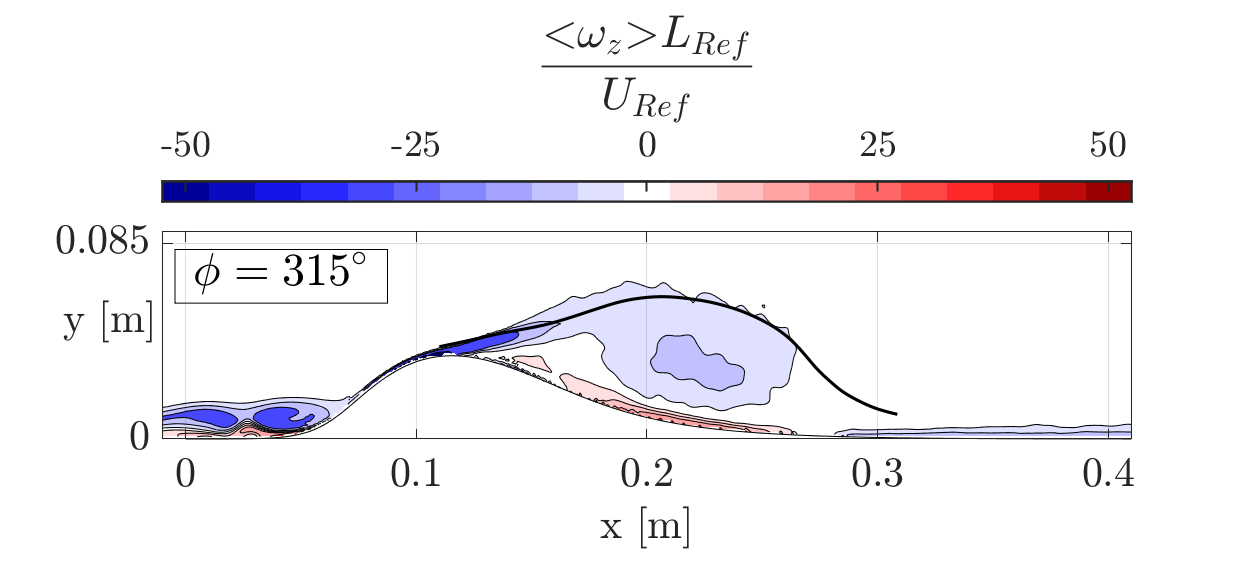} \\
\end{tabular}}
\caption{Phase-averaged spanwise vorticity. $A_{in}=0.1$ and $f_{in}^{*}=2$.}
\label{fig:phase_average_f2_A01}
\end{figure}

Figures \ref{fig:phase_average_f05_A001} to \ref{fig:phase_average_f2_A01} show the evolution of the phase-averaged spanwise vorticity fields at the midspan plane for the former two extreme cases ($A_{in}=0.01$, $f_{in}^* = 0.5$) and ($A_{in}=0.1$, $f_{in}^*=2$), and the intermediate case ($A_{in}=0.05$, $f_{in}^* = 1$). The thick black lines in the figures approximate the separation streamline at each phase. This line is computed by neglecting the spanwise velocity component in the phase-averaged flow and integrating $dx/\langle u \rangle = dy/ \langle v \rangle$ starting at the approximate location of the separation point.

For the case with the weakest inflow oscillation ($A_{in}=0.01$, $f_{in}^*=0.5$, Fig. \ref{fig:phase_average_f05_A001}), phase averaging successfully isolates the shear layer motion that is coherent with the inflow changes from the KH vortex shedding: the periodic motion of the shear layer towards and apart from the wall is captured in the phase-averaged field, but no imprint of individual vortices or details of the subsequent transition are captured.
Instead, the phase-averaged shear layer seems to diffuse as it evolves downstream, occupying the space where vortical structures are identified in the instantaneous flow. This region extends from the separated shear layer to the wall and presents a noticeable patch of positive vorticity adjacent to the wall for all phases; the latter is the imprint of vortical structures that are recirculated in the separation bubble.

The intermediate case ($A_{in}=0.05$, $f_{in}^* = 1$, Fig. \ref{fig:phase_average_f1_A005}) shows more clearly the vertical flapping motion of the shear layer. As opposed to the previous case, the positive vorticity region apparently disappears for phase angles corresponding to bulk flow acceleration ($\phi \approx 0^\circ$ in the figure) and becomes more intense around the peak deceleration ($\phi \approx 180^\circ$) where the wall-normal extension of the recirculation region is larger. 

Finally, Fig. \ref{fig:phase_average_f2_A01} shows the phase-averaged vorticity for the case with strongest inflow oscillations ($A_{in}=0.1$, $f_{in}^* = 2$). The phase-averaged field recovers the periodic formation and release of large patches of spanwise vorticity, coherent with the harmonic change of the bulk velocity. Phase $\phi = 270^\circ$ corresponds approximately to the conditions of minimum bulk velocity; for this phase, a vortex of size comparable to the bump is clearly defined downstream of the bump summit. As the flow re-accelerates, the vortex is released ($\phi \approx 0^\circ$) and advected downstream pushing the separation shear layer towards the wall, sensibly reducing the length of the separated flow region. 
Concurrently with this, a smaller two-dimensional vortex is formed upstream of the bump for $270^\circ \le \phi \le 45^\circ$, which is shed at $\phi \approx 45^\circ$ and reaches the bump summit at $\phi \approx 90^\circ$. This vortex interacts with the separated shear layer giving rise to two coherent vortices ($\phi \approx 180^\circ$) that subsequently break down into smaller structures, as shown in the instantaneous flow visualization of Fig. \ref{fig:insta_vort_A01_f2}. However, the phase-averaged field does not capture the evolution of these vortices after $\phi \approx 225^\circ$, indicating that their dynamics are chaotic and not reproduced from cycle to cycle. Comparing the instantaneous and phase-averaged fields (respectively Figs. \ref{fig:insta_vort_A01_f2} and  \ref{fig:phase_average_f2_A01}), it is observed that the vortical structures originated by the upstream vortex are completely entrapped in the recirculation region and contribute to its re-generation.

\subsection{Impact of the inflow conditions on streamwise acceleration and length of the separated flow}
\label{sec:length}

\begin{figure}
\centering{
\includegraphics[width=.35\textwidth,trim={3.57in 6.2in 0.65in 0in},clip] {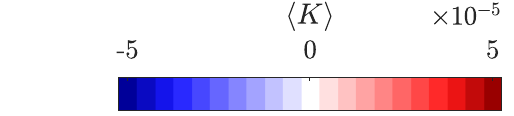}
\begin{tabular}{cccc}
\includegraphics[width=.30\textwidth,trim={0in 0in 3.9in 0.8in},clip] {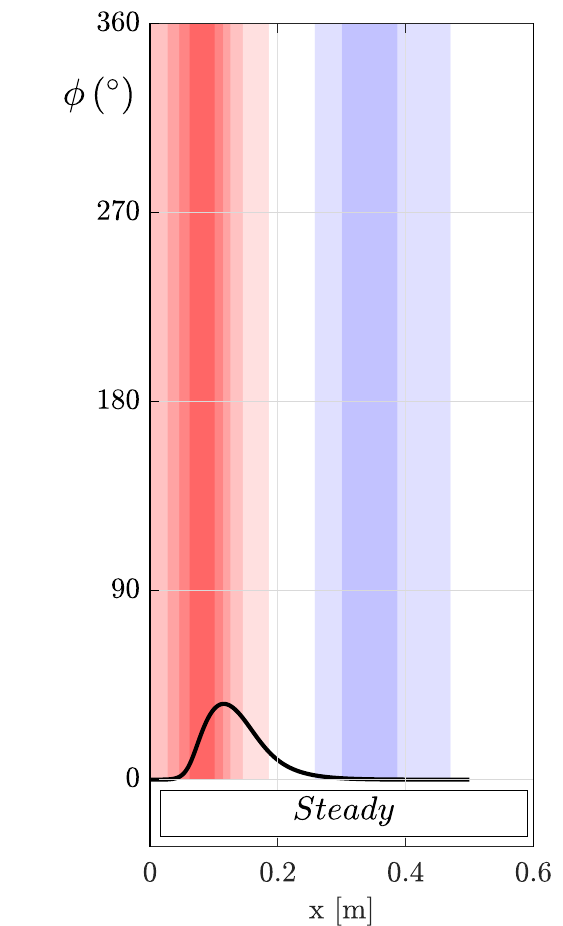} &
\includegraphics[width=.22\textwidth,trim={4.3in 0in 0.6in 0.8in},clip] {PhaseMean_KSumContent_Pub_A001_F05.eps} &
\includegraphics[width=.22\textwidth,trim={4.3in 0in 0.6in 0.8in},clip] {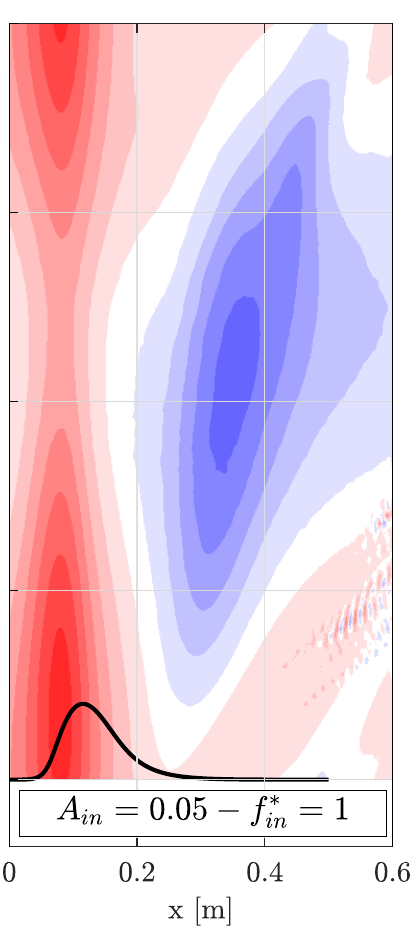} &
\includegraphics[width=.22\textwidth,trim={4.3in 0in 0.6in 0.8in},clip] {PhaseMean_KSumContent_Pub_A01_F2.eps} \\
\end{tabular}}
\caption{Phase-averaged streamwise acceleration parameter at $y=0.1$ m.
}
\label{fig:acceleration_midheight}
\end{figure}

\begin{figure}
\centering{
\includegraphics[width=.35\textwidth,trim={3.57in 6.17in 0.65in 0in},clip] {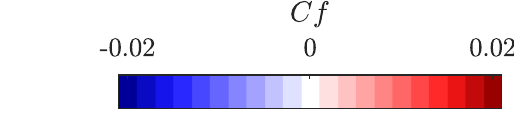}
\begin{tabular}{cccc}
\includegraphics[width=.30\textwidth,trim={0in 0in 3.9in 0.8in},clip] {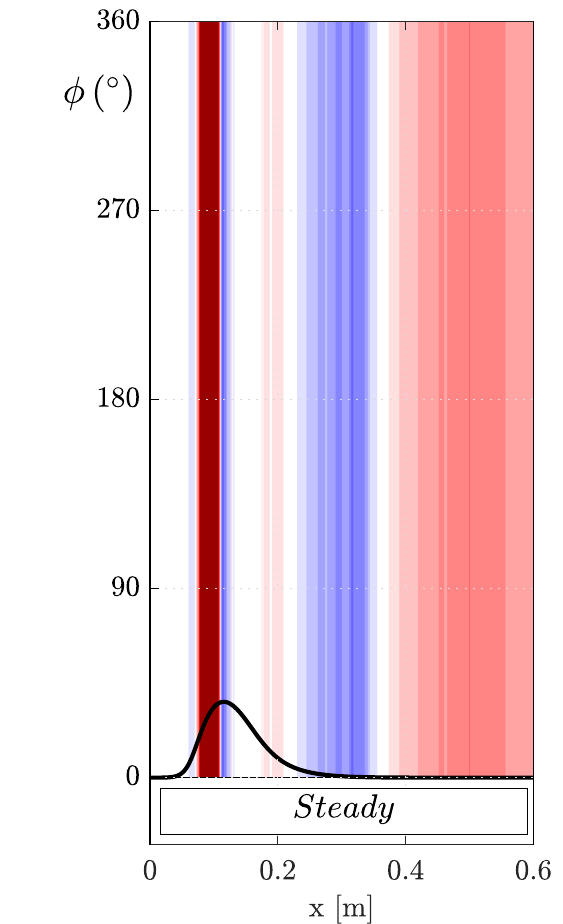} &
\includegraphics[width=.22\textwidth,trim={4.3in 0in 0.6in 0.8in},clip] {Fig_Cf_A001_F05_a.eps} &
\includegraphics[width=.22\textwidth,trim={4.3in 0in 0.6in 0.8in},clip] {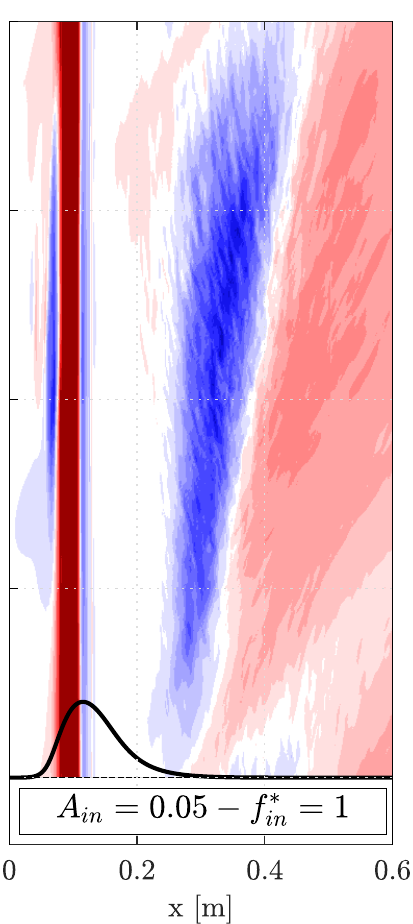} &
\includegraphics[width=.22\textwidth,trim={4.3in 0in 0.6in 0.8in},clip] {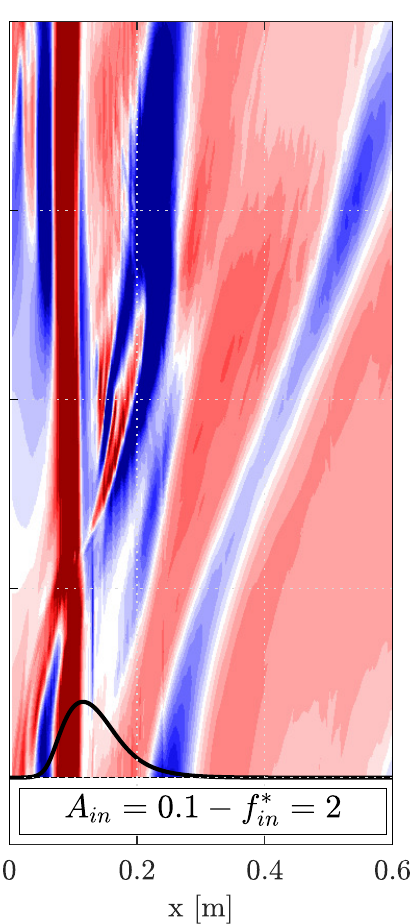} \\
\end{tabular}}
\caption{Phase-averaged skin friction at the mid-span of the bump surface. Left: $A_{in}= 0.01$, $f_{in}^{*}=0.5$. Center: $A_{in} = 0.05$, $f_{in}^* = 1$. Right: $A_{in}=0.1$, $f_{in}^* =2$.
}
\label{fig:cf}
\end{figure}

The phase-averaged fields provide relevant information regarding the impact of the harmonic inflow changes on the streamwise flow acceleration and deceleration and the resulting length of the separated flow region. In order to quantify the phase-dependent streamwise acceleration, the dimensionless  parameter

\begin{equation}
\langle K  \left( {\boldsymbol x}, \phi \right) \rangle  =  \bigg<
\dfrac{\nu}{u^3 \left( {\boldsymbol x}, \phi \right)} \left(\dfrac{\partial u \left( {\boldsymbol x}, \phi \right)}{\partial t} + u \left( {\boldsymbol x}, \phi \right) \dfrac{\partial u \left( {\boldsymbol x}, \phi \right)}{\partial x} \right)
\bigg>
\end{equation}

\noindent is used, that can be further divided into two contributions: the contribution from the local flow acceleration

\begin{equation}
<K_{\partial u / \partial t}  \left( {\boldsymbol x}, \phi \right)>= \bigg<
\dfrac{\nu}{u^3 \left( {\boldsymbol x}, \phi \right)} \dfrac{\partial u \left( {\boldsymbol x}, \phi \right)}{\partial t} \bigg>
\end{equation}

\noindent and the convective acceleration

\begin{equation}
<K_{\partial u / \partial x}  \left( {\boldsymbol x}, \phi \right)>= \bigg<
\dfrac{\nu}{u^2 \left( {\boldsymbol x}, \phi \right)} \dfrac{\partial u \left( {\boldsymbol x}, \phi \right)}{\partial x} \bigg> .
\end{equation}
 
 This parameter was introduced by \citet{Spalart:NTM86} as a pressure gradient parameter in studies of flow relaminarization under favorable pressure gradients ($K>0$)  but it is also used to quantify the flow deceleration associated with an adverse pressure gradient ($K<0$) \citep{Suzen:Jturb03, Saavedra18,Dellacasagrande:EF20,Ambrogi:JFM22} in boundary layer flows. 
 In the ideal scenario of a boundary layer that is unbounded on the wall-normal direction, the free-stream value of the streamwise velocity would be used and  Bernoulli's equation would relate it directly to the streamwise pressure gradient.  
In the geometry used herein, the upper wall of the channel prevents using this definition. As an approximation to the free-stream velocity, the streamwise velocity at the plane $y=0.1$ m is used, which corresponds approximately to the midpoint between the bump summit and the upper wall. The streamwise velocity on this plane is expected to be less affected by the instantaneous vortical structures and viscous effects on the separated flow region, though the irrotational flow is strictly not attained.
Figure \ref{fig:acceleration_midheight} shows the spatio-temporal evolution of the phase-averaged acceleration parameter $\langle K \rangle$, for the three representative cases of harmonic inflow oscillation. The steady inflow case is also shown for comparison. 
For the same cases, figure \ref{fig:cf} shows the temporal evolution of the dimensionless streamwise skin friction at the mid-span plane, defined as   

\begin{equation}
C\!f\left({\boldsymbol x},{\phi}\right) = \frac{\langle{{\tau_{w}}}({\boldsymbol x},\phi)\rangle}{p_{t}\left({\boldsymbol x}_{ref}, \phi\right)-p\left({\boldsymbol x}_{ref}, \phi \right)} ,
\label{eq:Cf}
\end{equation}

\noindent where ${\boldsymbol x}_{ref}$ corresponds to the location of the reference point (cf. Table \ref{tab:Ref_Sampling_Probes}).

The steady inflow case shows the flow acceleration-then-deceleration distribution generated by the bump. The minimum $K$ value is obtained at $x \approx 0.35$ m, which is coincident with the location of minimum $C\!f$. The time-averaged reattachment occurs a short distance downstream and can be identified as the coordinate where the skin friction changes from negative to positive. The streamwise acceleration parameter is still negative at reattachment, illustrating that the reattachment originated from unsteady flow entrainment rather than the action of a streamwise flow acceleration. Case ($A_{in}=0.01$, $f_{in}^*=0.5$) shows the same features as the steady inflow case, with a small amplitude modulation that follows the inlet frequency.  

The cases with increasingly stronger/faster inflow oscillations exhibit a pattern of phase-dependent deceleration-acceleration localized in the reattachment region that is repeated with each period. For the intermediate case ($A_{in}=0.05$, $f^*_{in}=1$), the minimum value of $\langle K \rangle$ occurs for $\phi \approx 180^\circ$, coincident with the peak bulk flow deceleration. As the flow re-accelerates for $\phi > 270^\circ$, the region of negative $\langle K \rangle$ is displaced downstream and reduced in size. Then, for the peak acceleration phase $\phi = 0^\circ$, $\langle K \rangle$ is positive around the time-averaged reattachment point. This evolution of $\langle K \rangle$ is followed by the $C\!f$ distribution. At $\phi=0^\circ$, the reattachment point is located around $x=0.25$ m; reattachment moves downstream gradually for increasing $\phi$, resulting in a longer recirculation bubble. During the flow re-acceleration, a new region of positive skin friction is formed upstream of the region of minimum $C\!f$, implying that a large coherent vortical structure has been released and a new one is being formed. 

Finally, the case ($A_{in}=0.1$, $f_{in}^* = 2$) shows new distinct features both in the acceleration parameter and the skin friction. For relatively elevated values of $A_{in}$ and $f^*_{in}$ the local flow acceleration $\langle K_{\partial u/\partial t} \rangle$ becomes comparable to the convective one. In consequence, $\langle K \rangle$ is alternatively positive or negative during about half of the period. The peak deceleration is now displaced to the later phase $\phi \approx 275^\circ$. However, its peak magnitude is increased substantially with respect to the case ($A_{in}=0.05$, $f^*_{in}=1$), and as a result the same negative values of $\langle K \rangle$ are reached before in the period. For instance, case ($A_{in}=0.05$, $f^*_{in}=1$) presents the minimum $\langle K \rangle$ value $-24.74\times 10^{-6}$ at $\phi$ slightly above $180^\circ$, while this value is attained in case ($A_{in}=0.1$, $f^*_{in}=2$) already at $\phi = 135^\circ$. 
The intense periodic acceleration-deceleration influences the $C\!f$ distribution notably, involving the formation of multiple recirculation regions that are related to the advection of the coherent vortex clusters shown in Fig. \ref{fig:phase_average_f2_A01}. Cross-comparison of the streamwise acceleration and skin friction (cf. Fig. \ref{fig:acceleration_midheight} and \ref{fig:cf}) suggests that the formation and release of the large vortical structures in the coherent flow component are associated to surpassing a threshold negative value of the acceleration parameter $\langle K \rangle$. The reasons for this will be further discussed in later sections. 

\begin{figure}
\centering{\begin{tabular}{ccc}
\includegraphics[width=.32\textwidth,trim={0in 0.3in 0in 0in},clip] {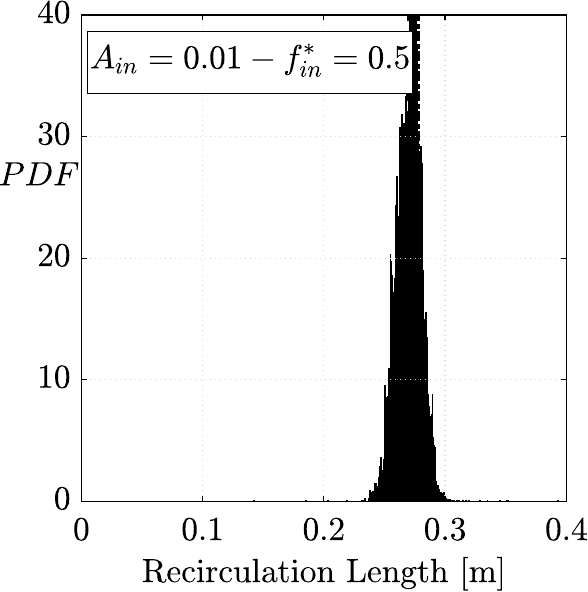} &
 \includegraphics[width=.32\textwidth,trim={0in 0.3in 0in 0in},clip] {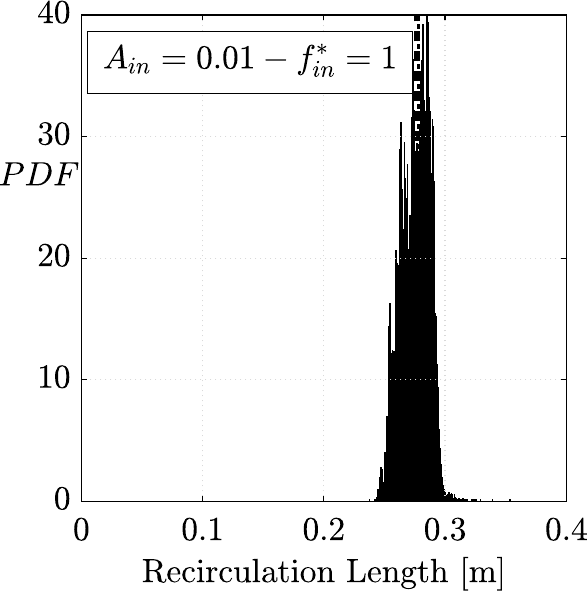} &
\includegraphics[width=.32\textwidth,trim={0in 0.3in 0in 0in},clip] {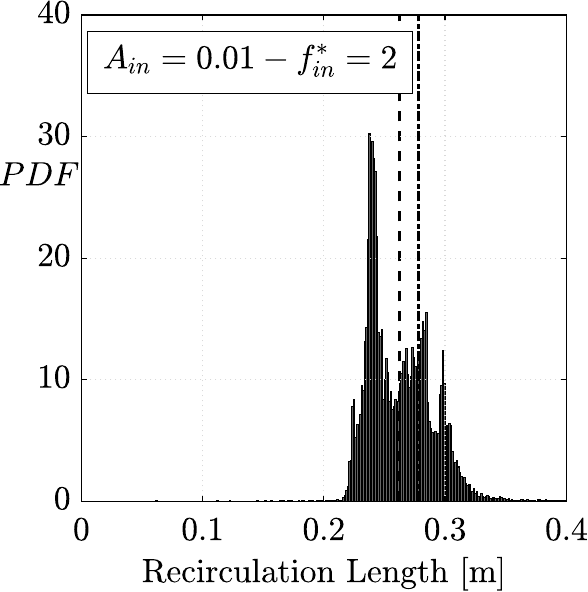} \\
\includegraphics[width=.32\textwidth,trim={0in 0.3in 0in 0in},clip] {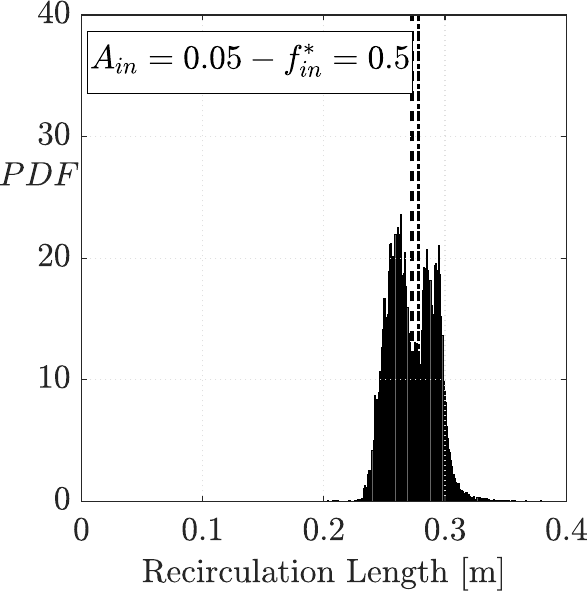} &
 \includegraphics[width=.32\textwidth,trim={0in 0.3in 0in 0in},clip] {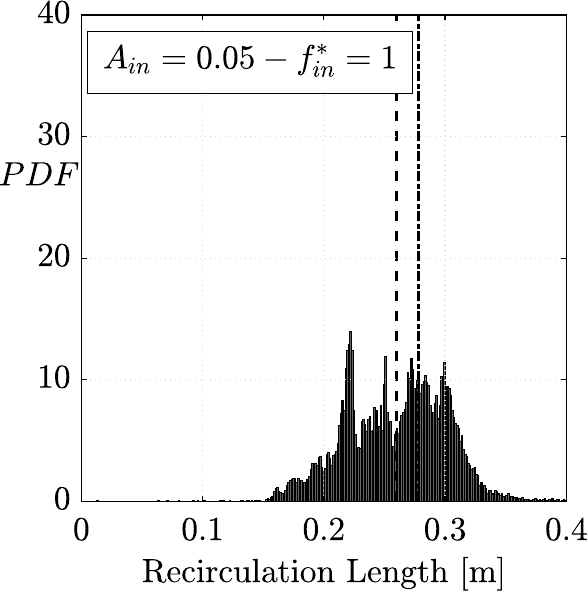} &
\includegraphics[width=.32\textwidth,trim={0in 0.3in 0in 0in},clip] {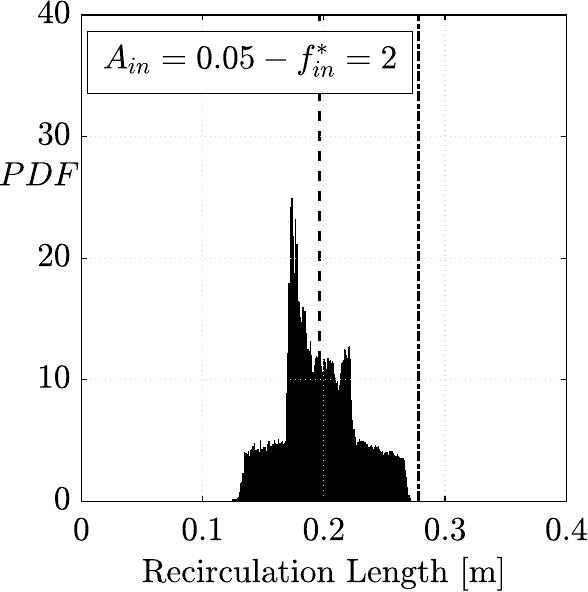} \\
\includegraphics[width=.32\textwidth,trim={0in 0.3in 0in 0in},clip] {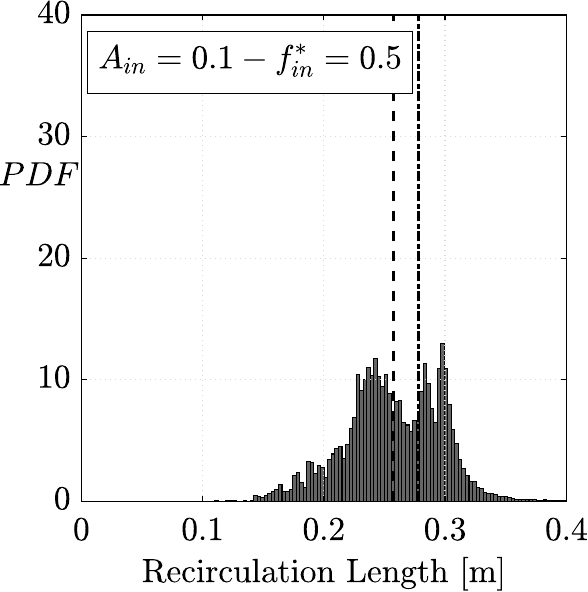} &
 \includegraphics[width=.32\textwidth,trim={0in 0.3in 0in 0in},clip] {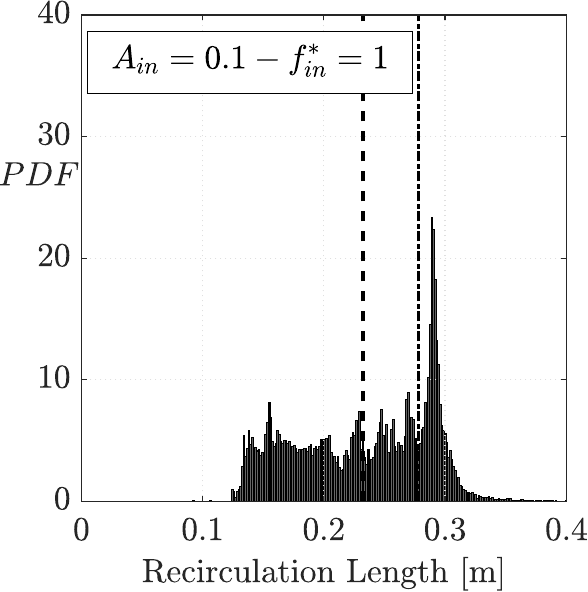} &
\includegraphics[width=.32\textwidth,trim={0in 0.3in 0in 0in},clip] {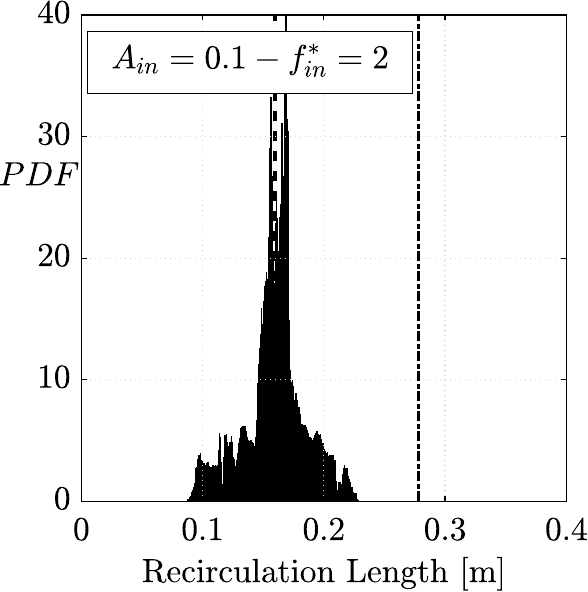} \\
\hspace{10pt} $\langle L_s \rangle$ &  \hspace{10pt} $\langle L_s \rangle$  & \hspace{10pt} $\langle L_s \rangle$\\
\end{tabular}}
\caption{Probability density function of the phase-averaged recirculation bubble length $\langle L_s \rangle$. (- - -) time-averaged value; (-$\cdot$-$\cdot$) time-averaged for the steady inflow case, $L_{s,steady}$.}
\label{fig:pdf_length}
\end{figure}

The streamwise length of the recirculation bubble $L_s$ is computed at each phase $\phi$ as the distance between the separation point near the bump summit and the first reattachment point downstream.  Figure \ref{fig:pdf_length} shows the probability density function (PDF) of $\langle L_s \rangle$ for all the simulated cases. The time-averaged length for each case and the time-averaged length for the steady inflow case are also shown. Table \ref{tab:simulations_details} tabulates the numeric values.  Cases with relatively low values of $A_{in}$ and $f_{in}^*$ (towards the upper left panels of the figure) show that the recirculation length remains close to that of the steady inflow case with a very narrow distribution range. As $A_{in}$ or $f_{in}^*$ are increased individually (e.g., $A_{in}=0.05$, $f_{in}^*=1$), the PDF becomes significantly broader, indicating large changes in $\langle L_s \rangle$ over the period. The time-averaged recirculation length is reduced in all cases, but the distribution is not centered around it and the length at some phases can be substantially longer or shorter than $L_{s,steady}$. This intense temporal variation of $\langle L_s \rangle$ is expected to be associated with periodic changes in the aerodynamic forces exerted on the bump, which may be highly undesirable in the practical scenario of a low-pressure turbine \citep{Curtis97}.
Finally, the cases with larger values of $A_{in}$ and $f_{in}^*$ (towards the bottom right panels of Fig. \ref{fig:pdf_length}, and particularly $A_{in}=0.1$, $f_{in}^* = 2$) present a PDF which is again centered on the mean $L_s$ value. This value is remarkably reduced with respect to $L_{s,steady}$. Notably, the PDF tail falls to zero for recirculation lengths below $L_{s,steady}$, implying that the separation length is reduced for all phases, including those in which the bulk flow is decelerated. This is also relevant in practical scenarios, as the detrimental effects of flow separation would be consistently reduced with respect to the steady inflow case.


\subsection{Incoherent vorticity and vortex dynamics}
\label{sec:Vortex_dynamics}

\begin{figure}
\centering{\begin{tabular}{ccc}
\includegraphics[width=.333\textwidth,trim={0in 2.95in 0in 0in},clip] {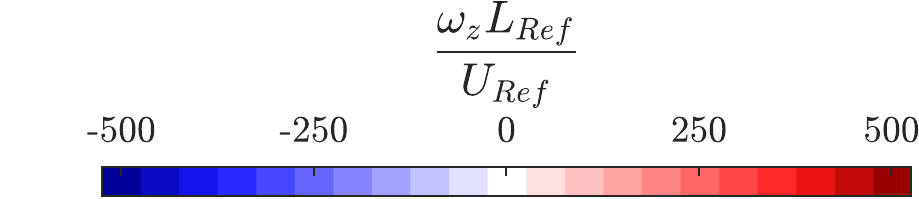} &
\includegraphics[width=.30\textwidth,trim={0.6in 2.95in 0in 0in},clip] {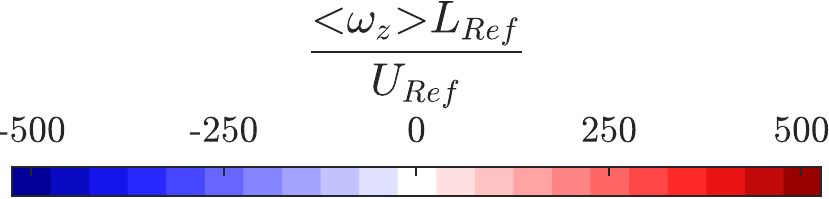} &
\includegraphics[width=.30\textwidth,trim={0.6in 2.95in 0in 0in},clip] {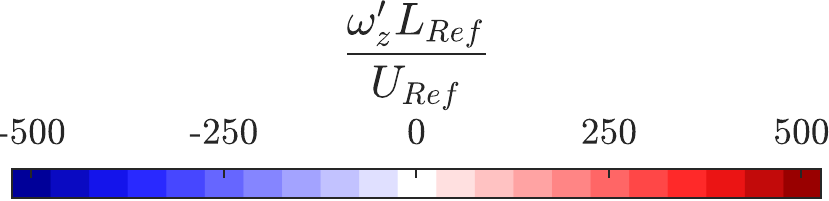} \\

\includegraphics[width=.333\textwidth,trim={0.25in 1.0in 0.62in 1.6in},clip] {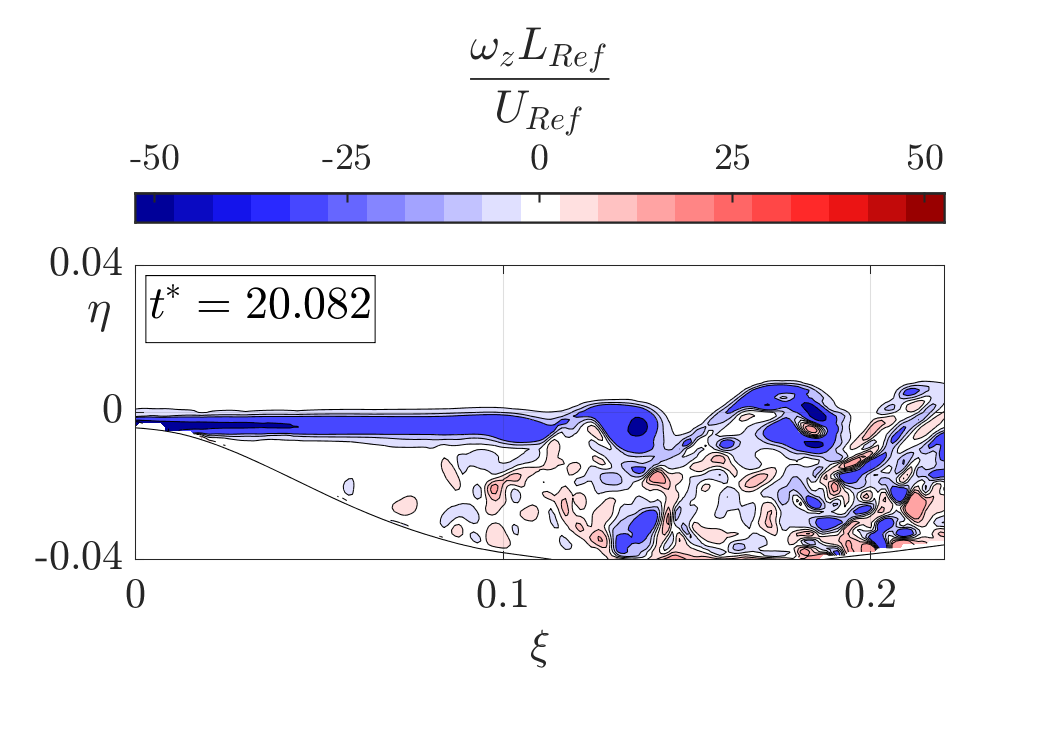} &
\includegraphics[width=.30\textwidth,trim={0.85in 1.0in 0.62in 1.6in},clip] {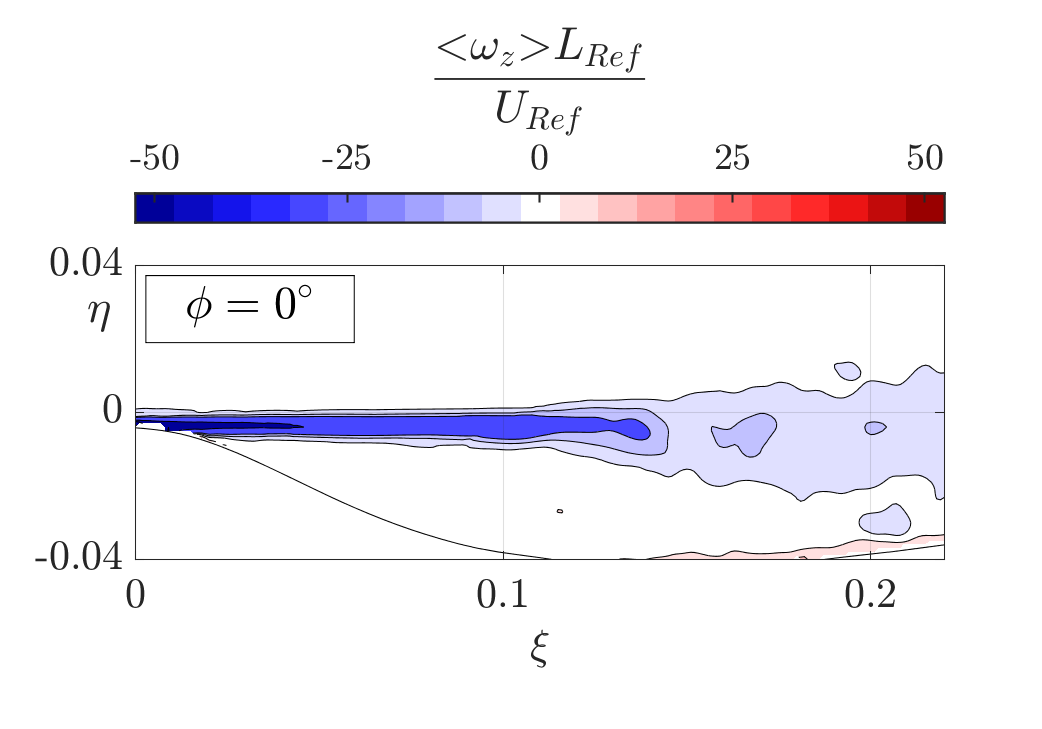} &
\includegraphics[width=.30\textwidth,trim={0.85in 1.0in 0.62in 1.6in},clip] {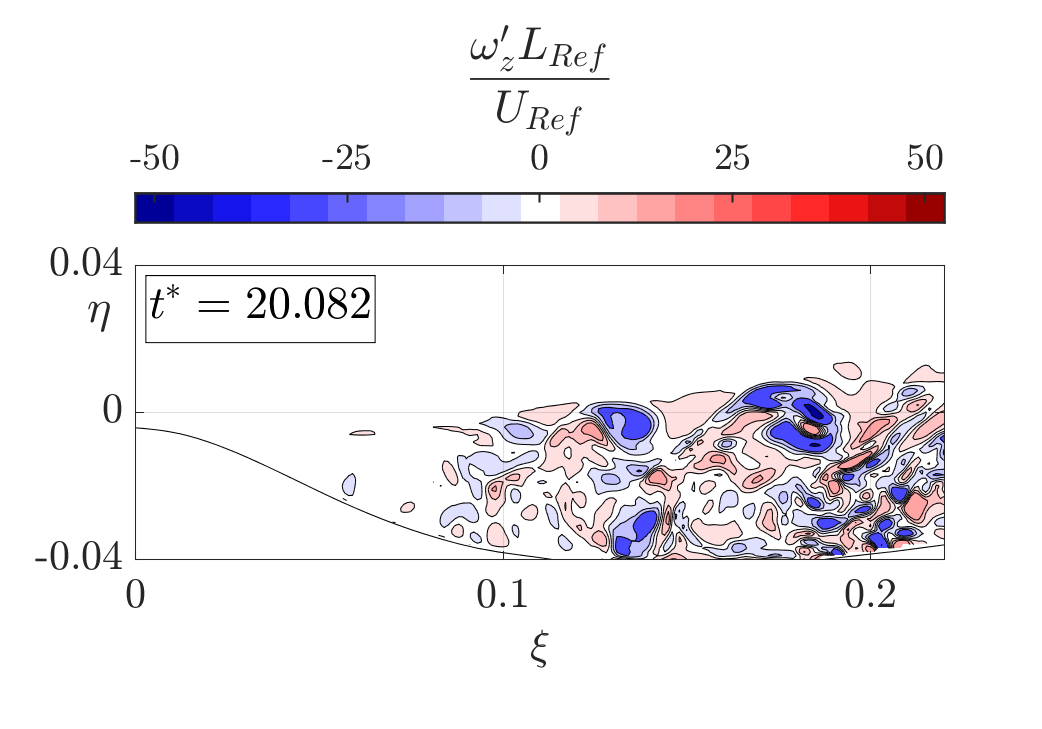} \\
\includegraphics[width=.333\textwidth,trim={0.25in 1.0in 0.62in 1.6in},clip] {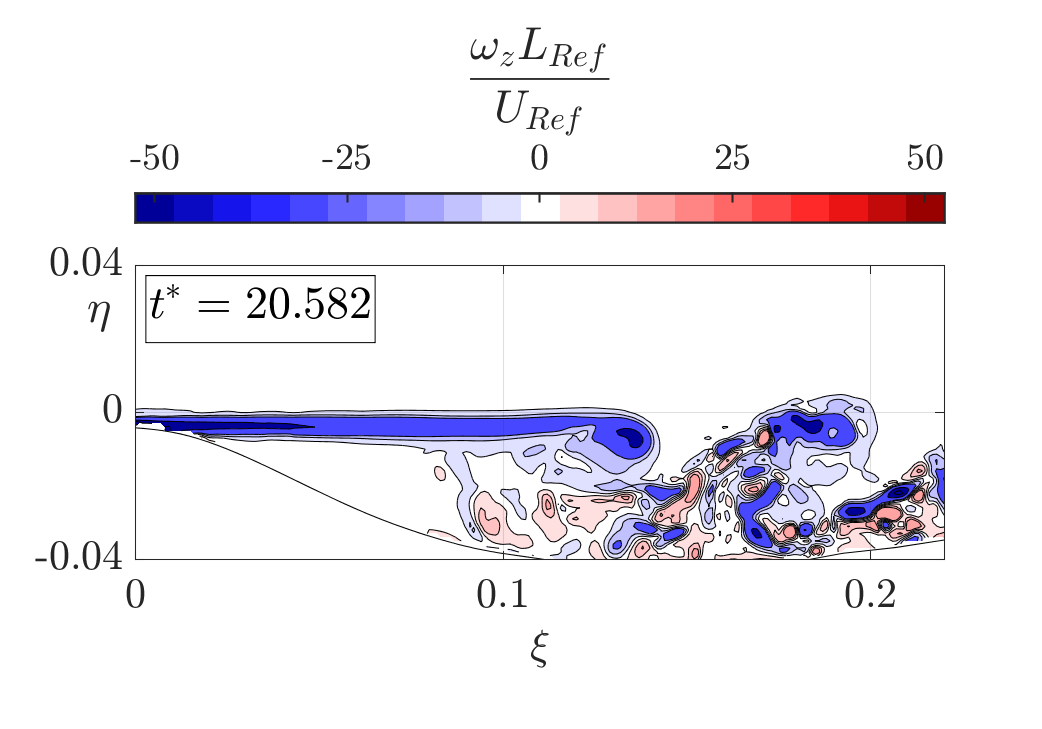} &
\includegraphics[width=.30\textwidth,trim={0.85in 1.0in 0.62in 1.6in},clip] {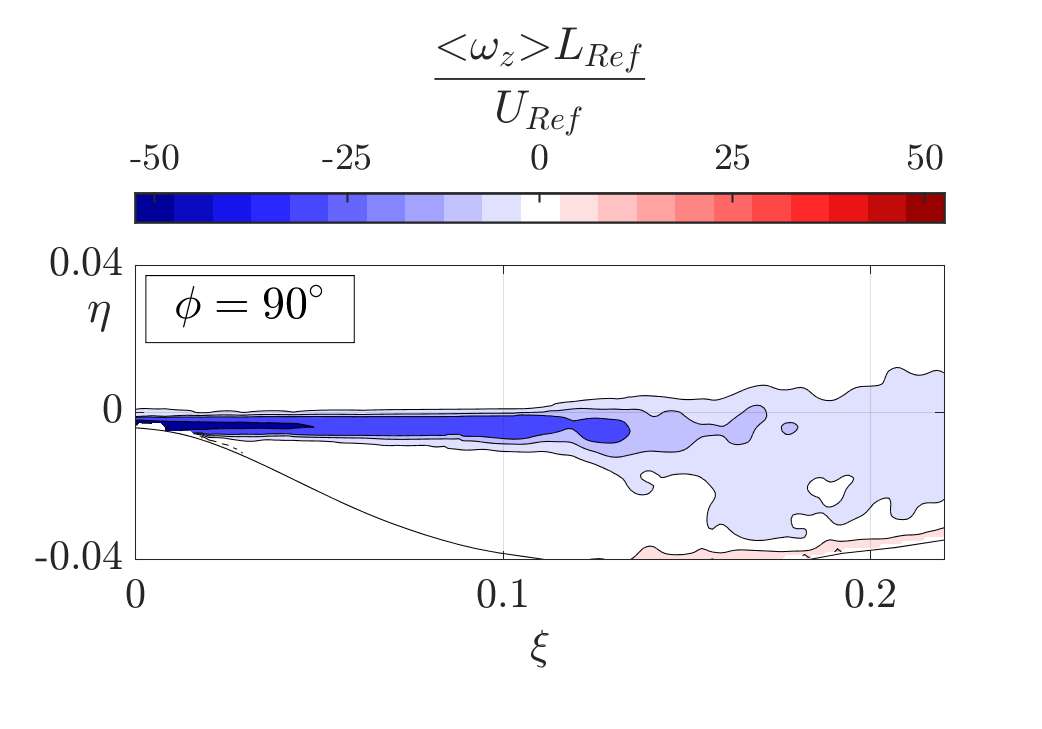} &
\includegraphics[width=.30\textwidth,trim={0.85in 1.0in 0.62in 1.6in},clip] {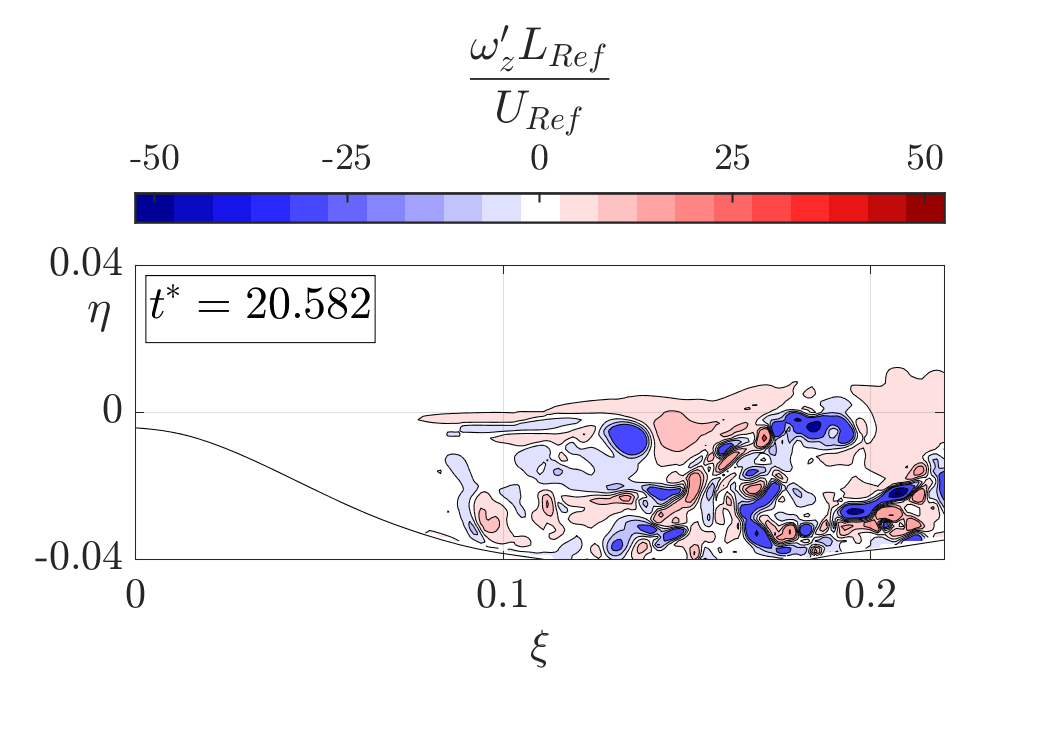} \\
\includegraphics[width=.333\textwidth,trim={0.25in 1.0in 0.62in 1.6in},clip] {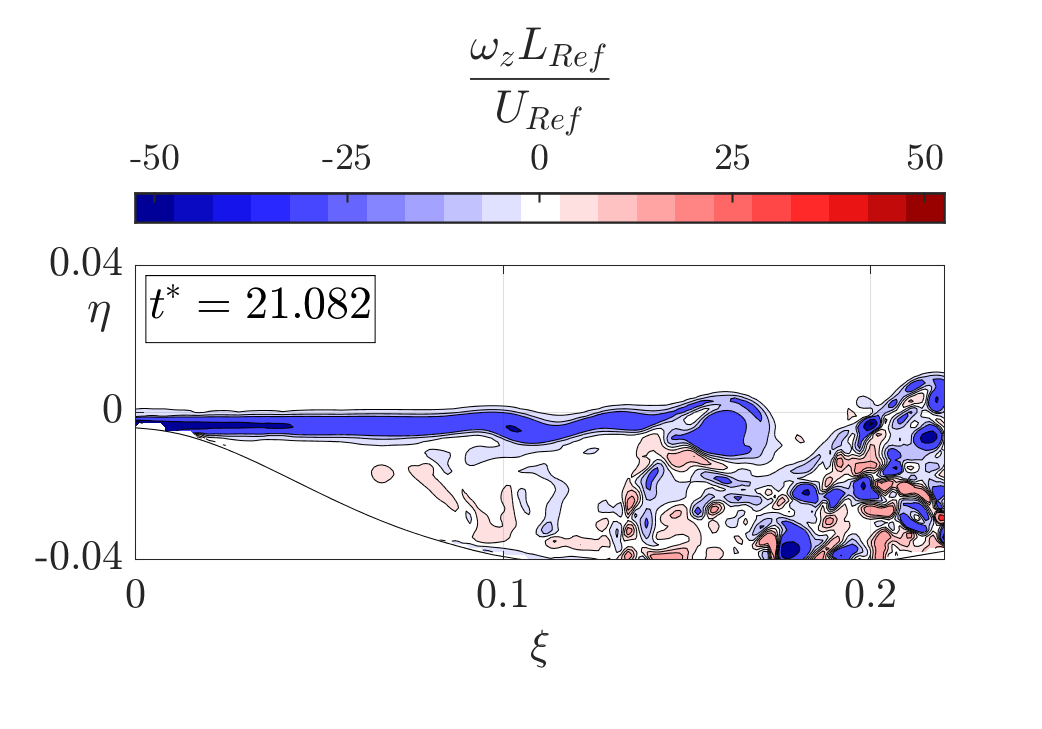} &
\includegraphics[width=.30\textwidth,trim={0.85in 1.0in 0.62in 1.6in},clip] {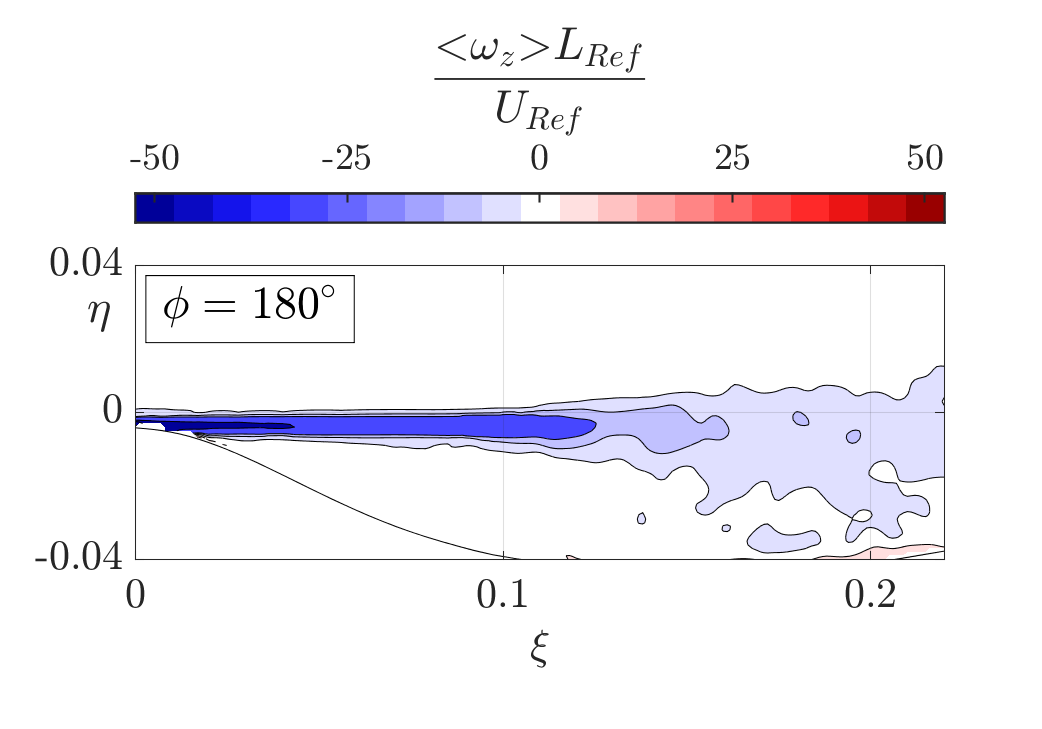} &
\includegraphics[width=.30\textwidth,trim={0.85in 1.0in 0.62in 1.6in},clip] {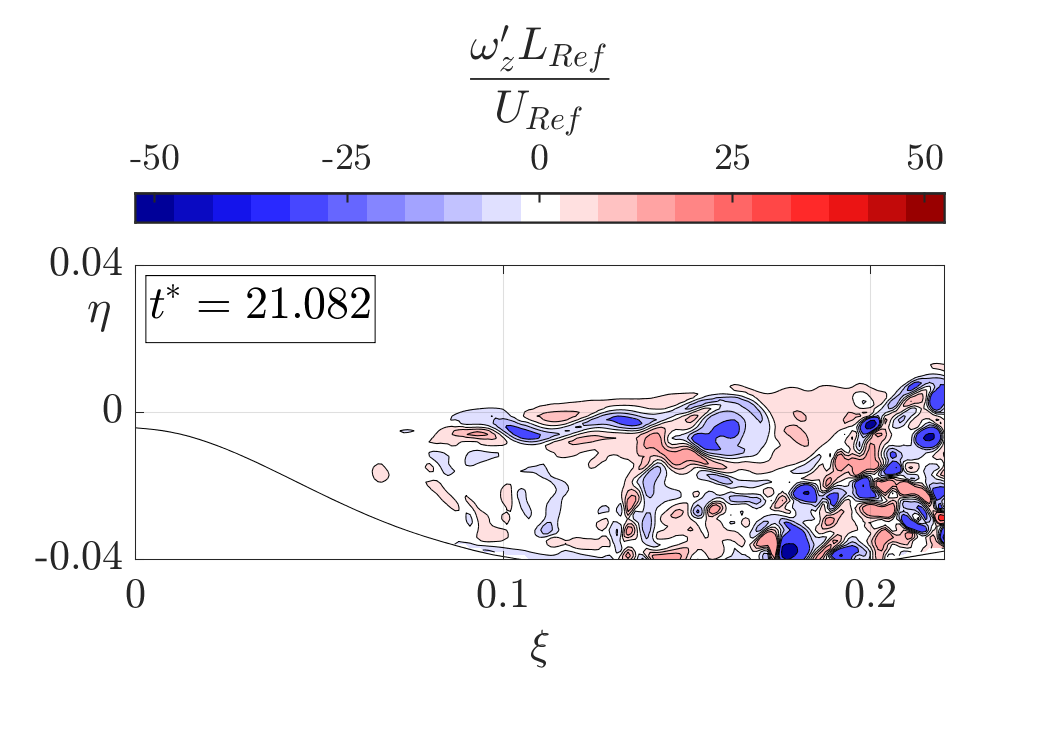} \\
\includegraphics[width=.333\textwidth,trim={0.25in 0.2in 0.62in 1.6in},clip] {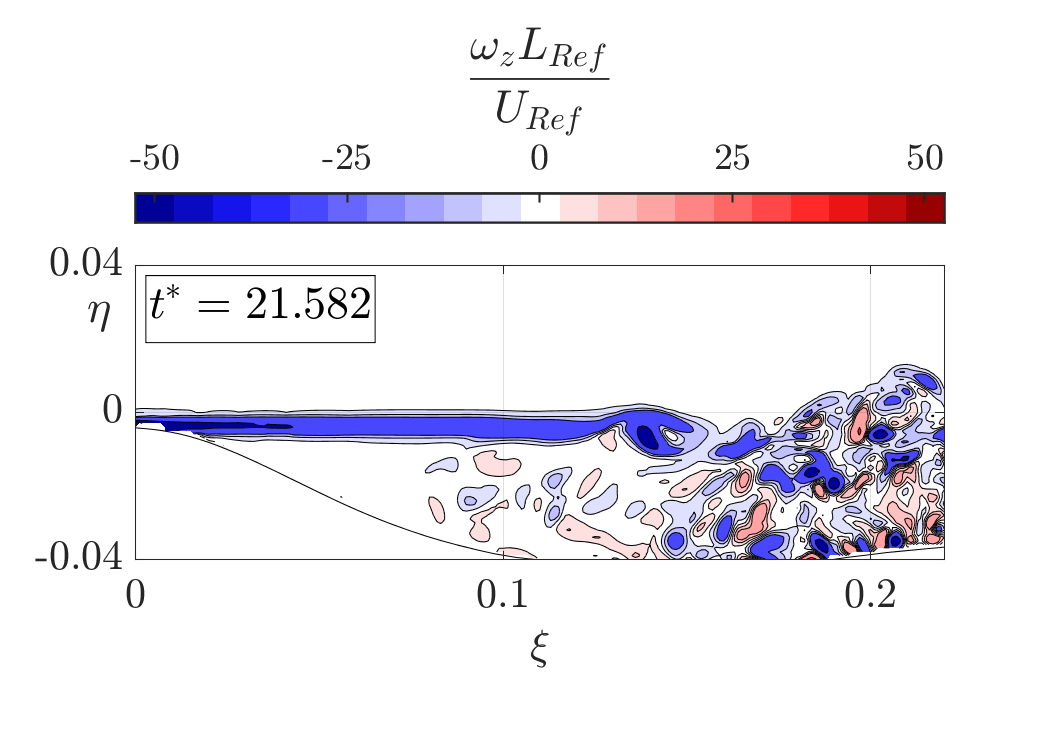} &
\includegraphics[width=.30\textwidth,trim={0.85in 0.2in 0.62in 1.6in},clip] {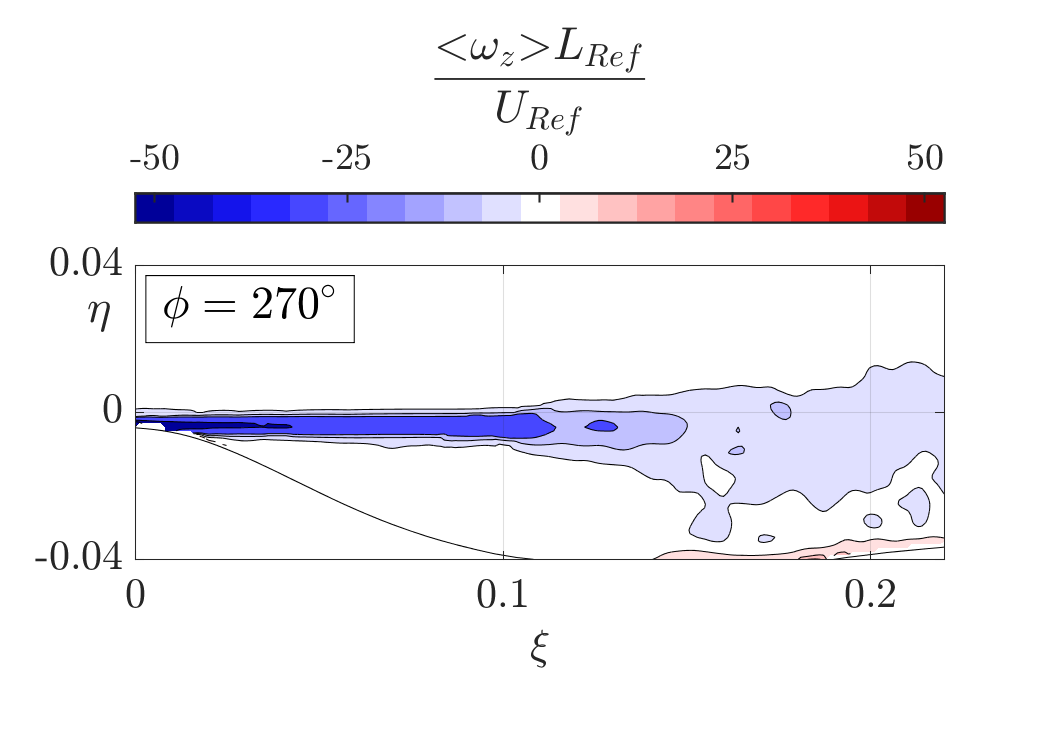} &
\includegraphics[width=.30\textwidth,trim={0.85in 0.2in 0.62in 1.6in},clip] {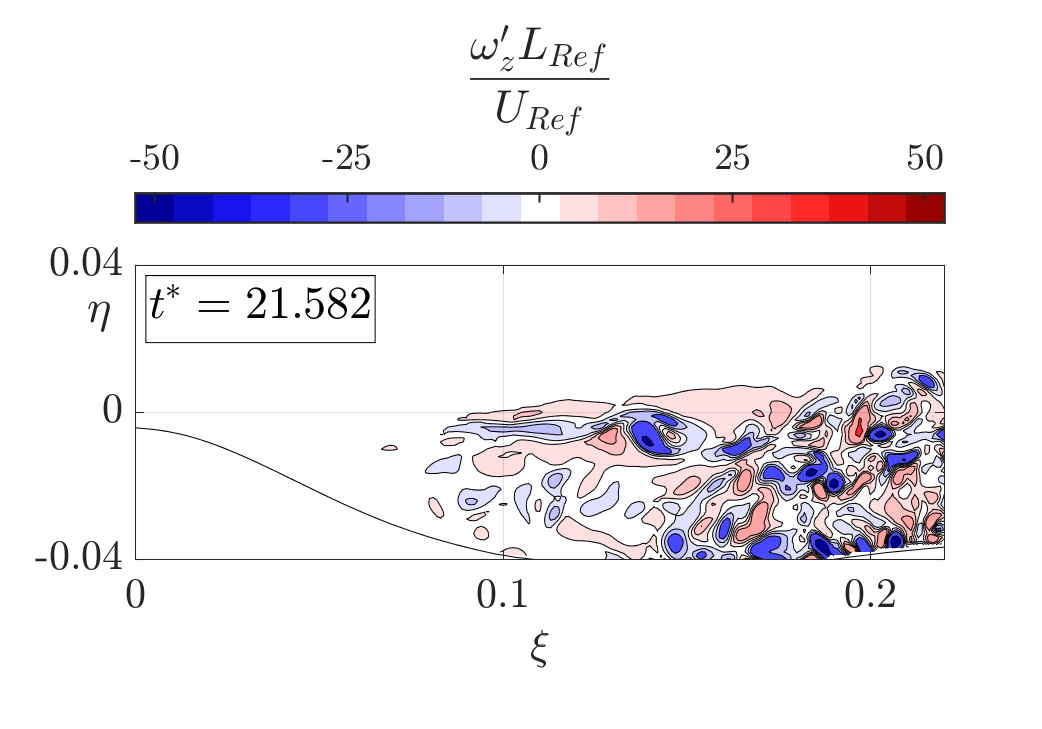} \\

\end{tabular}}
\caption{Triple decomposition of spanwise vorticity. $A_{in}= 0.01$ and $f_{in}^{*}=0.5$.}
\label{fig:td_A001_f05}
\end{figure}

\begin{figure}
\centering{\includegraphics[width=.25\textwidth,trim={0in 10.4in 0in 0in},clip] {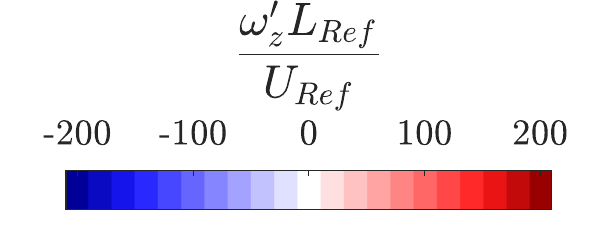}
\begin{tabular}{cccc}
\includegraphics[width=.227\textwidth,trim={0in 0in 0.14in 1.5in},clip] {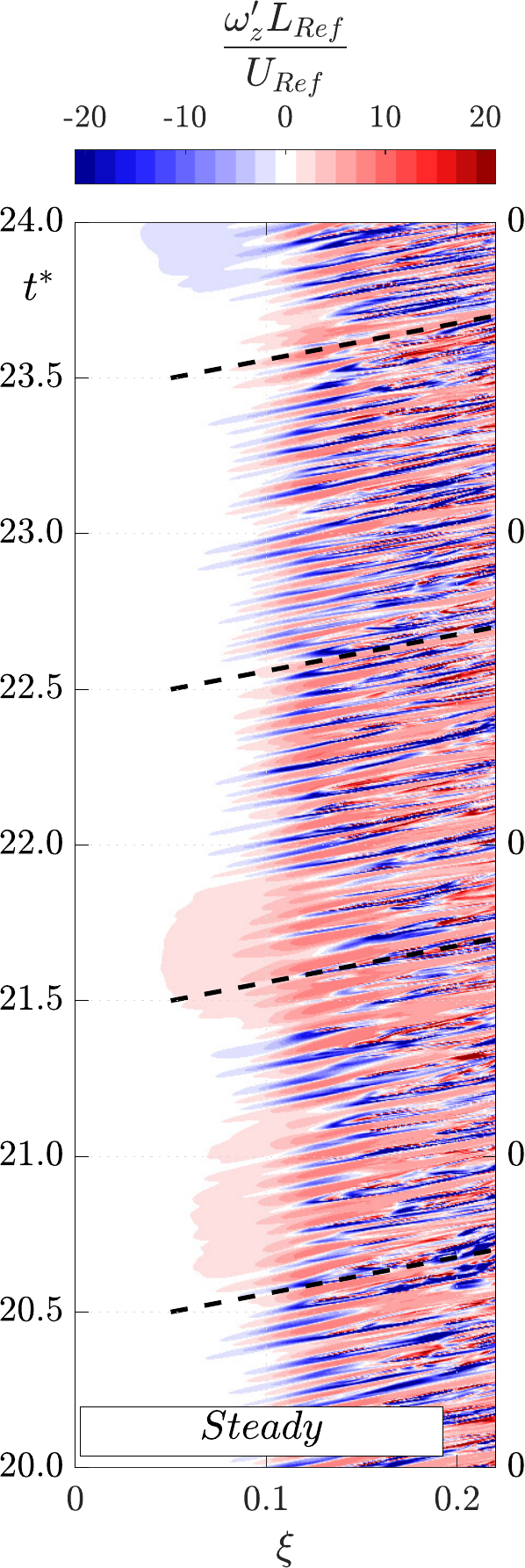} &
\includegraphics[width=.24\textwidth,trim={0in 0in 0in 1.5in},clip] {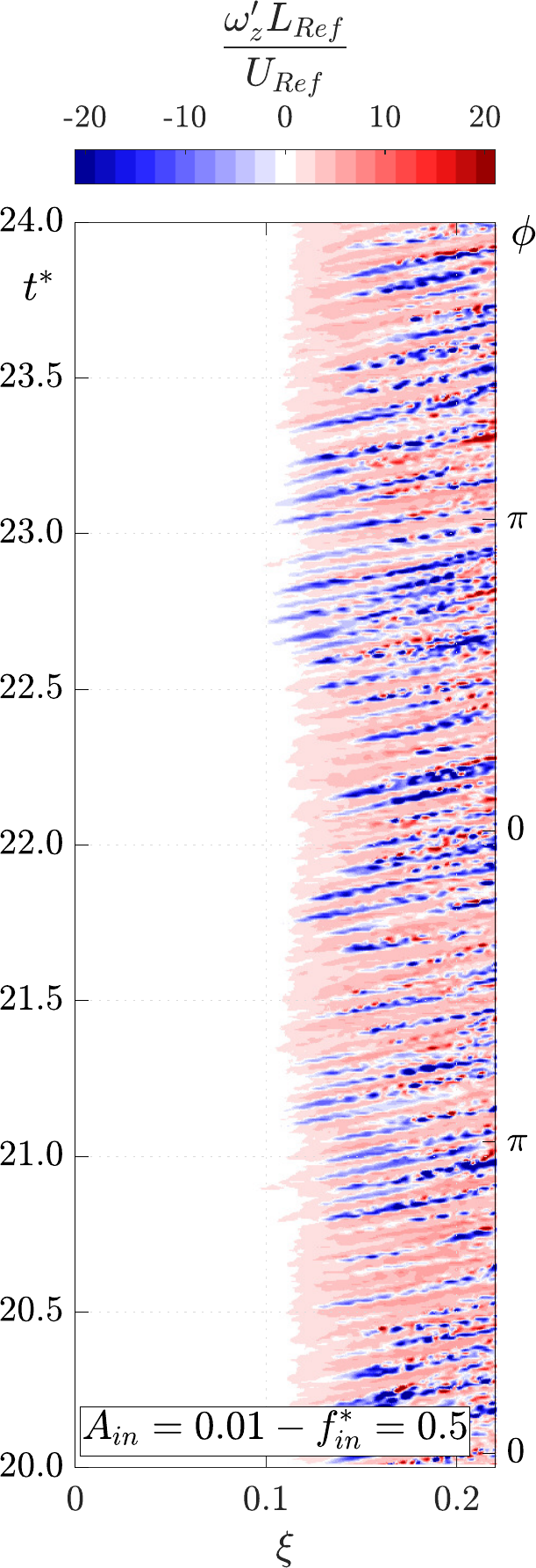} & 
\includegraphics[width=.24\textwidth,trim={0in 0in 0in 1.5in},clip] {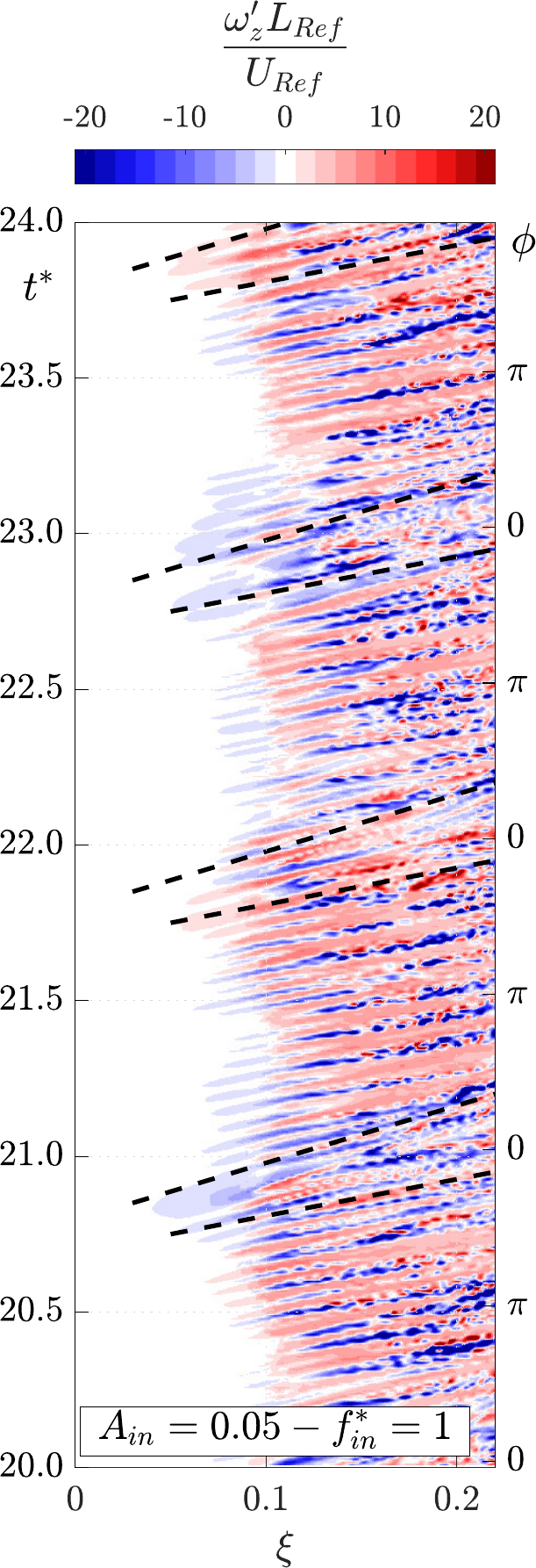} &
\includegraphics[width=.24\textwidth,trim={0in 0in 0in 1.5in},clip] {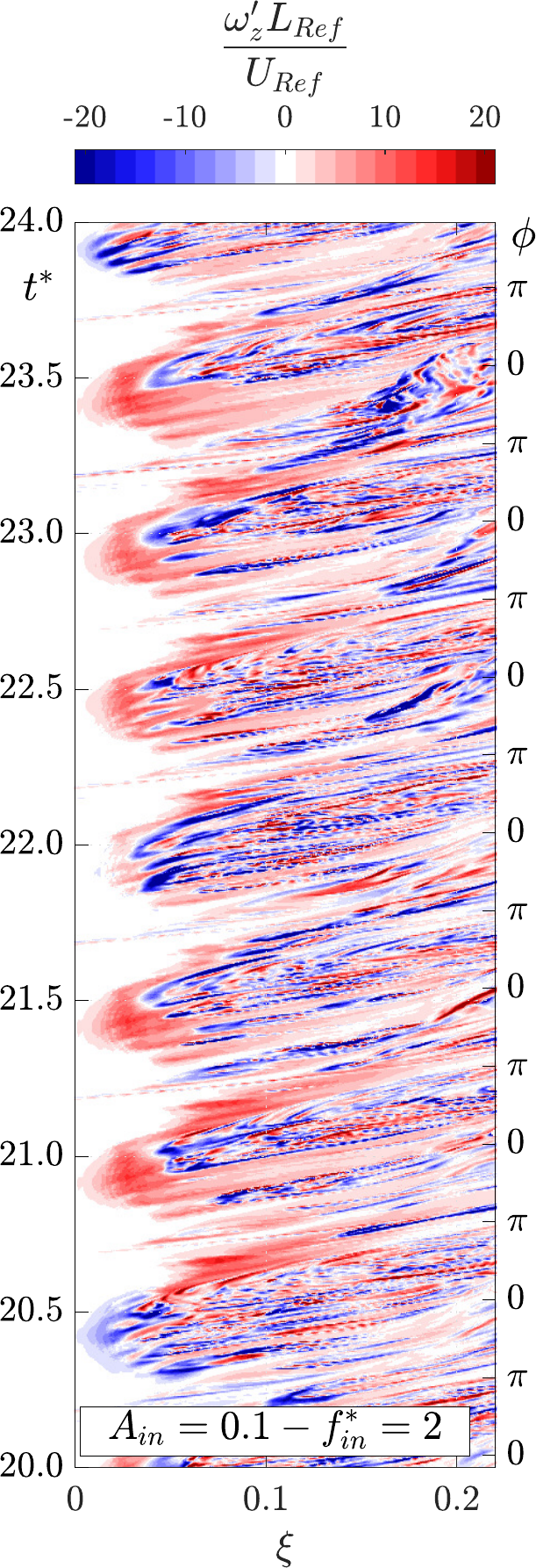} \\
\end{tabular}}
\caption{Incoherent spanwise vorticity at $\eta=0$ for the steady inflow case and the three representative cases with harmonic inflow. }
\label{fig:st_vort}
\end{figure}

The phase-averaged results in the previous section illustrate the overall dynamics of the shear layer and the behaviour of the separated flow on account of the harmonic inflow changes. The length of the recirculation region and the impact of the inflow oscillation frequency and amplitude are, in turn, a consequence of changes in the vortex dynamics induced by the transient changes in the streamwise acceleration. For comparatively low values of $A_{in}$ and $f_{in}^*$, vortex rolls are formed at the separated shear layer resulting from the KH instability, which initiates the laminar-to-turbulent transition. The spread of the shear layer, entrainment of fluid into the recirculation region and ultimately flow reattachment are governed by the complex dynamics of these vortices. These processes are not captured by the phase-averaged fields, which only recover their consequences. 

For low $f_{in}^*$, the characteristic period in which the shear-layer motion occurs is much longer than the characteristic time for the formation and advection of the vortices, as will be shown below. In consequence, the vortex dynamics are completely recovered in the incoherent flow component ${q}'$ of the triple decomposition. The vortices are formed and initially move along the separated shear layer, whose motion is captured by the phase-averaged component. To separate the vortex dynamics from the shear-layer motion, the following procedure is adopted. First, an orthogonal curvilinear coordinate system $(\xi,\eta)$ is defined, as shown in Fig. \ref{fig:phase_average_f05_A001}. The line $\eta=0$ corresponds to the phase-averaged separation streamline and $\xi$ is the curvilinear coordinate measured along it. Then, an inverse transformation is performed to map the curvilinear-coordinates grid points to cartesian coordinates with the streamline as the centerline axis. Details of the inverse transformation process are described by \citet{Legleiter06}. 

This shear-layer fitted system of coordinates is inspired by the one used by \citet{Himeno21} to study the vortex dynamics along the steady shear layer formed within a slat cove.
In the present case, the curvilinear coordinates depend on the phase angle, following the motion of the phase-averaged separation streamline. Figure \ref{fig:td_A001_f05} shows the total (left column), the phase-averaged (middle column) and the incoherent (right column) components of the spanwise vorticity in the curvilinear mesh, for the case representative of the weaker inflow oscillations ($A_{in} = 0.01$, $f^* _{in} = 0.5$). Note that the phase-averaged separation streamline corresponds to the horizontal line $\eta = 0$, and the bump wall is mapped to a curve line with $\eta<0$ whose geometry changes with the phase. 

 The incoherent component portrays a complex arrangement of vortices typical of a transitional shear layer \citep{Marquillie03,DiwanRamesh2009,McAuliffeJTurb2010,Marxen:JFM13}. The process is initiated with the formation of organized pockets of spanwise vorticity in the initial part of the shear layer. These vortices grow in amplitude as they travel downstream along $\eta=0$, soon reaching non-linear amplitudes and then interacting and merging with the recirculating vortical structures adjacent to the wall. 

  To visualize the spatio-temporal dynamics of these vortices, the incoherent spanwise vorticity $\omega'_z$ at $\eta = 0$ is plotted in the $(\xi,t^*)$ plane for the steady inflow case and the three cases representative of harmonic inflow in Fig. \ref{fig:st_vort}. Note that $\omega'_z$ is a disturbance superimposed upon the phase-averaged flow and consequently its positive and negative values do not directly imply vortical structures rotating both in clockwise and counter-clockwise directions.

  The steady inflow case shows the continuous formation of vortical structures of similar amplitude and apparent shedding frequency that propagate downstream at a nearly constant speed along the mean shear layer. With the choice of contour levels used, they become observable around $\xi = 0.1$ m, which for the steady case corresponds to $x\approx 0.2$ m. Case ($A_{in}=0.01$, $f_{in}^*=0.5$) presents a very similar picture, but the location where the vortical structures are first seen now oscillates very mildly following the inflow phase; bulk flow deceleration displaces the observable incoherent vorticity upstream and \textit{vice versa}. The modulation of the vortex shedding location is increased by increasing either $A_{in}$ or $f_{in}^*$.

  The intermediate case ($A_{in}=0.05$, $f_{in}^* = 1$) presents an additional feature for certain regions of the $(\xi,\phi)$ plane. As in the previous cases, regular periodic positive and negative streaks of vorticity are visible, with their initial $\xi$ location following the inflow fluctuation phase. 
  However, for a certain phase range (between $\phi = 3\pi/2 = 270^\circ$ and $\phi=2\pi=360^\circ$) this pattern is replaced by a ``wedge'' of distinct behaviour, as schematized in the corresponding panel of Fig. \ref{fig:st_vort}.
  The wedge originates at the phase $\phi \approx 3\pi/2$, for which $\omega'_z$ reaches observable amplitudes sensibly upstream than for the preceding phases. From this point, two rays depart at different downstream speeds that enclose a region where the vorticity presents a disorganized behaviour. 

  The wedge's origin is coincident with the phase of minimum bulk velocity. This is interpreted in the following manner: during most of the inflow period, the vortex dynamics follow the same qualitative picture as for the lower $A_{in}$ and $f_{in}^*$ cases, characterized by a regular formation of KH vortices for which the shedding location follows the inflow phase. At some instant during the phase-average deceleration ($90^\circ < \phi < 270^\circ$), the streamwise deceleration parameter $\langle K \rangle$ surpasses a threshold value in the region neighbouring the reattachment point. As a result, the KH vortices and other shear layer eddies are not released from the rear part of the reversed flow region but are entrapped in the recirculation region and initiate the formation of a large cluster of vortical structures. Immediately following the beginning of the acceleration phase, the large vortex cluster is released, similar to the shedding of leading-edge vortices in oscillating airfoils (e.g. \citet{Lind:PF16}).

  The vortex cluster is advected downstream during the initial part of the acceleration phase and subsequently, the regular shedding of KH vortices resumes. While a large vortex cannot be clearly discerned in the phase-averaged flow for this case (Fig. \ref{fig:phase_average_f1_A005}), the presence of the vortex lump can be inferred by the changes in the streamline curvature between $\phi = 315^\circ$ and $360^\circ$. 

  Finally, the incoherent vorticity for the case ($A_{in}=0.1$, $f_{in}^*=2$) does not present a regular shedding of KH vortices akin to the steady or low inflow frequency cases. Instead, it is characterized by the continuous appearance of wedges, corresponding to the periodic formation and shedding of large vortex clusters following the inflow changes. With the large vortex cluster being recovered in the phase-averaged flow (Fig. \ref{fig:phase_average_f2_A01}), the incoherent component consists of smaller size eddies that are trapped and evolve inside the recirculation region during the deceleration part of the period and are convected downstream when the vortex cluster is released. Owing to their chaotic nature, these structures are not repeated from cycle to cycle and hence are not coherent with the inflow changes, but their presence in the $\omega'_z$ field allows the location and tracking of the coherent vortex clusters.

\subsection{Frequency spectra}
\label{sec:spectra}

\begin{table*}
\caption{Temporal sampling parameters: $N_s$: total number of snapshots; $\Delta t^*_{s}$: time-step between snapshots used in the analysis; $N_{ss}$: number of snapshots per segment.}
\label{tab:welch_parameter_harmonic}
\begin{ruledtabular}
\begin{center}
\def~{\hphantom{0}}
\begin{tabular*}{.8\textwidth}{@{\extracolsep{\fill}}crcrc}
$f^*_{in}$ & $N_s$ \hspace{4pt} & $\Delta  t_{s}^*$& $N_{ss}$ \hspace{4pt}  & Overlap \\ [1pt] \hline
$steady$ & 2 410 000  & $3\times10^{-5}$   & 66 666 & 50\% \\
0.5&  50 000  &$1\times10^{-3}$  &  2000 &  0\%  \\
 1 &  50 000  & $1\times10^{-3}$   &  2000 &  0\%  \\
2 &  100 000  & $5\times10^{-4}$  &  4000 &  0\%  \\
\end{tabular*}
\end{center}

\end{ruledtabular}
\end{table*}

\begin{figure}
\centering{
\includegraphics[width=0.66\textwidth,trim={0in 0in 0in 0in},clip] {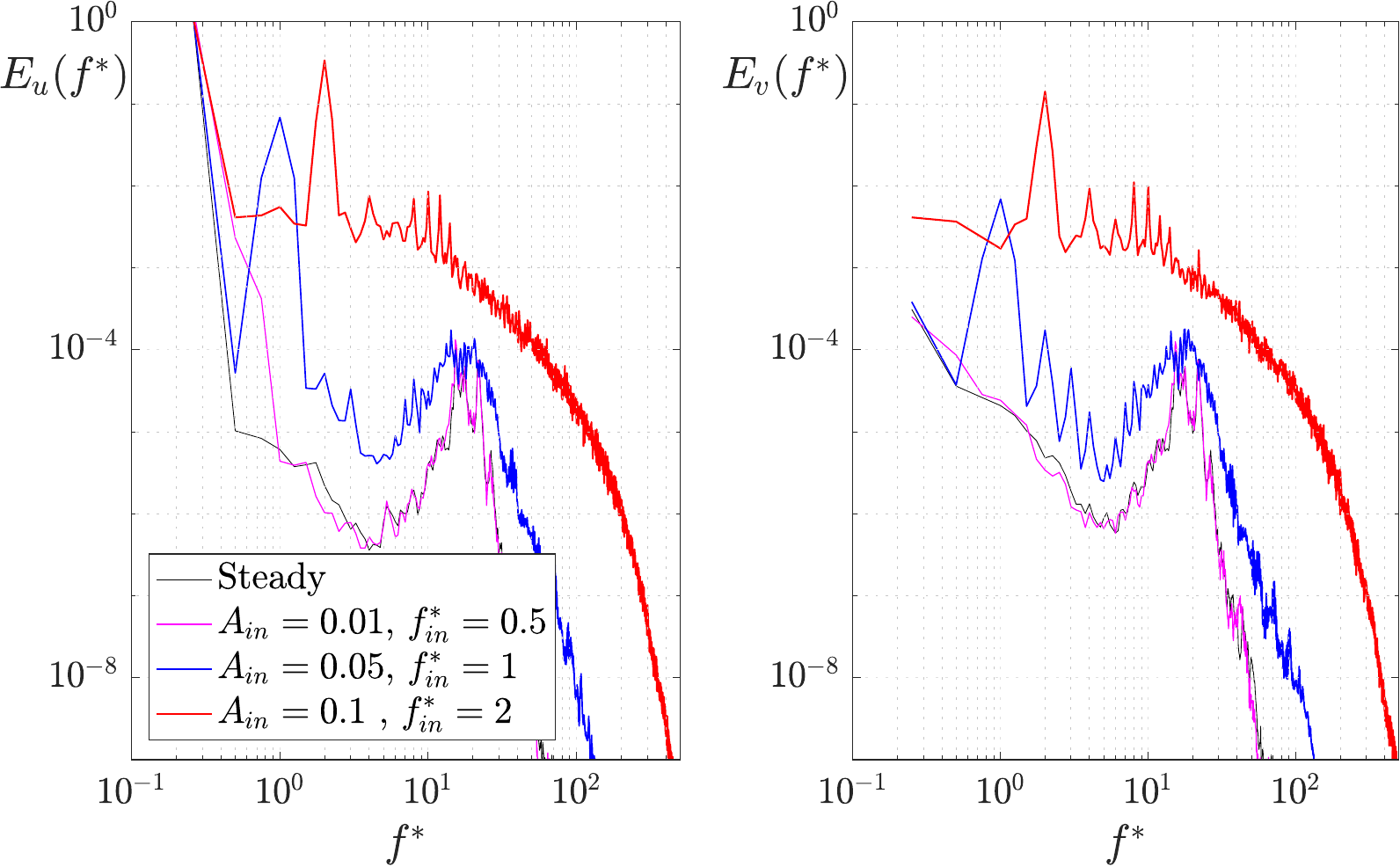} \\
}
\caption{Power spectral density of the streamwise and wall normal velocity at Probe 4, $(x,y)=(0.2,0.05)$. }
\label{fig:psd_ins_uv}
\end{figure}

\begin{figure}
\centering{
\includegraphics[width=1\textwidth,trim={0in 0in 0in 0in},clip] {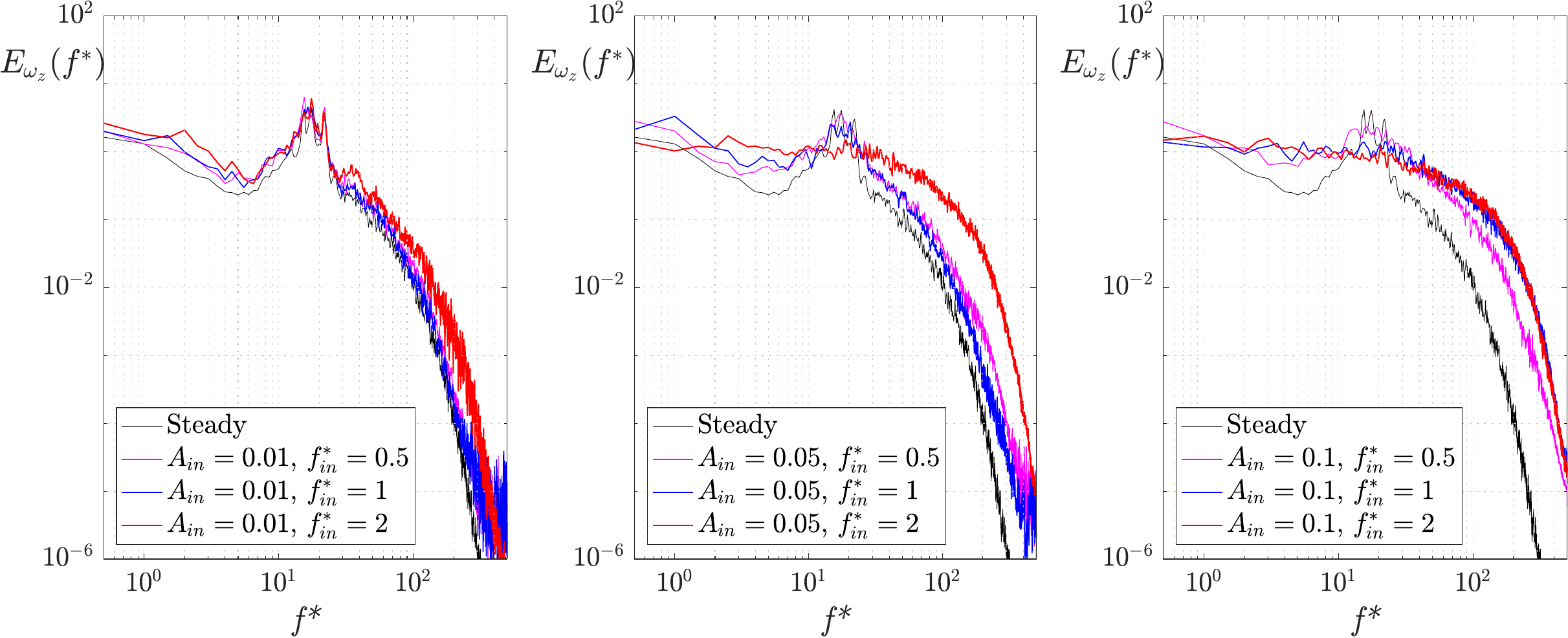} \\
}
\caption{Power spectral density of the incoherent spanwise vorticity $\omega'_z$ at the shear layer, $\xi=0.12$.}
\label{fig:psd_turb_wz}
\end{figure}

Frequency spectra are studied in this section to further the characterization of the impact of the inflow oscillations on the separated flow dynamics. The procedure for the calculation of the power spectral densities (PSD) for the steady inflow case is described in Section \ref{sec:Numerics_convergence}. For the harmonic inflow cases, the procedure is adapted to account for the periodic nature of the data. The sampling parameters are summarized in Table \ref{tab:welch_parameter_harmonic}. To allow for comparisons, the same averaging period is used for all cases, corresponding to the lowest of the inflow frequencies, $\Delta f^* = 0.5$. Thus, the averaging is done over segments comprising one inflow period for $f^*_{in} = 0.5$, two periods for $f^*_{in} = 1$ and four periods for $f^*_{in} = 2$. For all cases, the PSD is averaged over 25 segments of length $1/ \Delta f^*  = 1/0.5 = 2$.

Figure \ref{fig:psd_ins_uv} shows the spectra of the streamwise and wall-normal velocity components at Probe 4 (see Fig. \ref{fig:probes} and Table \ref{tab:Ref_Sampling_Probes}). For the steady inflow case, this probe is located just outside of the separated shear layer at the streamwise location where the first KH vortices are formed. The spectra for the steady inflow and the weak inflow oscillation ($A_{in} = 0.01$, $f^*_{in}=0.5$) cases are identical, presenting a narrowband peak at frequencies $f^* = 15-21$ with a maximum at $f^* \approx 18$. As shown in Appendix \ref{sec:appendix_KH}, this frequency corresponds to KH instability and follows accurately the scaling proposed by \citet{DiwanRamesh2009}. The intermediate case ($A_{in}=0.05$, $f^*_{in}=1$) also exhibits the narrowband peak corresponding to the KH vortices; the peak amplitude and frequency remain comparable to that of the steady forcing case, but the sidebands are broader. However, new peaks appear for the inflow frequency and its harmonics. The $f^*_{in}$ peak has an amplitude that is two orders of magnitude larger than the peak KH frequency. The spectra for the case with the strongest inflow oscillation ($A_{in}=0.1$, $f^*_{in}=2$) also contain the peaks corresponding to the inflow frequency and its harmonics. However, the narrowband peak corresponding to KH vortices is not observable in this case. The amplitude for all frequencies is increased above those corresponding to KH vortices. In consequence, KH vortices could still exist but be shadowed by other, more energetic, fluctuations.

To shed light on the last point, the PSD of the incoherent spanwise vorticity $\omega'_z$ is calculated at a location at the shear layer corresponding to $\xi=0.12$ for all the cases simulated and shown in Fig. \ref{fig:psd_turb_wz}. Left, centre and right subfigures correspond to the lower, intermediate and higher amplitude of the inlet oscillation $A_{in}$, respectively. For each of them, the three frequencies are shown. The spectra for the steady inflow case are also shown in the three figures for reference. 

The spectra for the lower $A_{in}$ cases (Fig. \ref{fig:psd_turb_wz}, left) are qualitatively identical, being dominated by KH vortices. Roughly the same amplitudes are obtained for the frequencies associated with the KH vortices for the three values of $f^*_{in}$, while the higher frequency range, corresponding to the turbulent cascade, is found to be slightly more energetic with increasing $f^*_{in}$.

Conversely, the spectral for the intermediate $A_{in}$ (Fig. \ref{fig:psd_turb_wz}, middle) shows qualitative changes that occur gradually as the inflow frequency $f^*_{in}$ is increased. The amplitude of the KH narrowband peak is reduced while it is increased for all the other frequencies. For the largest inflow frequency, $f^*_{in}=2$, the KH peak is not present anymore and the amplitude at the corresponding frequency is lower than for the cases with KH vortices. This shows that the KH vortex shedding is not shadowed by more energetic fluctuations, but rather eliminated. Finally, the spectral for the largest $A_{in}$ (Fig. \ref{fig:psd_turb_wz}, right) follows the same trend as the intermediate $A_{in}$ ones.

The changes observed in the PSD with increasing $A_{in}$ and $f^* _{in}$ are consistent with the vortex dynamics discussed in Section \ref{sec:Vortex_dynamics}. As the large vortex cluster is formed, most of the vortical structures present in the separated shear layer get trapped in the recirculation region instead of being shed and advected downstream. Non-linear interactions between the recirculating structures lead to their merging and progressive breakdown into smaller structures, which leads to a more energetic and flatter spectral. In turn, the recirculation of random eddies of diverse scales prevents the formation of well-defined KH vortices in the separated shear layer.

\begin{figure*}
\centering{\includegraphics[width=.78\textwidth,trim={0in 0.4in 0in 0in},clip] {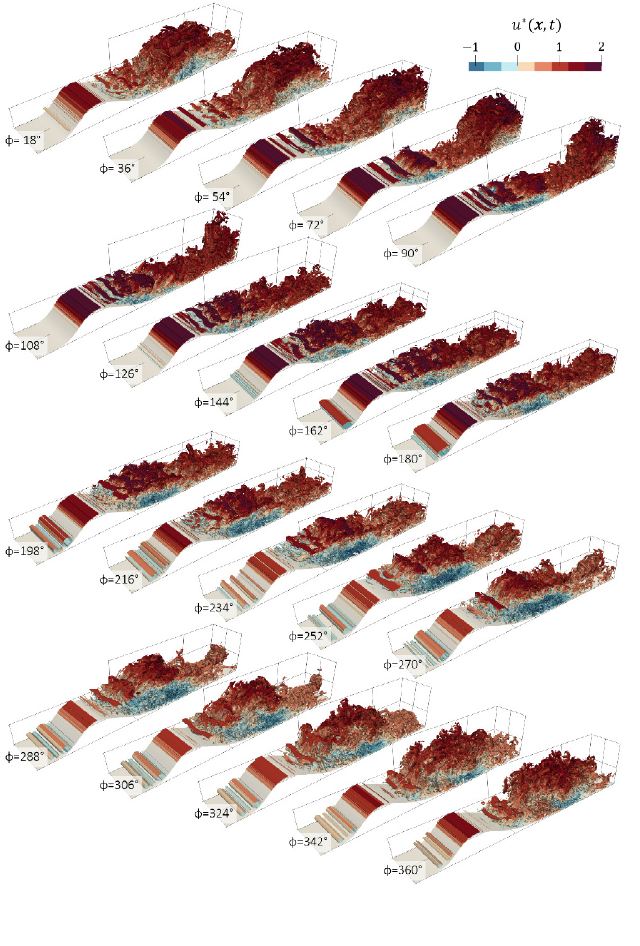}}
\caption{$Q(+)$ 
isosurface coloured by streamwise velocity. $A_{in}=0.1$ and $f_{in}^*=1$.}
\label{fig:Qcriterion_A01_F1}
\end{figure*}


\section{Discussion and conclusions}
\label{sec:Conclusion}

This paper studies the impact of harmonic oscillations of the inflow velocity, imposed via the total pressure, on the flow over a wall-mounted bump geometry. This geometry gives rise to a streamwise pressure gradient distribution with similar features to those encountered in LPT blades. The harmonic inflow oscillation roughly models the effect of the passage of the wake due to the previous stage of blades, periodically creating a velocity deficit followed by acceleration, and consequently a periodic modification of streamwise velocity gradients.

Under steady inflow conditions, the laminar-to-turbulent transition is initiated by a self-excited KH instability with a well-defined vortex-shedding frequency ($f^*_{KH} \approx 18$). These vortices are subject to secondary instabilities and interactions with turbulent structures that are recirculated within the separated flow and progressively break down into smaller eddies as they travel downstream. The associated entrainment leads to the reattachment of the mean flow.  

The impact of the inflow oscillations on the flow dynamics is strongly dependent on the frequency and amplitude of the oscillations. The cases studied involve oscillations of the inflow total pressure $A_{in}$ between 1\% and 10\% of the mean value, and frequencies $f^*_{in} = 0.5 - 2$, substantially lower than $f^*_{KH}$ but comparable to the wake-passing frequency in LPT turbines. Three different scenarios have been identified (Appendix \ref{sec:appendix_scenarios} shows the classification of the cases):

Scenario (i): \emph{Inflow-modulated Kelvin-Helmholtz vortex shedding}, corresponding to comparatively weak inflow oscillations (lowest values of $A_{in}$ and $f^*_{in}$, cf. Fig \ref{fig:Qcriterion_A001_F05}). The transition process is qualitatively the same as for the steady inflow case. However, the harmonic changes of the bulk velocity lead to a periodic vertical displacement of the separated shear layer, phase-locked to the inflow oscillation. While the self-sustained vortex shedding originated by KH instability remains, its spatial amplification is modulated by the changes in the shear layer properties: vortex-shedding takes place upstream for the part of the period in which the bulk flow is decelerated and downstream for the part in which it is accelerated. Consequently, the phase-averaged length of the recirculation region $L_s$ changes gradually over the period, but the mean length (i.e. averaged over the period) is approximately the same as for the steady case. 

Scenario (ii): \emph{Alternation between KH vortex shedding and formation/release of a large vortex cluster}, corresponding to intermediate inflow oscillations (see Fig. \ref{fig:Qcriterion_A01_F1} and supplementary movies 3 and 4). For a portion of the period, KH-initiated vortex shedding is dominant, identical to scenario (i).
However, as the bulk flow decelerates, the phase-averaged streamwise acceleration parameter $\langle K \rangle$ transiently surpasses a threshold negative value in the region neighbouring flow reattachment, giving rise to new dynamics: the vortical structures formed in the shear layer are not shed and advected downstream. Instead, they are entrapped in the recirculation region, which accordingly grows in size forming a large vortex cluster formed by eddies of a wide range of scales. When the bulk flow is re-accelerated, instead of reducing the recirculation region gradually through the shedding of KH vortices (as in scenario (i)), the large vortex cluster is released and advected downstream. The advection of the vortex cluster pulls the stagnant fluid, transiently reducing the separated flow extent. Subsequently, the recirculation region regenerates and the KH vortex shedding re-starts.
The periodic formation and release of the large vortex cluster impact the phase-averaged length of the reversed flow region drastically: the time-averaged $L_s$ is reduced with respect to the steady inflow case, but the deviations from the mean value (see the PDFs in Fig. \ref{fig:pdf_length}) become considerably broader including lengths longer than those for the steady inflow. Important hysteresis effects appear between the accelerating and decelerating parts of the period. This scenario would translate into undesirable conditions regarding practical LPT blades, involving strong oscillatory loads without a substantial reduction of the mean separation length.

Scenario (iii): \emph{Phase-locked formation and release of large vortex clusters}, corresponding to strong inflow oscillations (the largest values of $A_{in}$ and $f^*_{in}$, cf. Fig. \ref{fig:Qcriterion_A01_F2}). The dynamics of the formation and release of the large vortex cluster following the inflow oscillations occupy all periods: before one vortex cluster has time to travel downstream a distance sufficient for the recirculation region to re-initiate the KH vortex shedding, the threshold value of the streamwise acceleration parameter is reached and a new vortex cluster is being formed. The deviation of the separated flow length over the period is larger than that of scenario (i) but less than that of scenario (ii). More importantly, the phase-averaged $L_s$ is smaller than the mean length for the steady inflow case for all the phases and the time-averaged $L_s$ is remarkably reduced, above a 40\% for the case ($A_{in}=0.1$, $f^*_{in}=2$). Regarding a practical LPT application, this scenario would be preferable over the other two.

In order to clarify if our conclusions regarding the three different scenarios are general or particular to wall-mounted bump geometry considered so far, Appendix \ref{sec:appendix_NASA_hump} briefly presents an analogous study considering the related setup of the NASA hump. The same three scenarios are recovered, while the combination of values of $A_{in}$ and $f^*_{in}$ for each of them is changed.

It is to be noted that the impact of the inflow oscillations is not related to the individual parameters $A_{in}$ or $f^*_{in}$ but to a combination of both. In all the cases $A_{in}$ is too large to be considered a linear flow disturbance, and $f^*_{in}$ is an order of magnitude lower than the natural frequency of the Kelvin-Helmholtz instability, $f^*_{KH}$. The transition between the different scenarios is thus not associated with the excitation of the KH instability. The transition between scenarios (i) and (ii) is related to the existence of a threshold value of the acceleration parameter $\langle K \rangle$. When this value is exceeded transiently in the region towards the end of the bump, a large vortex cluster is formed and eventually released. 
According to the cases simulated, the threshold value is bounded between $\langle K \rangle = -16.71\times 10^{-6}$ and $-22.43\times 10^{-6}$ (see Appendix \ref{sec:appendix_scenarios}). However, the numerical value of this parameter is particular to the definition  of $\langle K \rangle$ used, which is based on an arbitrary $y$ coordinate. On the other hand, the transition between scenarios (ii) and (iii) is related to the ratio between the inlet oscillation period and the time required for the recirculation region to regenerate and re-start the KH vortex shedding after the release of one vortex cluster.  

Our results cannot ascertain if a further increase in the inflow frequency would lead to further reductions in the time-averaged separated flow length. Within scenario (iii), the optimal time-averaged $L_s$ reduction would be achieved by a balance between increasing the amount of recirculating fluid advected with one vortex cluster (i.e. the size of the phase-averaged vortex) and increasing the frequency of release of such clusters. However, the impact of inflow oscillations on the dynamics and time-averaged length of separated flows described in this work presents similarities with studies of active flow control via harmonic suction/blowing or moving parts reviewed by \citet{GreenblattPAS2000} and with the experiments by \citet{hasan_1992, sigurdson_1995}. These works report the existence of an optimal forcing frequency that scales with the global size of the recirculating flow region and the free-stream inflow velocity. Forcing at this frequency range promotes the phase-locked formation and shedding of large vortices similar to present scenario (iii), altering the dynamics from those of the KH-initiated transition (scenario (i)). \citet{GreenblattPAS2000} concluded that the optimal forcing frequency scales with the streamwise length of the unforced recirculating flow region, $L_s$. Conversely, \citet{hasan_1992, sigurdson_1995} argue that the natural frequency for the ``shedding of the entire bubble'' (i.e. release of the large coherent vortex cluster) scales with its wall-normal height $h$, in analogy to the von Karman vortex street, and the optimum forcing would act at this frequency. \citet{sigurdson_1995} proposed a dimensionless frequency $St_{shedding} = f h / U_s = 0.07 - 0.08$ (where $U_s$ is the free-stream velocity at the separation point) and demonstrated that it correlates well with a variety of geometry-induced separation bubbles. Translated to the dimensionless form used herein, the global shedding frequency results $f^*_{shedding} = 2.9- 3.3$, which is only slightly higher than the largest $f^*_{in}$ considered.
 
On the other hand, forcing with higher frequencies, aimed at exciting the -local- KH instability, may lead to a faster transition via scenario (i) \cite{Embacher:JFM2014} or by inducing pairing of KH-vortices \cite{Marxen:JFM13,Kurelek:PRF2019}. This certainly leads to reductions in the time-averaged separation length, but in view of present results, these reductions are expected to be smaller than the ones achieved by exciting the global dynamics of the separated flow.


\begin{acknowledgments}{
This project has received funding from the European Union's Horizon 2020 research and innovation programme under the Marie Skłodowska Curie grant agreements No 955923-SSECOID and 101019137-FLOWCID. D.R. also acknowledges funding by the Government of the Community of Madrid within the multi-annual agreement with Universidad Polit\'ecnica de Madrid through the Program of Excellence in Faculty (V-PRICIT line 3).

The authors gratefully acknowledge the ``Centro de Supercomputaci\'on y Visualizaci\'on de Madrid'' (CESVIMA) and Universidad Polit\'ecnica de Madrid for providing computing resources on Magerit Supercomputer.}
\end{acknowledgments}


\appendix

\section{Frequency of the Kelvin-Helmholtz instability}
\label{sec:appendix_KH}
\begin{figure}
\centering{
\includegraphics[width=0.8\textwidth,trim={0in 0in 0in 0in},clip] {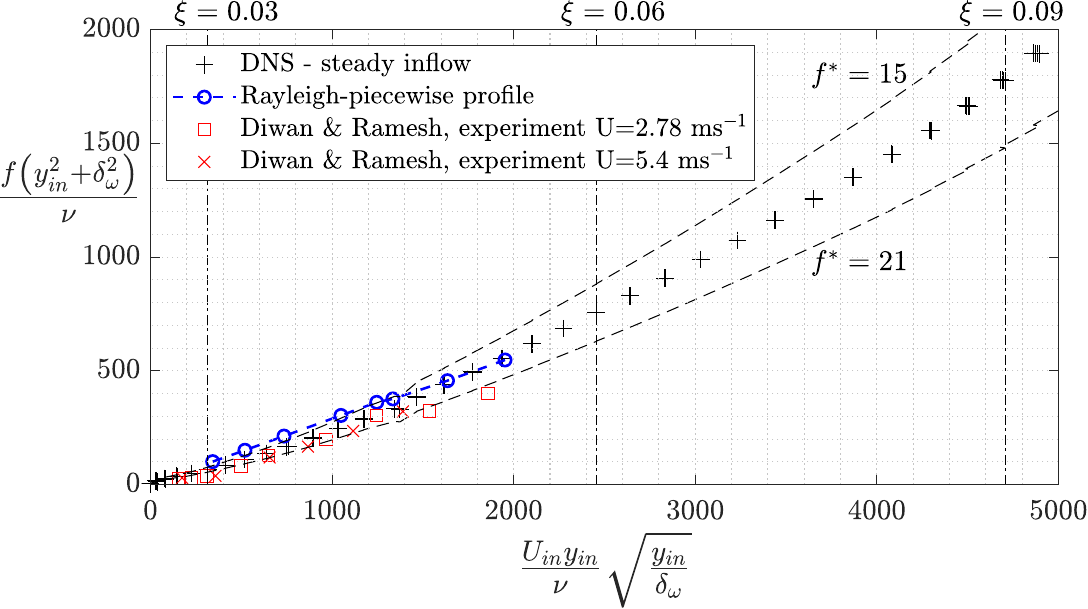} \\
}
\caption{Scaling of the most amplified frequency along the separated shear layer.}
\label{fig:scaling}
\end{figure}

\citet{DiwanRamesh2009} proposed a scaling of the frequency corresponding to the most amplified KH instability based on the linear instability analysis of piecewise linear velocity profiles resembling separated shear layers in the presence of a wall. This scaling combines the local (i.e. at individual streamwise locations) properties of the flow: the wall-normal coordinate and streamwise velocity of the inflection point ($y_{in}$ and $U_{in}$), the vorticity thickness $\delta_\omega$, and the most amplified frequency $f$, into two dimensionless parameters: the dimensionless frequency $f\left(y^2_{in} + \delta^2 _{\omega} \right)/\nu$ and a modified Reynolds number $\bar{R}=\left(U_{in} y_{in} / \nu \right)\sqrt{y_{in}/\delta_{\omega}}$. A linear dependence between the two numbers is postulated. In a practical flowfield, the separated shear layer evolves downstream, which translates into increasing values of $\bar{R}$. As the scaling stems from inviscid linear instability, its predictions are expected to hold up to limited $\bar{R}$ values. \citet{DiwanRamesh2009} demonstrated that the scaling correctly predicts the KH frequency of several different experiments, up to $\bar{R} = 1400$. 

To ascertain that the narrowband frequency peak of $f^* \approx 18$ identified in the PSD for the steady inflow case corresponds to KH instability, the scaling is applied to our simulation data and shown in Fig. \ref{fig:scaling}. The theoretical and experimental results of \citet{DiwanRamesh2009} are also shown for comparison. This comparison concludes that $f^* = 18$ indeed corresponds to the KH instability. 

\section{Classification of the harmonic inflow cases.}
\label{sec:appendix_scenarios}

\begin{figure}
\centering{
\includegraphics[width=.4\textwidth,trim={0in 7.4in 0.0in 0in},clip] {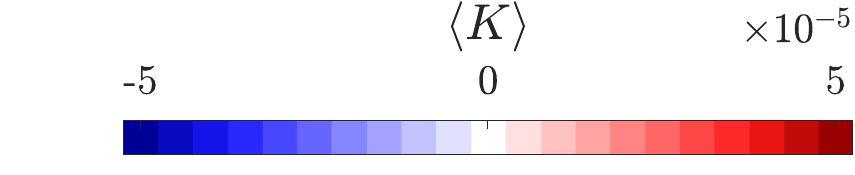} \\
 \vspace{5mm}
\begin{tabular}{ccc}
\includegraphics[width=.342\textwidth,trim={0in 0.9in 0in 1.4in},clip] {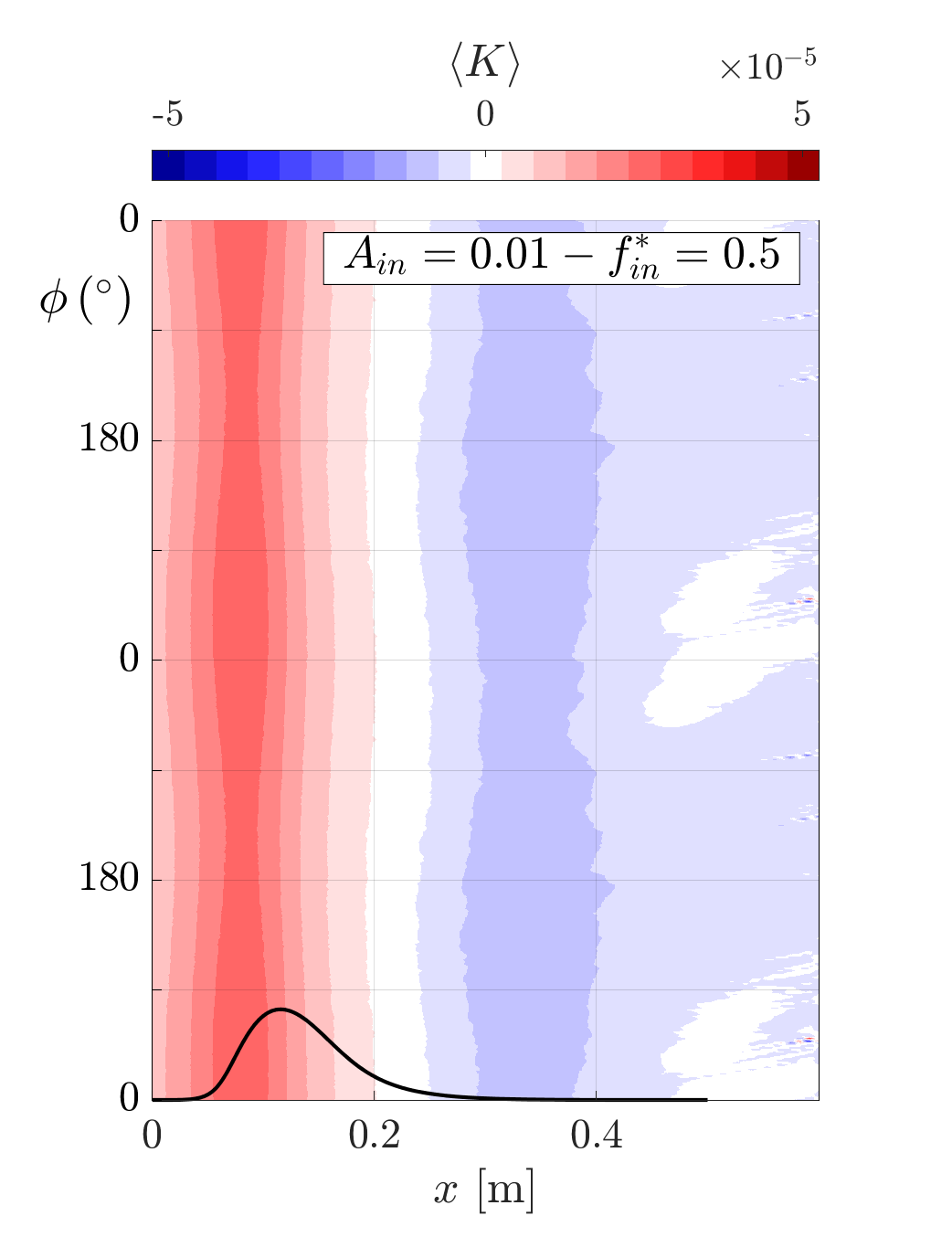}  &
\includegraphics[width=.29\textwidth,trim={1.05in 0.9in 0in 1.4in},clip] {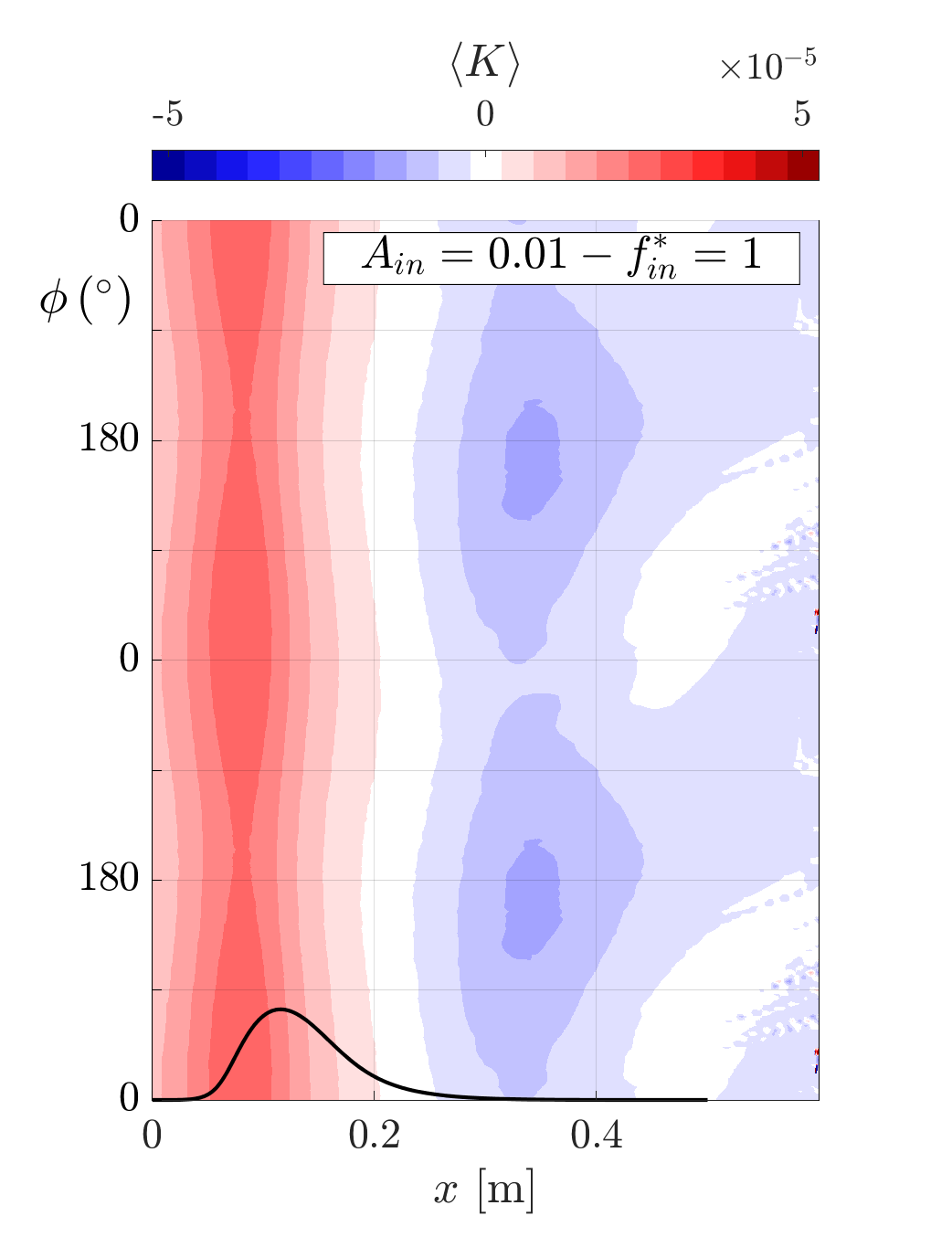}  &
\includegraphics[width=.29\textwidth,trim={1.05in 0.9in 0.0in 1.4in},clip] {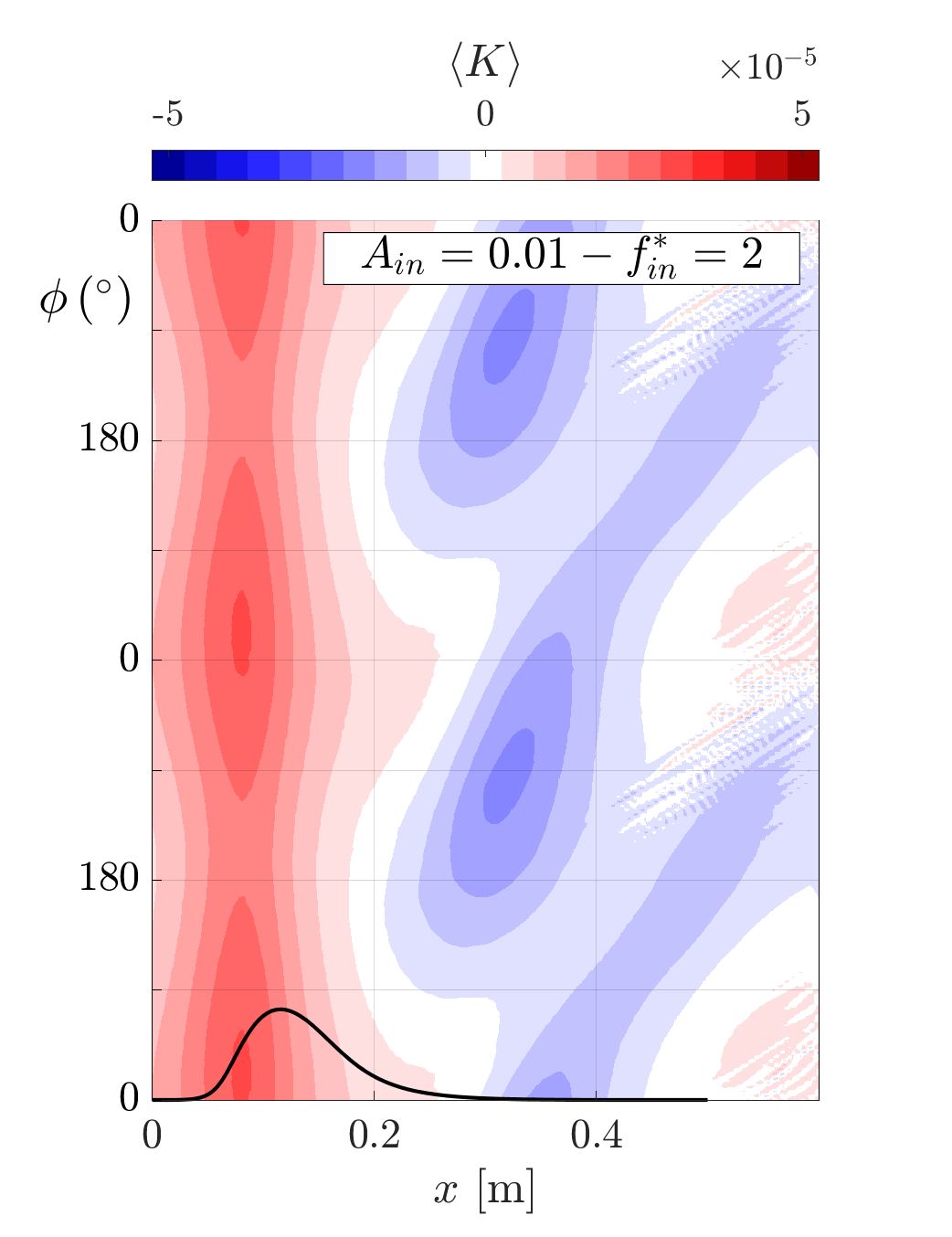} \\
\vspace{1mm} \\
\includegraphics[width=.342\textwidth,trim={0in 0.9in 0in 1.4in},clip] {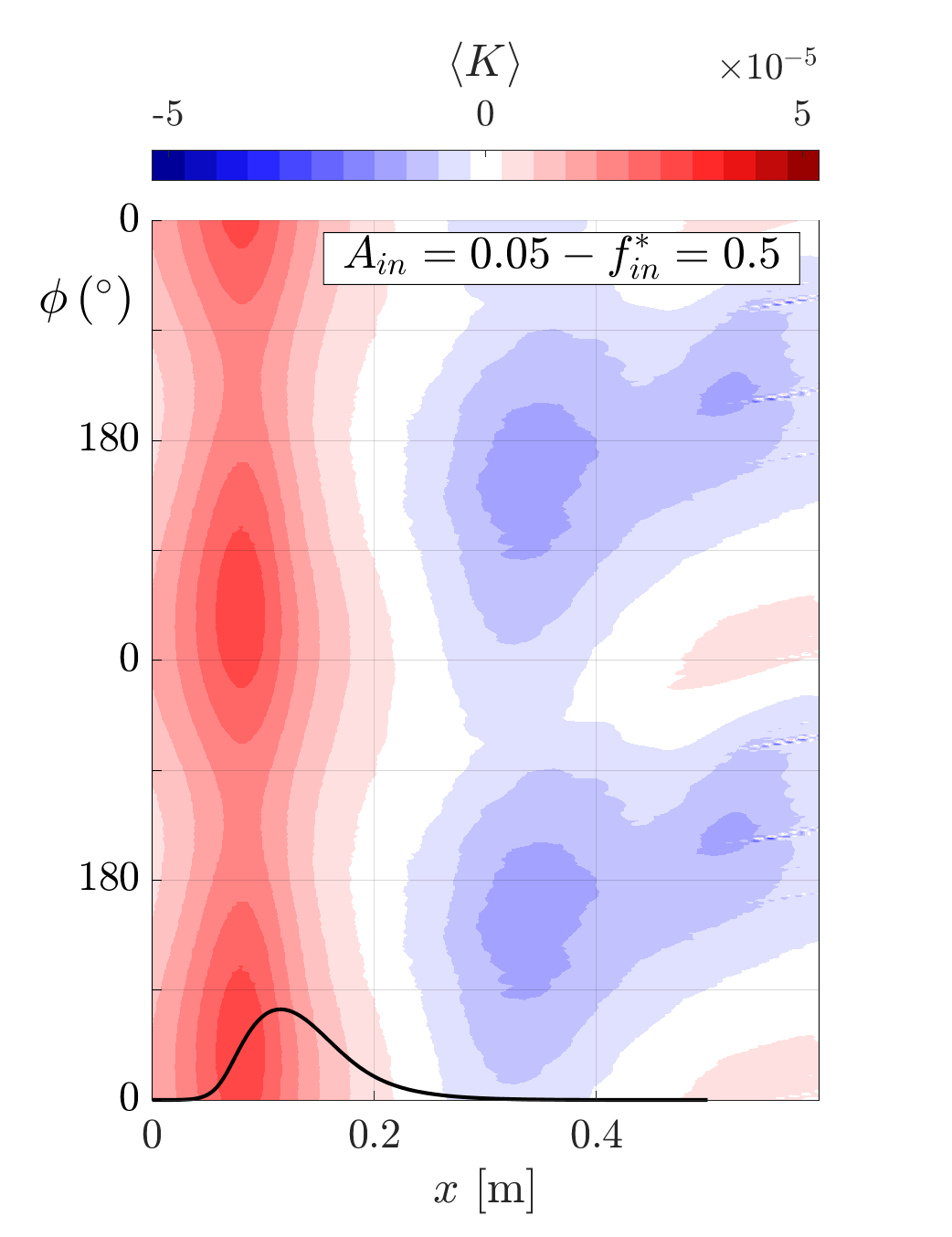}  &
\includegraphics[width=.29\textwidth,trim={1.05in 0.9in 0in 1.4in},clip] {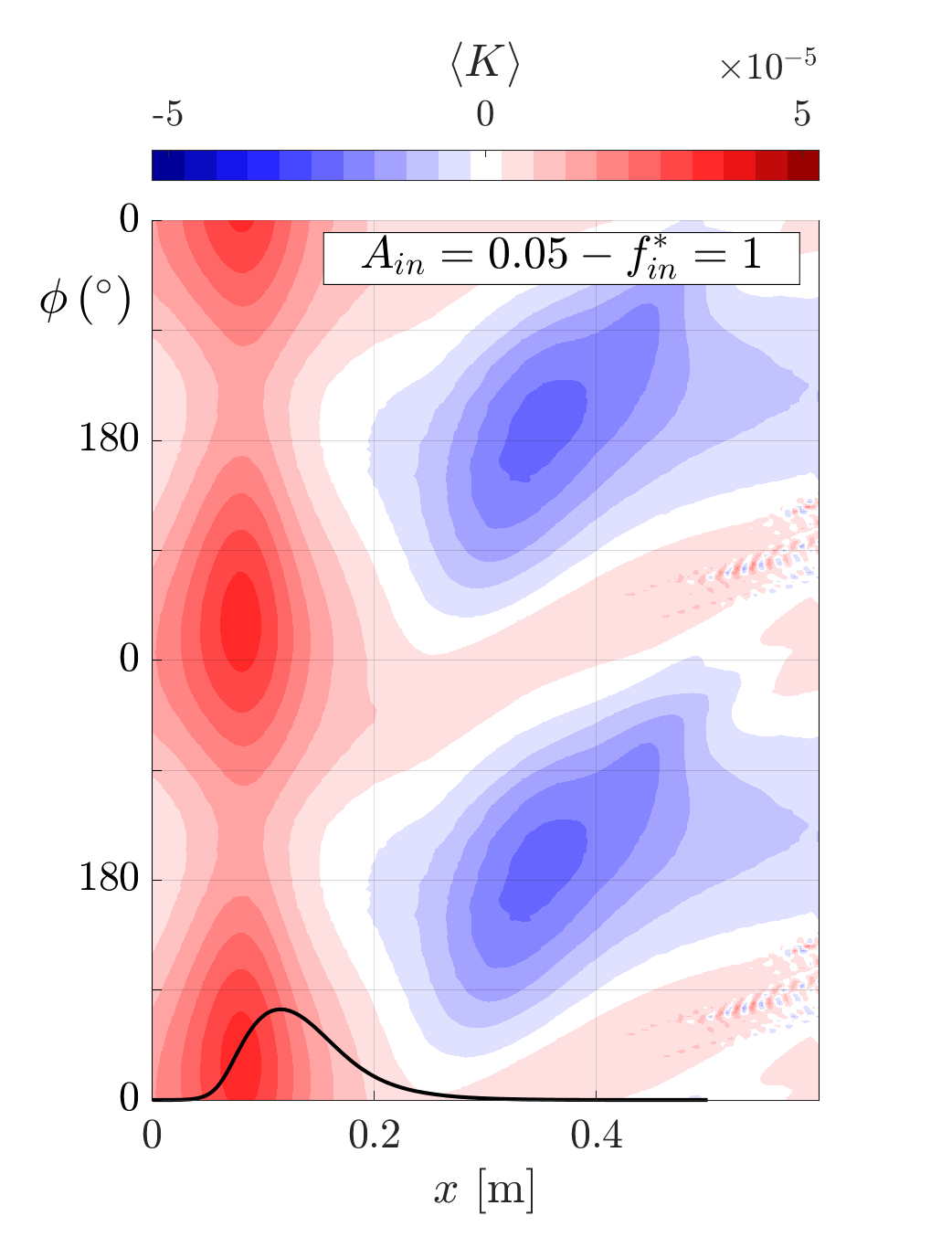} &
\includegraphics[width=.29\textwidth,trim={1.05in 0.9in 0.0in 1.4in},clip] {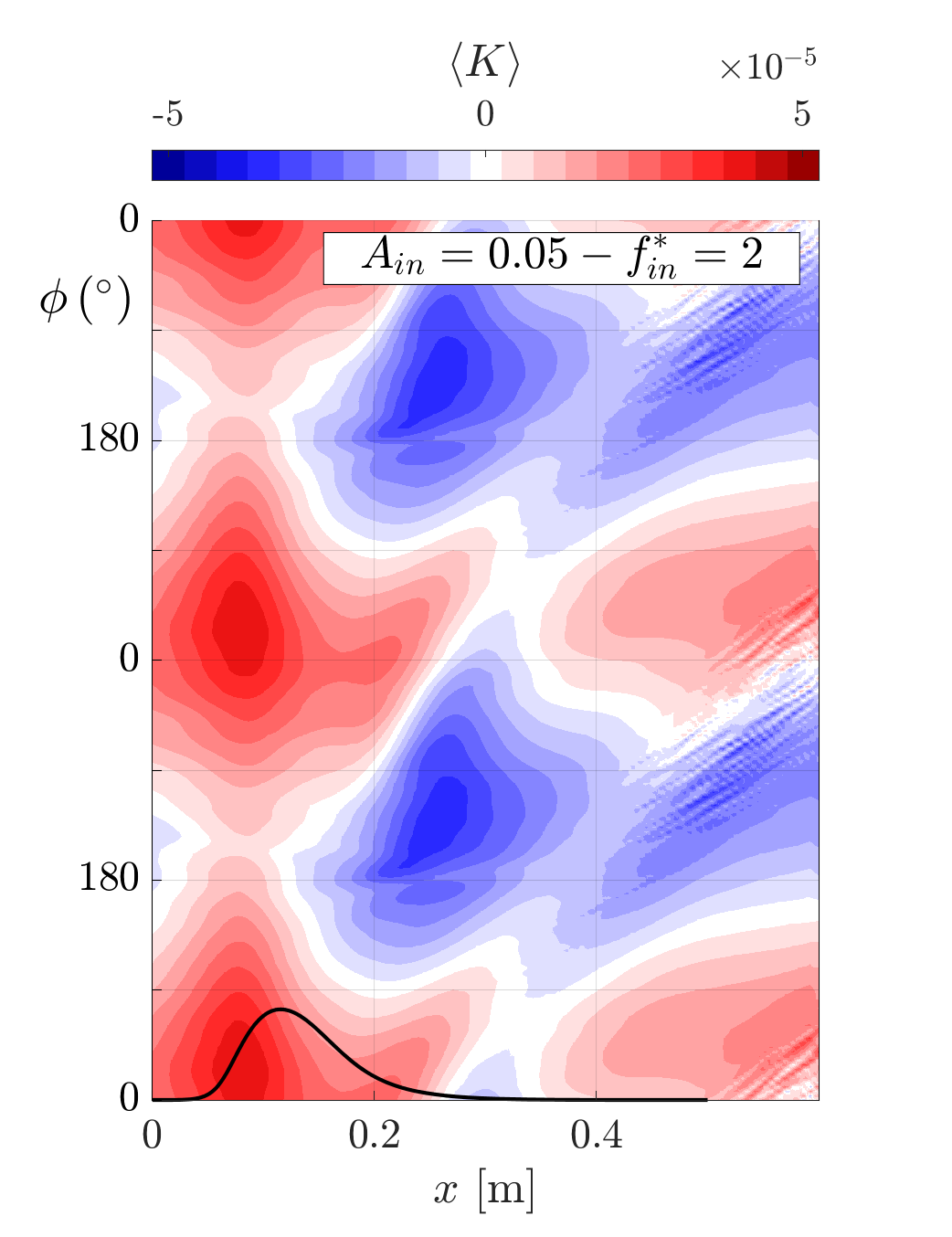} \\
\vspace{1mm} \\
\includegraphics[width=.342\textwidth,trim={0in 0in 0in 1.4in},clip] {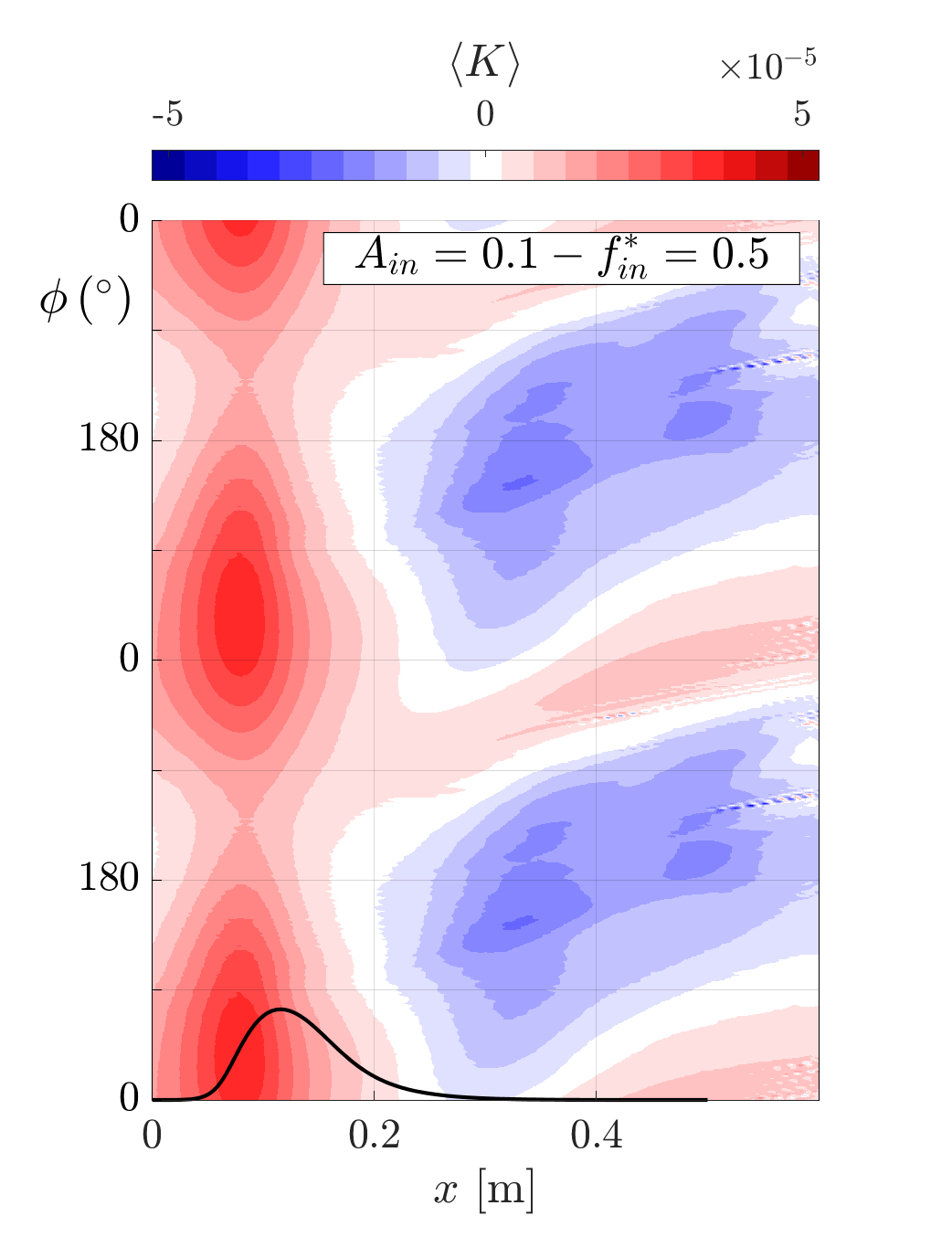}  &
\includegraphics[width=.29\textwidth,trim={1.05in 0in 0in 1.4in},clip] {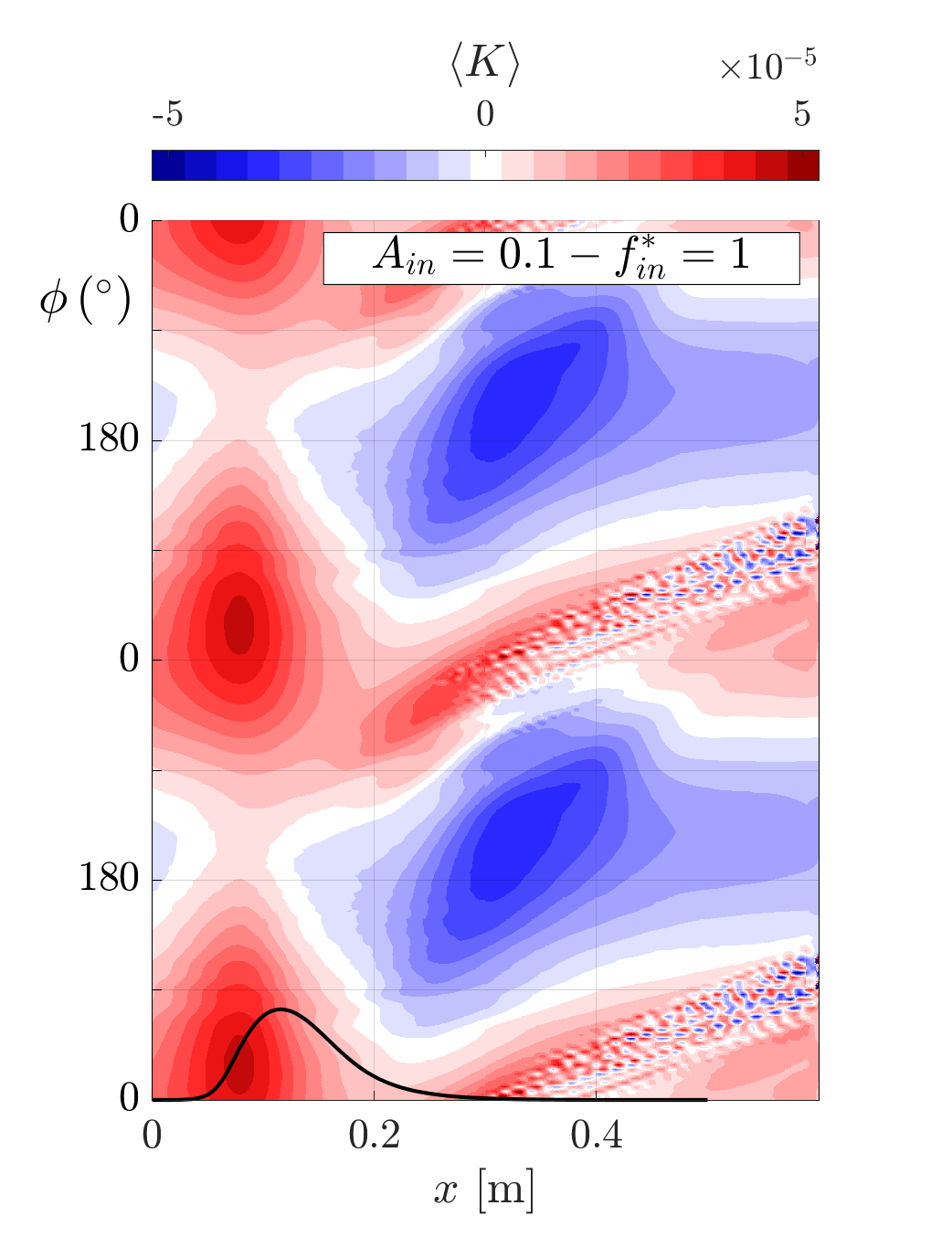}  &
\includegraphics[width=.29\textwidth,trim={1.05in 0in 0.0in 1.4in},clip] {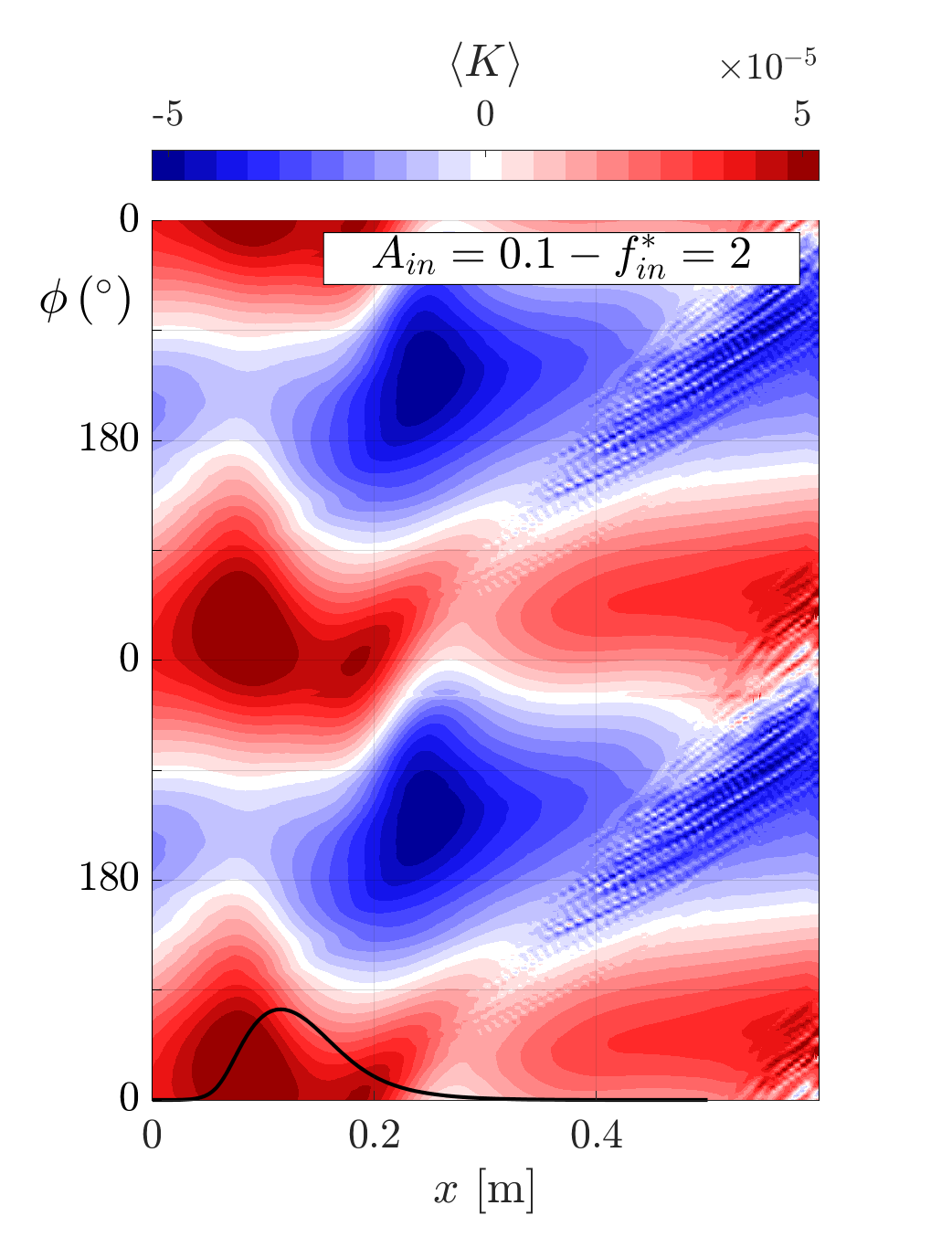} \\
\end{tabular}}
\caption{Phase-averaged streamwise acceleration parameter.
}
\label{fig:appendix_acceleration_sum}
\end{figure}

\begin{figure}
\centering{
\includegraphics[width=.44\textwidth,trim={0in 7.4in 0.0in 0in},clip]  {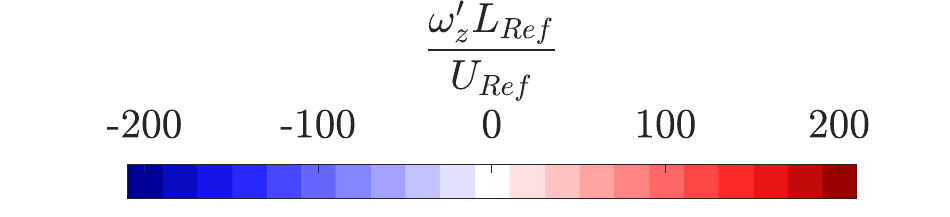} \\
 \vspace{2mm}
\begin{tabular}{ccc}
\includegraphics[width=.342\textwidth,trim={0in 0.9in 0in 1.4in},clip] {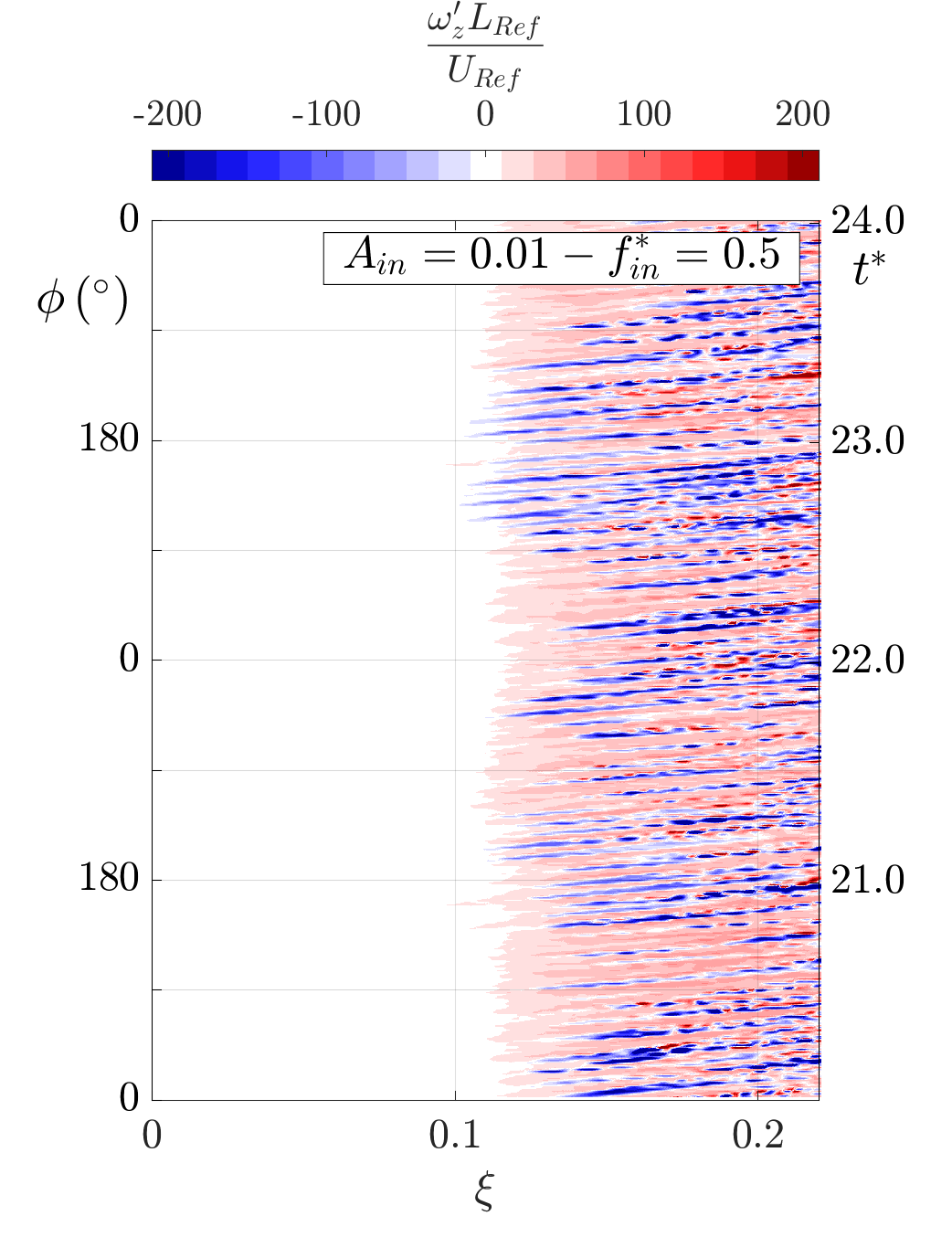}  &
\includegraphics[width=.29\textwidth,trim={1.05in 0.9in 0in 1.4in},clip] {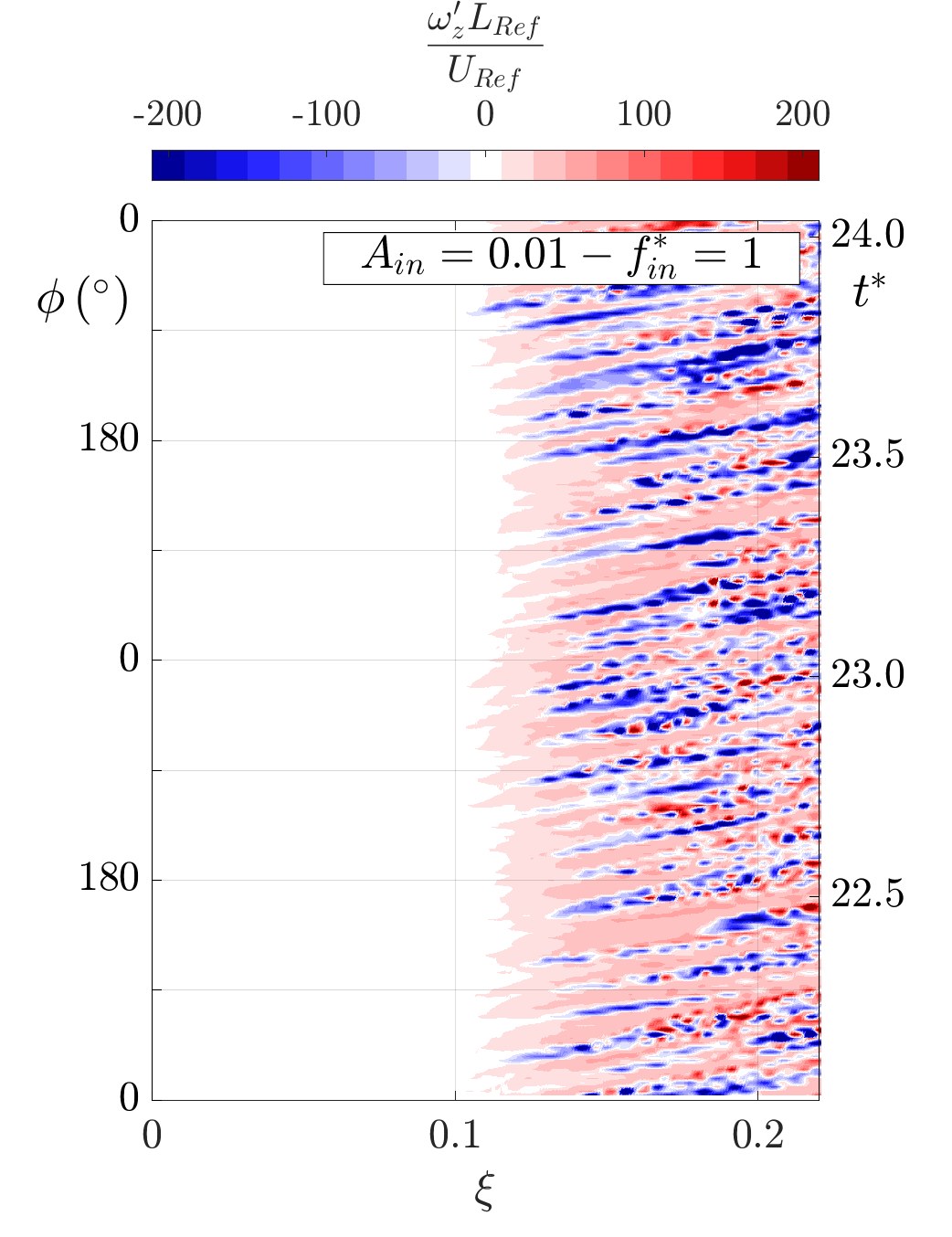}  &
\includegraphics[width=.29\textwidth,trim={1.05in 0.9in 0.0in 1.4in},clip] {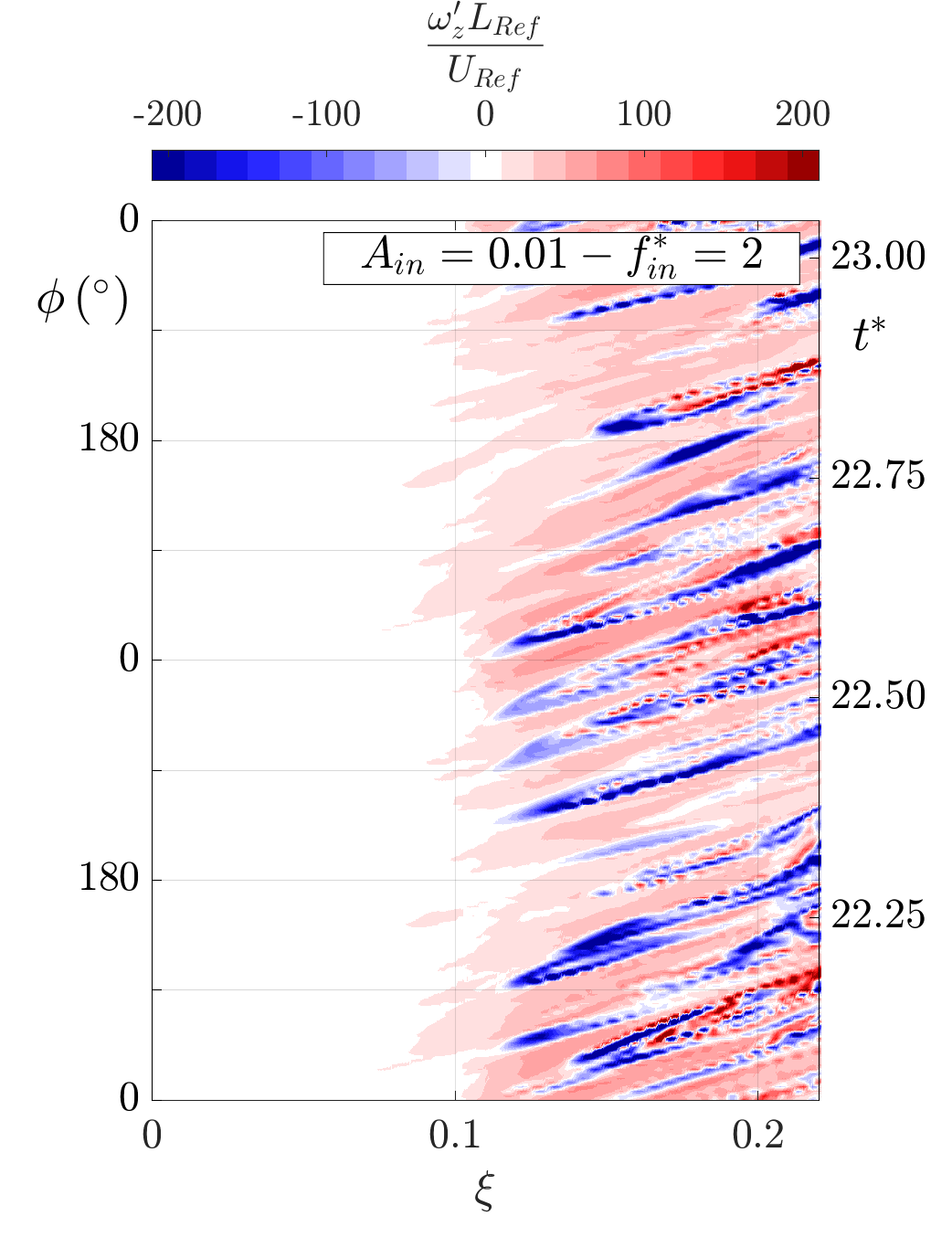} \\
\vspace{1mm} \\
\includegraphics[width=.342\textwidth,trim={0in 0.9in 0in 1.4in},clip] {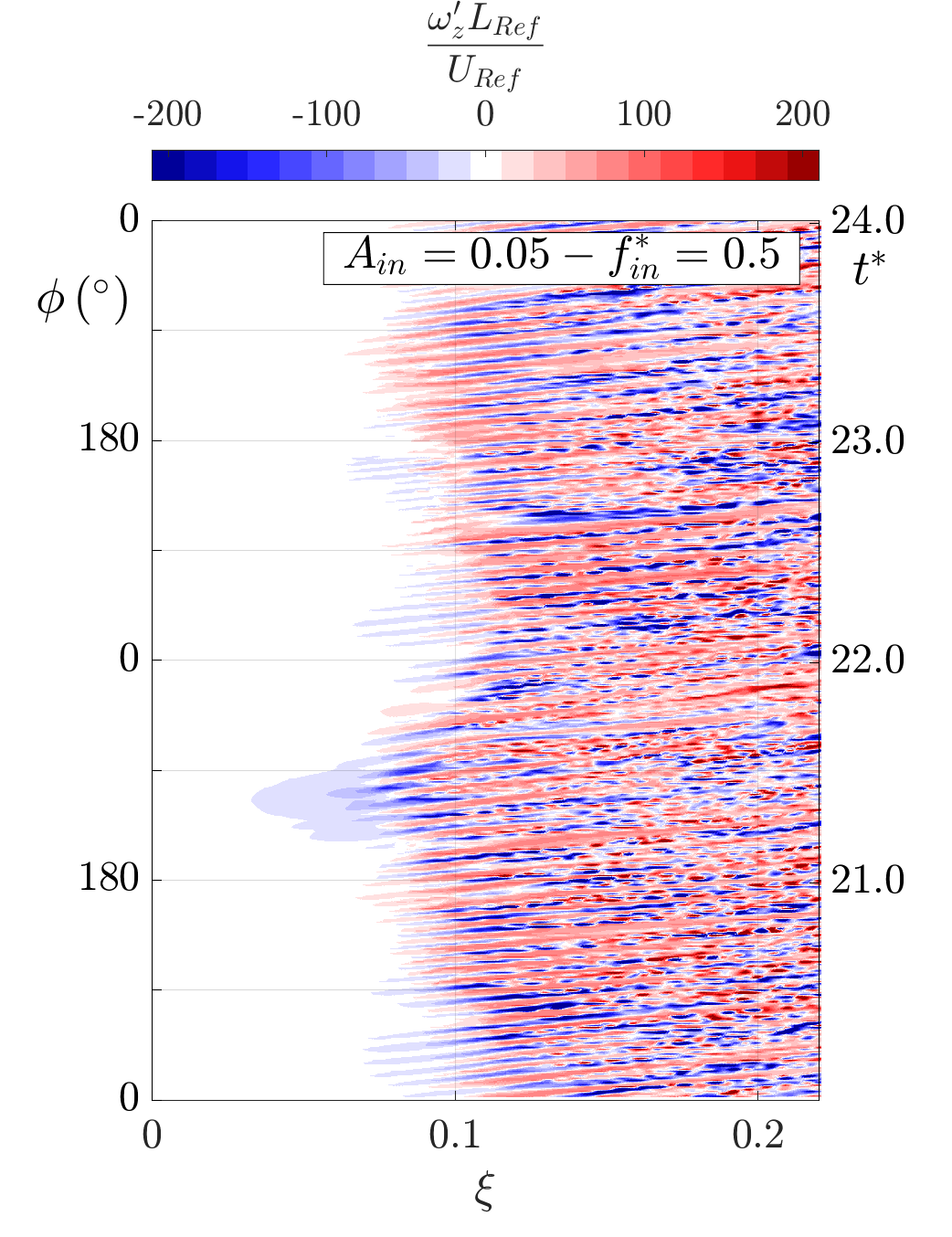}  &
\includegraphics[width=.29\textwidth,trim={1.05in 0.9in 0in 1.4in},clip] {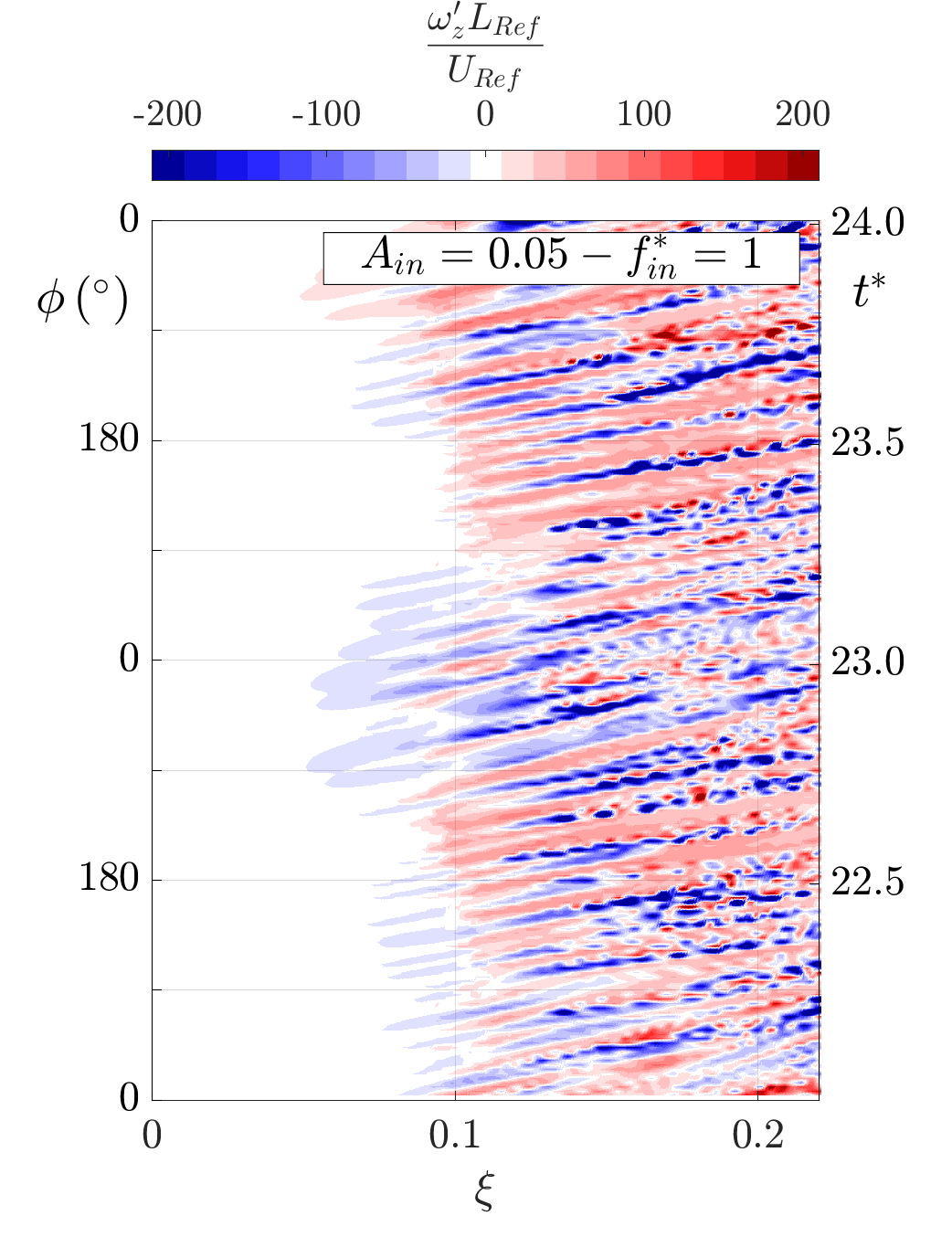} &
\includegraphics[width=.29\textwidth,trim={1.05in 0.9in 0.0in 1.4in},clip] {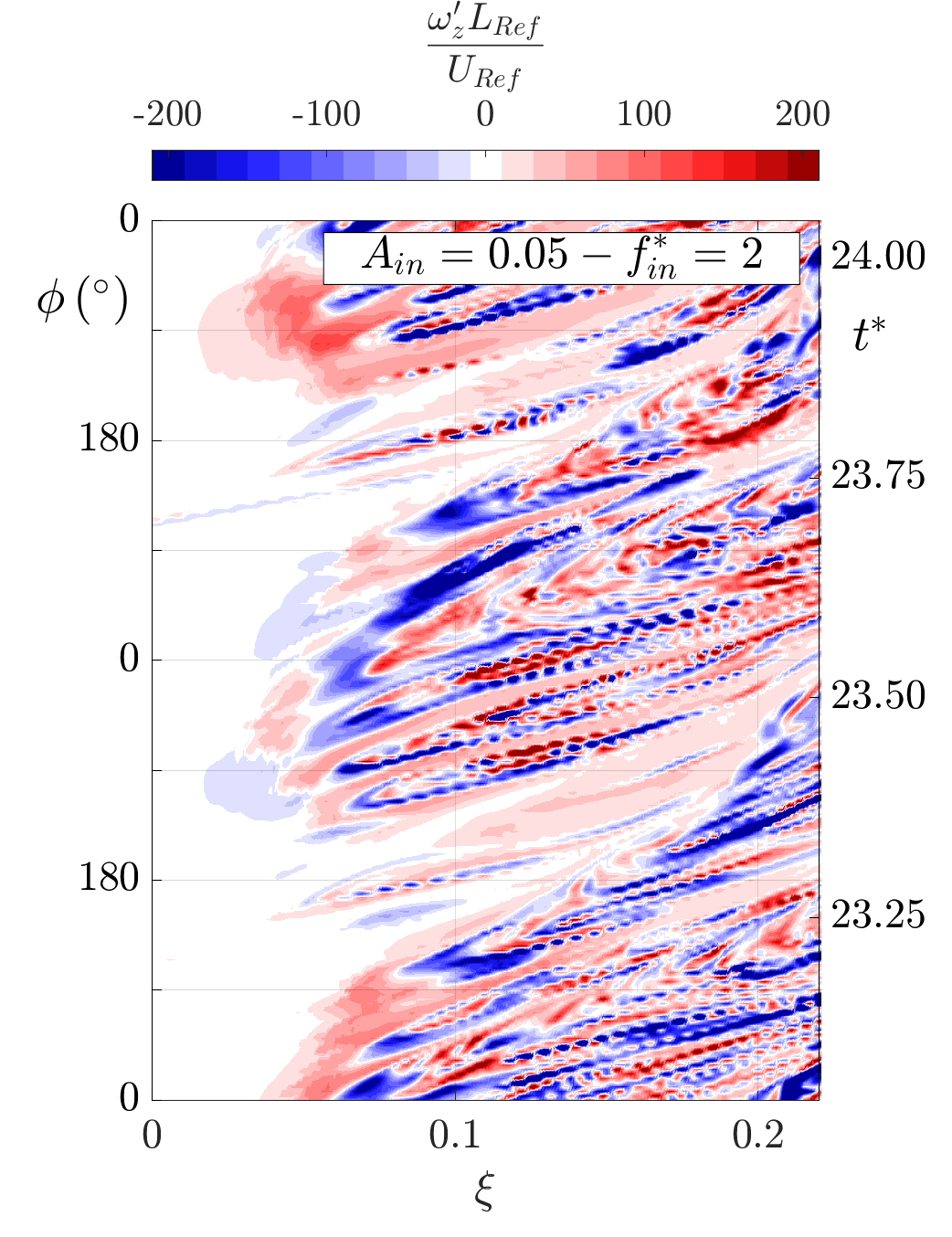} \\
\vspace{1mm} \\
\includegraphics[width=.342\textwidth,trim={0in 0in 0in 1.4in},clip] {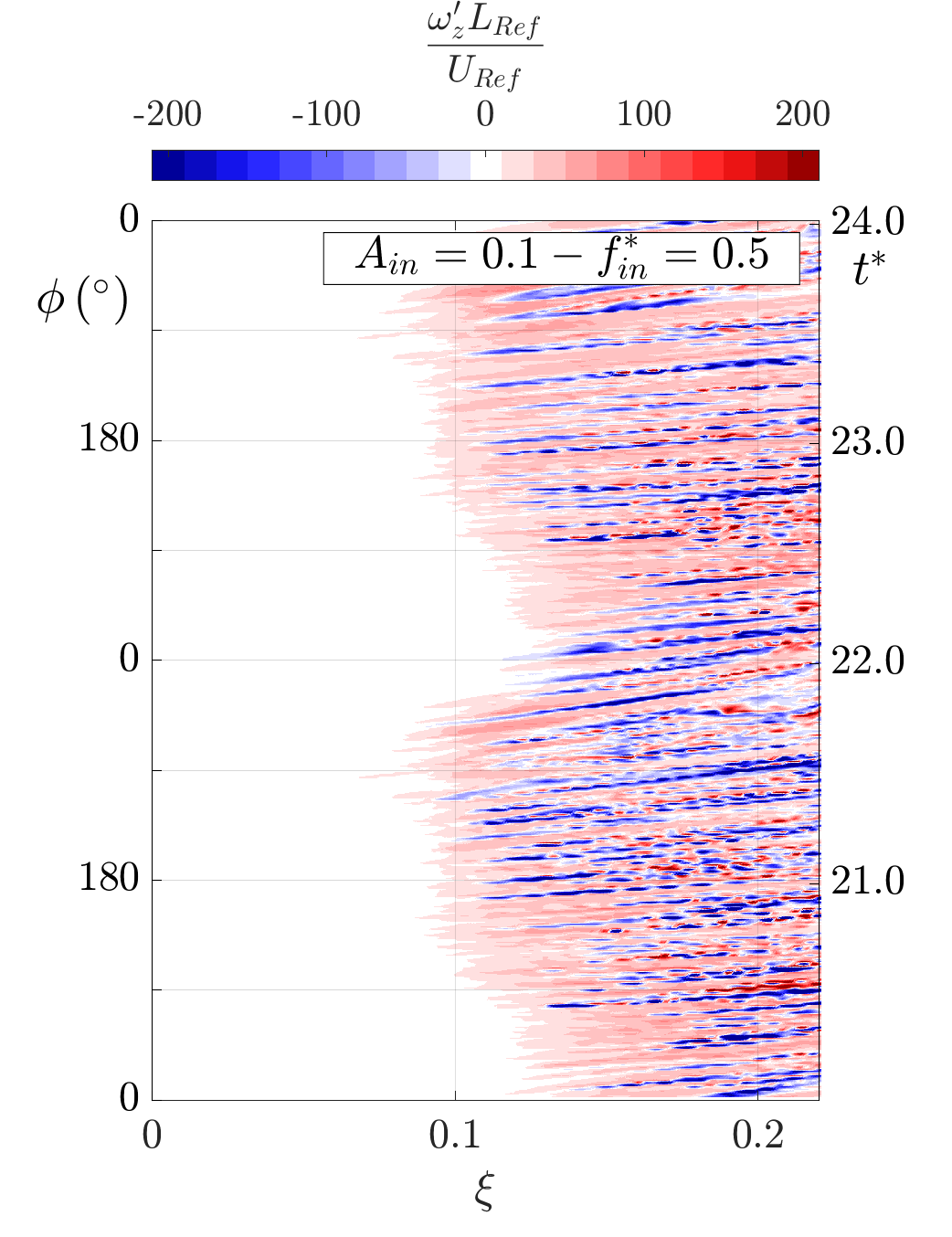}  &
\includegraphics[width=.29\textwidth,trim={1.05in 0in 0in 1.4in},clip] {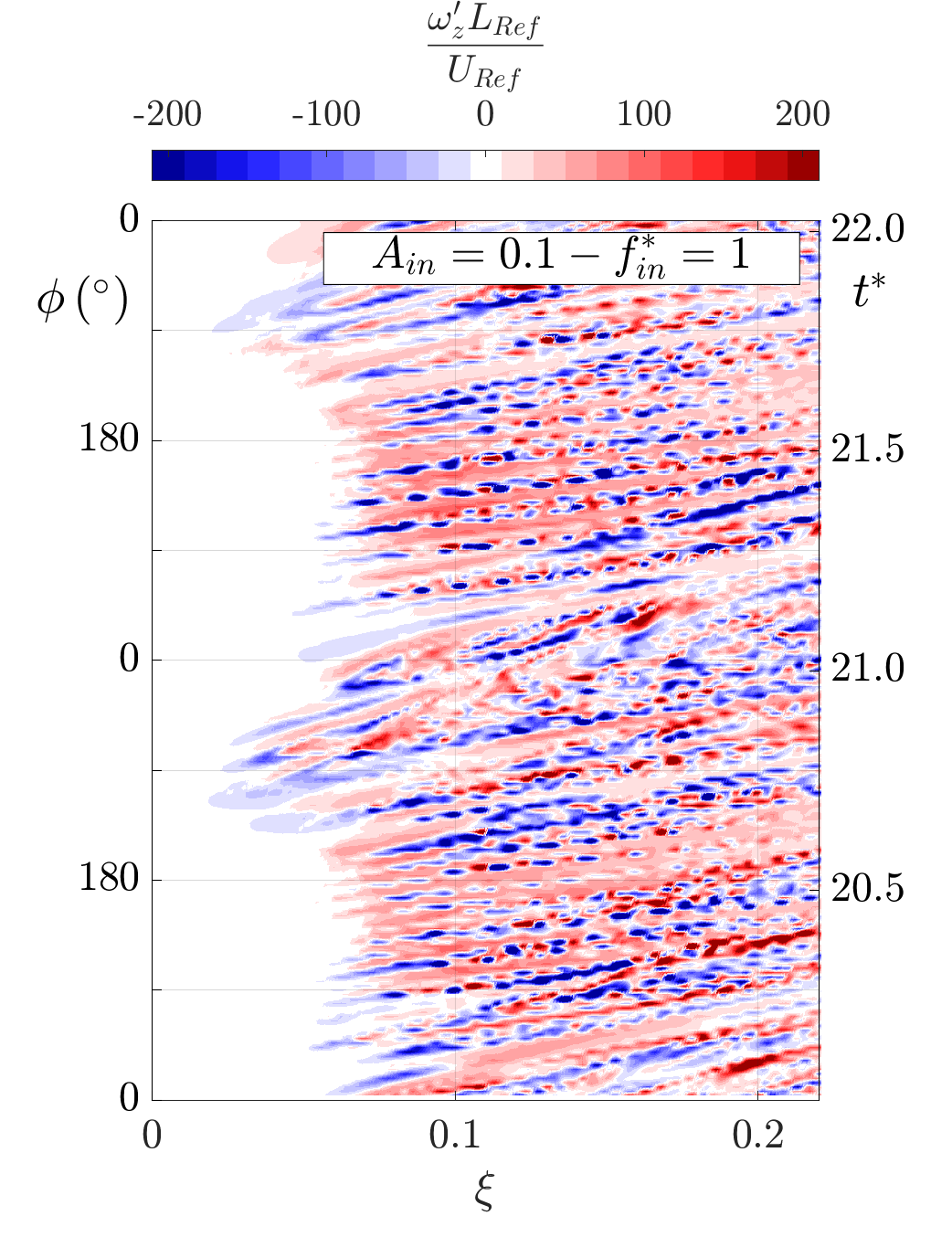}  &
\includegraphics[width=.29\textwidth,trim={1.05in 0in 0.0in 1.4in},clip] {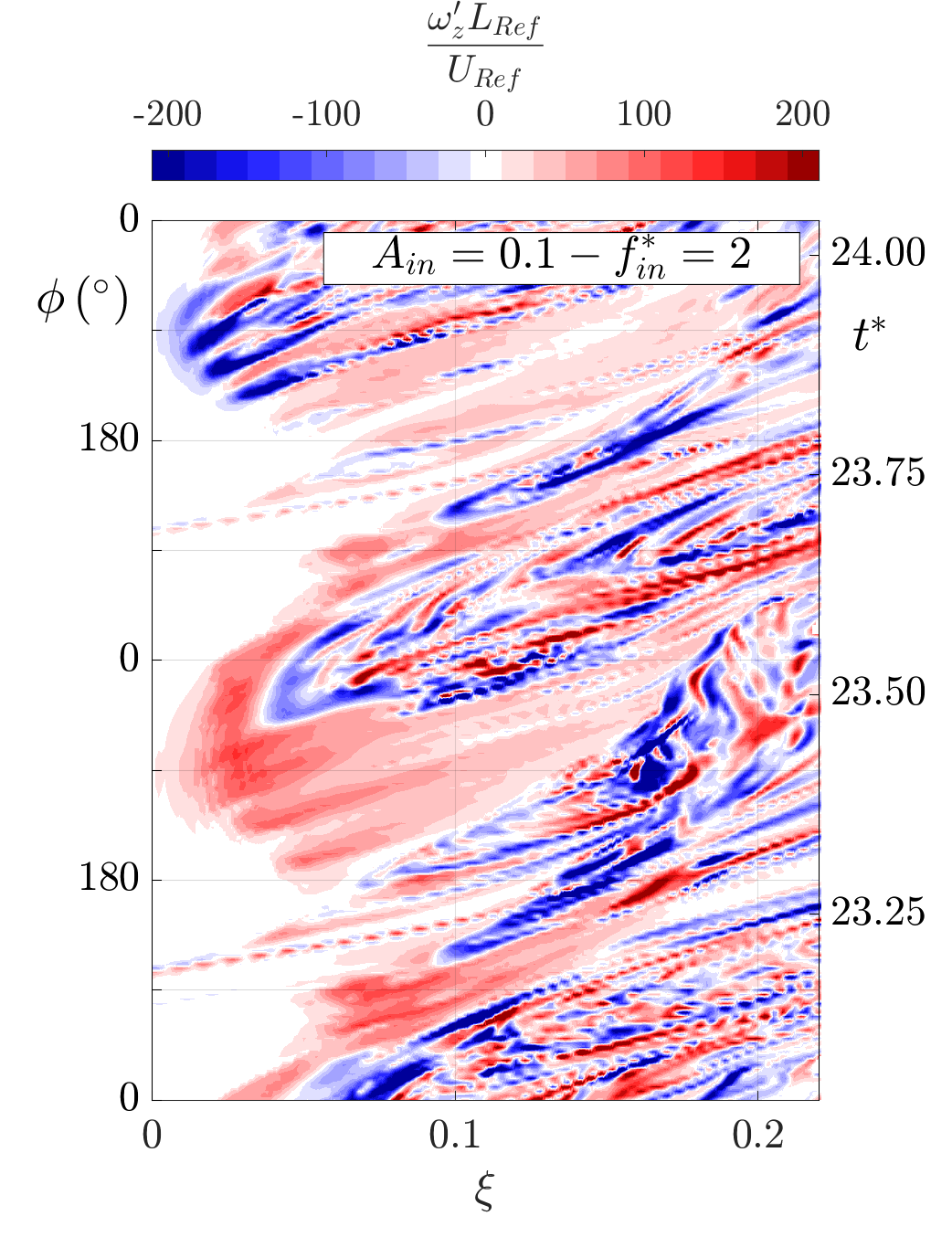} \\
\end{tabular}}
\caption{Incoherent spanwise vorticity at $\eta=0$ for the all nine cases with harmonic inflow. }
\label{fig:st_vort_all}
\end{figure}

\begin{table*}
\caption{Classification of the cases. Minimum values of the time-averaged and phase-averaged streamwise acceleration parameter and reduction of the time-averaged length of the separated flow region. The first row corresponds to the steady inflow case}
\label{tab:K_downstream}
\begin{ruledtabular}
\begin{center}
\def~{\hphantom{0}}
\begin{tabular}{cclcccccc} 
$A_{in}$ & $f_{in}^*$ & $A_{in} f_{in}^*$ & ${\overline{K}}_{min}$  &  $  <K> _{min} $  & $<K_{\partial u / \partial x}>_{min} $ & $<{K}_{\partial u / \partial t}>_{min}$  & $\Delta L_s / L_{s,steady}$ & Scenario \\  
& & & $\times 10^{-6}$& $\times 10^{-6}$& $\times 10^{-6}$& $\times 10^{-6}$ & & \\ [1pt] \hline
& & & & & & & & \\
- & - & - & -9.76 & - & - & - & - & (i) \\
 \\
0.01 & 0.5 & 0.005 & -10.62 &  -12.44 & -13.22  & -2.37   & 0 & (i) \\
0.01 &  1  & 0.01  & -10.54 &  -13.60 & -13.96  & -2.44   & -0.0007 & (i) \\
0.01 &  2  & 0.02  & -11.45 &  -22.43 & -24.45  & -4.51   & -0.0564 & (ii) \\
\\
0.05 & 0.5 & 0.025 & -10.24 &  -16.71 & -16.04  & -6.46   & -0.0198 & (i) \\
0.05 & 1   & 0.05  & -10.48 &  -24.74 & -22.20  & -11.53  & -0.0654 & (ii) \\
0.05 & 2   & 0.1   & -14.10 &  -36.12 & -39.93  & -22.87  & -0.2933 & (iii) \\
\\
0.1 & 0.5  & 0.05  & -9.11  &  -23.34 & -22.75  & -12.82  & -0.0744 & (ii) \\
0.1 & 1    & 0.1   & -12.16 &  -36.44 & -36.14  & -22.61  & -0.1650 & (ii) \\
0.1 & 2    & 0.2   & -17.86 &  -53.37 & -50.99  & -40.87  & -0.4490 & (iii) \\
\end{tabular}
\end{center}
\end{ruledtabular}
\end{table*}

Table \ref{tab:K_downstream} shows the correspondence between the simulated cases and the scenario observed for the vortex dynamics, as described in Section \ref{sec:Conclusion}. This classification is based on cross-comparing the time histories of the phase-averaged streamwise acceleration parameter (Fig. \ref{fig:appendix_acceleration_sum}), 
the incoherent spanwise vorticity along the phase-averaged shear layer (Fig. \ref{fig:st_vort_all}), the PDFs of the phase-averaged length of the separated flow region (Fig. \ref{fig:pdf_length}) and the frequency spectra \ref{fig:psd_turb_wz}. For reference, the evolution of the three-dimensional flow represented as in Fig. \ref{fig:Qcriterion_A01_F1} is also checked. Animations showing the flowfield evolution for representative cases are provided as supplementary material.

The table shows, for each simulated case, the minimum values of the acceleration parameter for the corresponding time-averaged flow $\bar{K}$, the phase-averaged flow $\langle K \rangle$ and its two components. The time-averaged values are considerably lower than the phase-averaged ones and are not useful in the classification of the cases. The minimum of $\langle K \rangle$ decreases independently with the amplitude and frequency of the inlet oscillations. The convective acceleration $\langle K_{\partial u/\partial x} \rangle$ dominates for low $A_{in}$ and/or $f^*_{in}$ values.  As a rough approximation, the temporal acceleration $\langle K_{\partial u/\partial t} \rangle \sim A_{in} f^*_{in}$, and this component becomes comparable to the convective acceleration for the largest values of the product $A_{in} f^*_{in}$. 

\section{Simulations considering the NASA hump}
\label{sec:appendix_NASA_hump}

\begin{figure}[h]
\includegraphics[width=.9\textwidth,trim={0in 0.4in 0.5in 2.6in},clip] {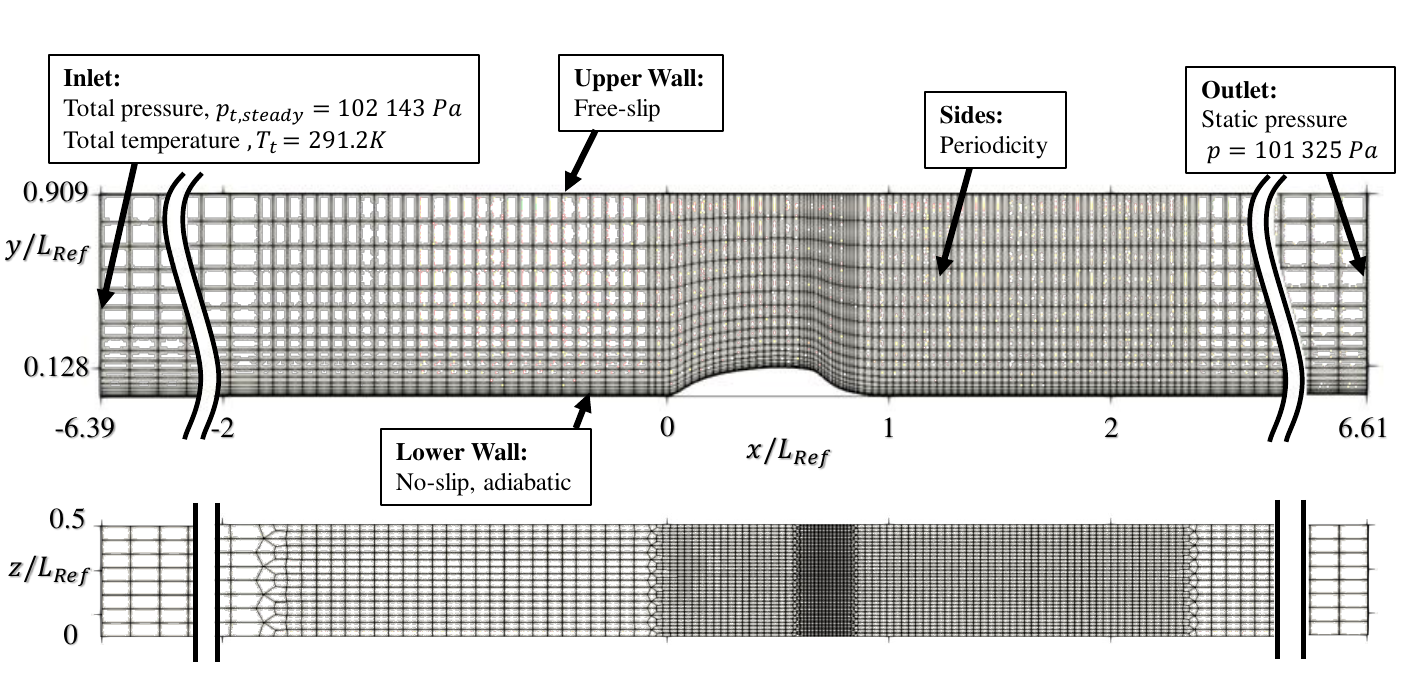}
\caption{Computational domain, representative mesh, and boundary conditions for NASA Hump cases\
.}
\label{fig:mesh_NASA_Hump}
\end{figure}

To demonstrate the generality of the physical phenomena with respect to other wall-bounded bump geometries, an analogous study is conducted on a wall-mounted modified Glauert hump model \citep{Seifert:AIAAJ2002,GreenblattAIAAJ2006}, often referred to in the literature as the NASA hump. The setup of the simulation is shown in Fig. \ref{fig:mesh_NASA_Hump}. 
The domain of the simulation is $\left(13\times0.909\times0.5 \right)$ m in the longitudinal, normal, and spanwise directions, respectively. A scaling of the hump geometry allows the reference chord length of the model, $L_{Ref}$ to be unity. The ratio of the hump height to the channel height is identical to \citet{GreenblattAIAAJ2006}, which is 0.1407. The domain is discretized into 77,616 high-order elements and simulated with $3^{rd}$-order polynomial (4,967,424 DOFs). The fringe region starts from $x_{start}=3$ m with $\Delta_{rise}=2$ m and $\overline{\lambda}=400$.

The boundary conditions of the domain are nearly identical to the aforementioned study but a free-slip condition is imposed at the upper boundary. In the harmonic inflow cases, the total pressure variation with time follows Eq.\ref{eq:Ptotal}, with $p_{t,steady}=102  143$ Pa. Simulations are conducted with the inflow air density $\rho=1.184$ kg/m$^3$, and the dynamic viscosity $\mu=1.366 \times 10^{-3}$ Pa.s. In these conditions, the flow has a characteristic Reynolds $Re=30,000$ and Mach 0.1 at the inlet for the steady inflow. The Reynolds number based on the maximum height of the recirculation region and respectively the length of the separated region are $Re_{h}\approx 3840$ and $Re_{L_s}=37380$. These numbers are comparable with those of the simulations presented in the paper. The Reynolds number based on the boundary layer momentum thickness at the separation point is $Re_{\theta_s}=171$. Based on the proposed scaling on \citet{hasan_1992} and \citet{sigurdson_1995}, the natural frequency of shedding of the KH vortices is $f^*_{KH}\approx 2.1$ and the vortex cluster shedding frequency is $f^*_{shed} \approx 0.6-0.7$.

\begin{figure}[h!]
\centering{
\includegraphics[width=.5\textwidth,trim={0.0in 2.3in 0in 0in},clip] {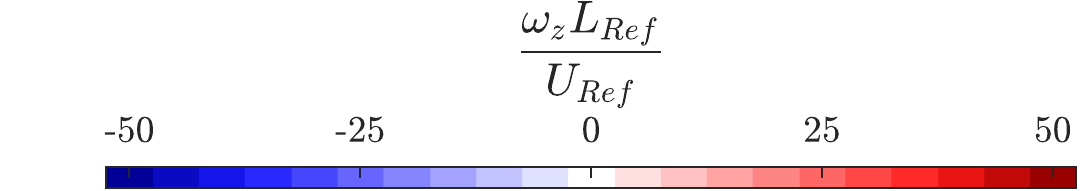}
\begin{tabular}{cc}
\includegraphics[width=.48\textwidth,trim={0.35in 1.1in 0.75in 1.7in},clip] {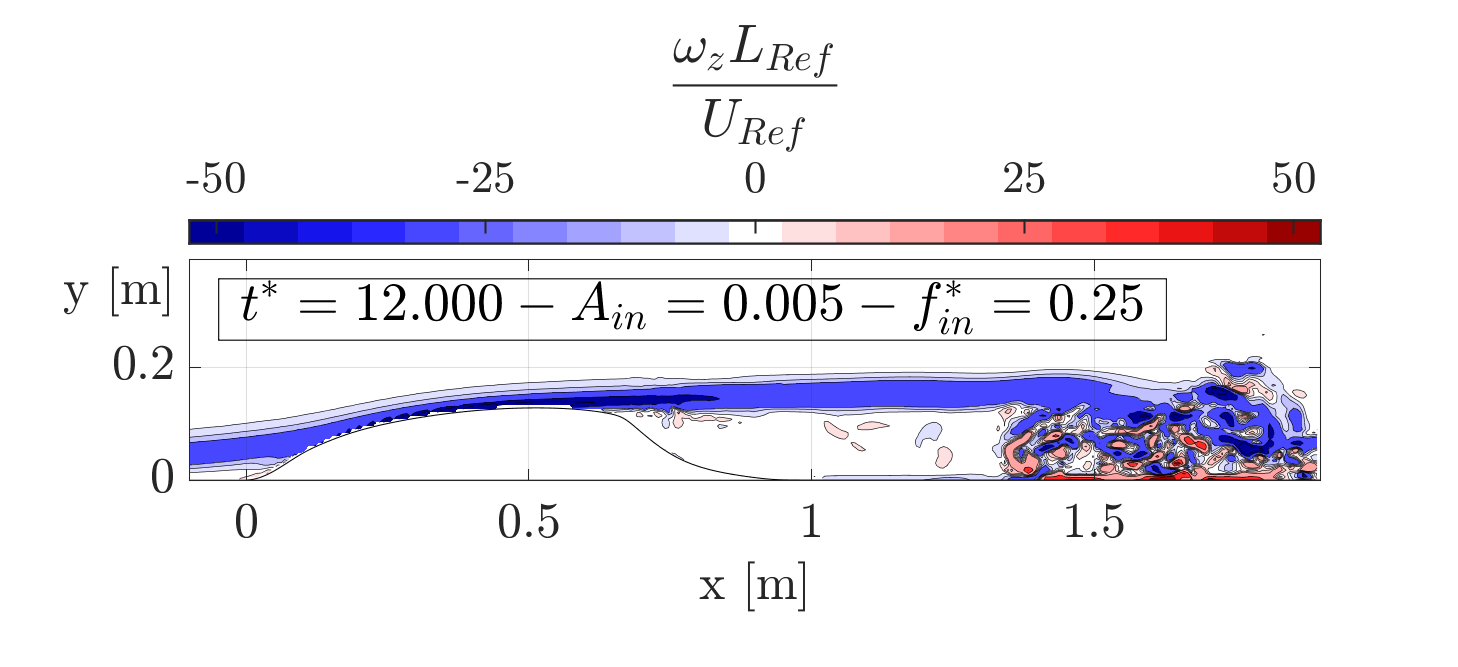} &
\includegraphics[width=.435\textwidth,trim={1.2in 1.1in 0.75in 1.7in},clip] {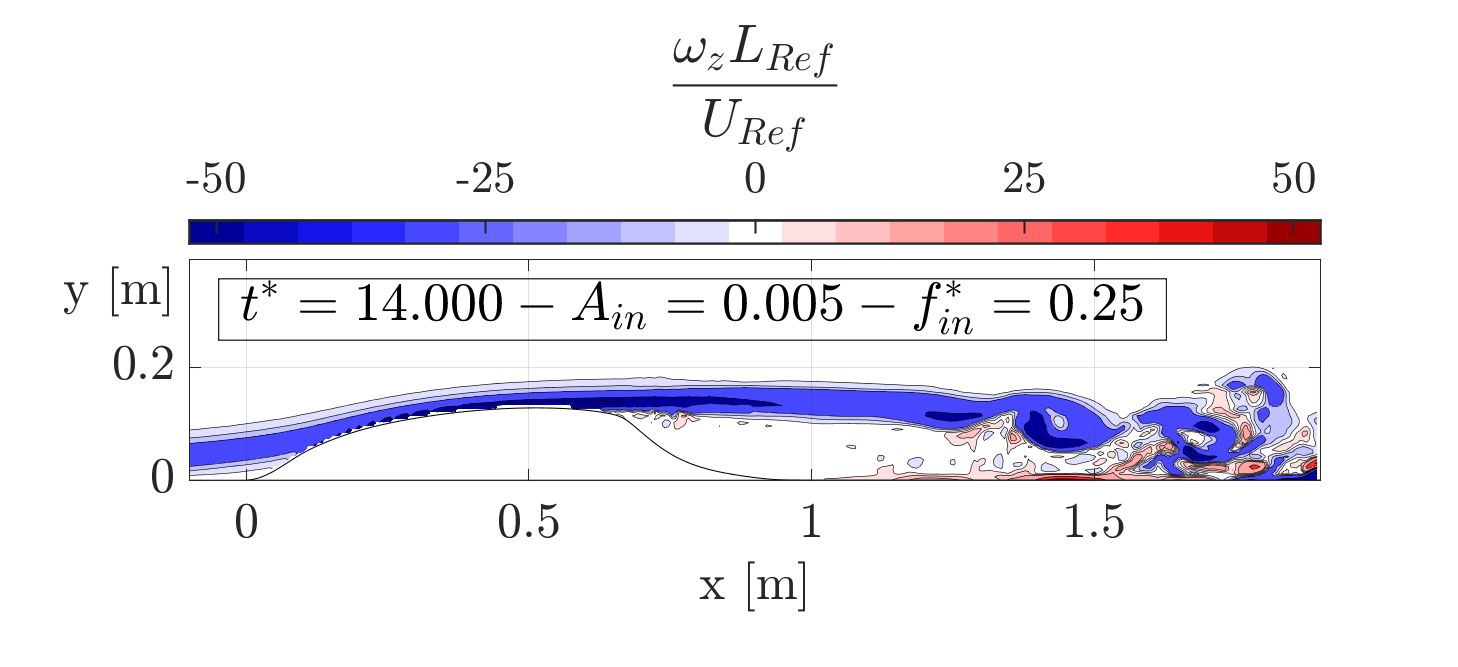} \\
\includegraphics[width=.48\textwidth,trim={0.35in 1.1in 0.75in 1.7in},clip] {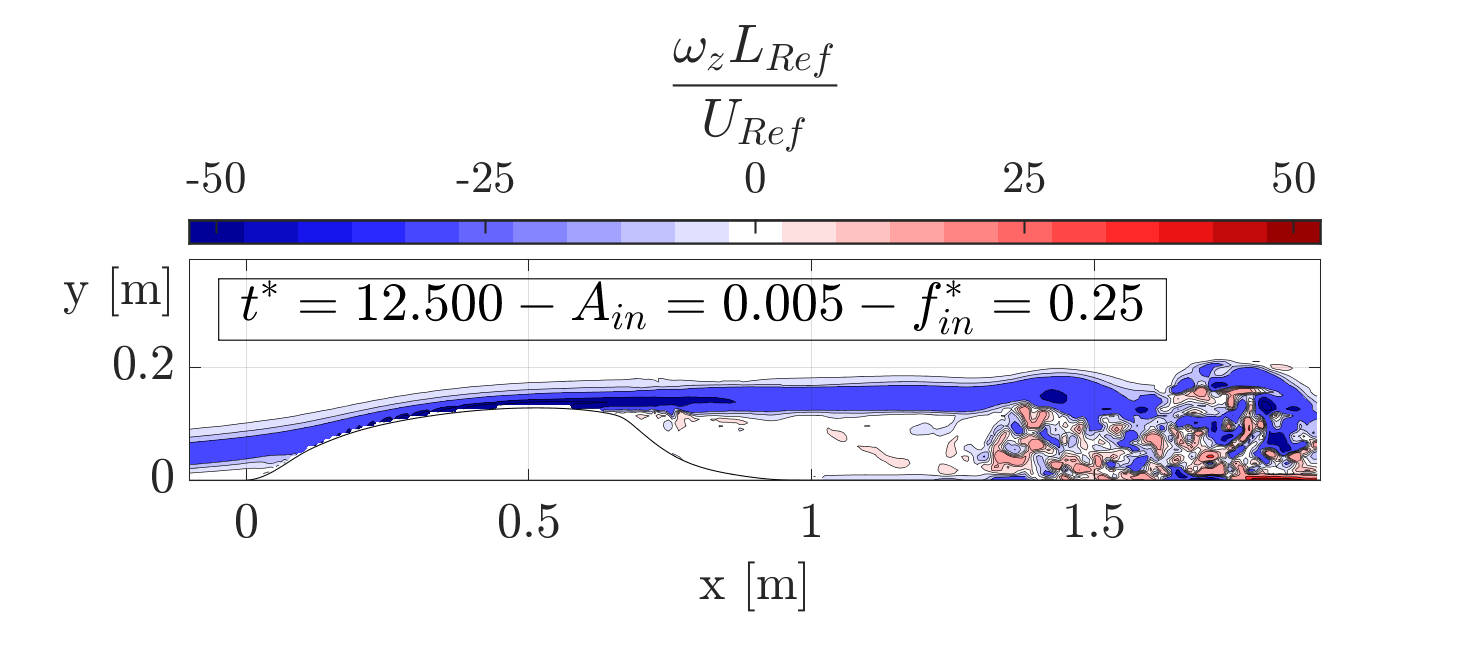} &
\includegraphics[width=.435\textwidth,trim={1.2in 1.1in 0.75in 1.7in},clip] {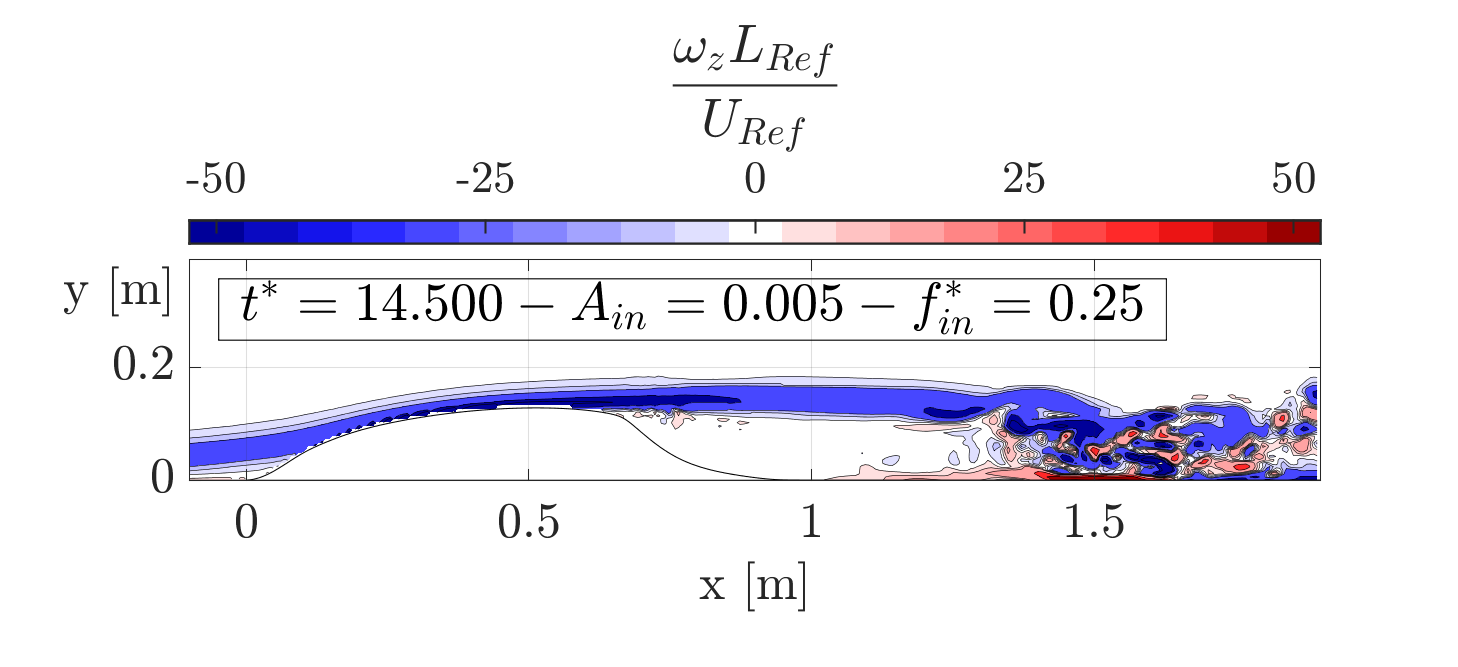} \\
\includegraphics[width=.48\textwidth,trim={0.35in 1.1in 0.75in 1.7in},clip] {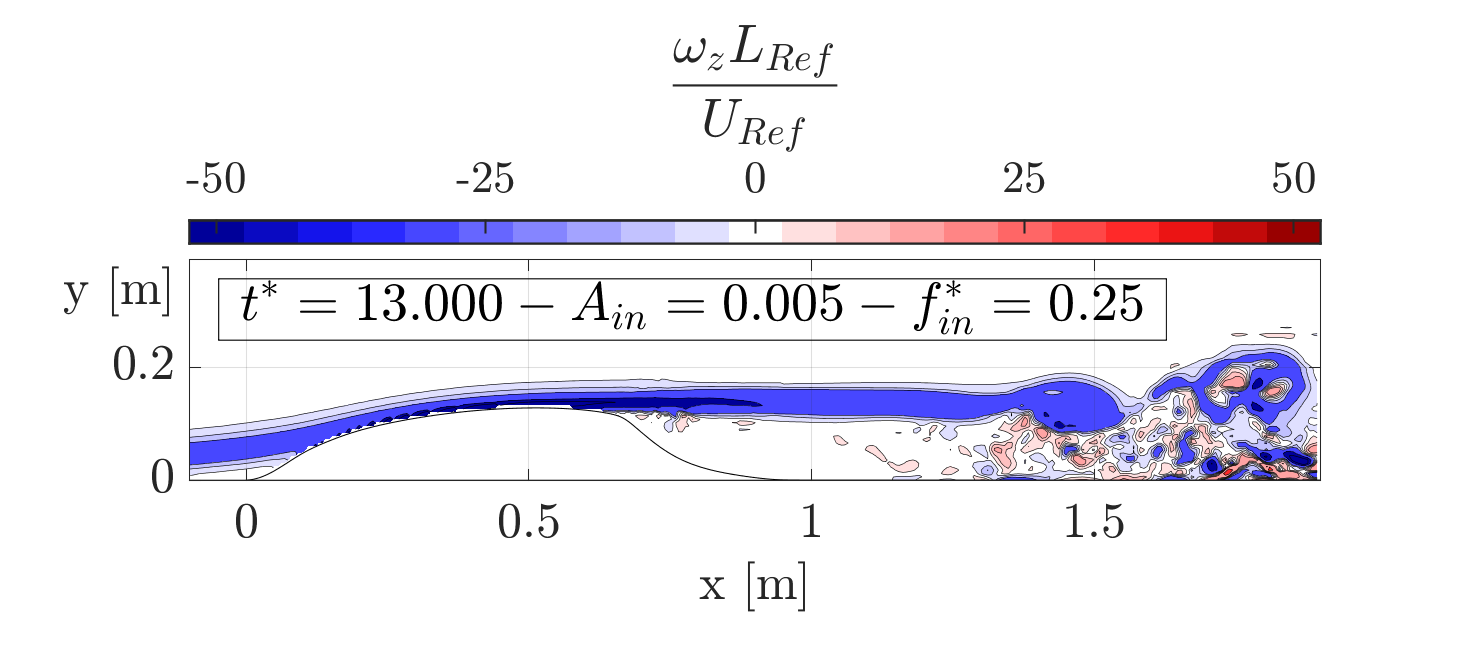} &
\includegraphics[width=.435\textwidth,trim={1.2in 1.1in 0.75in 1.7in},clip] {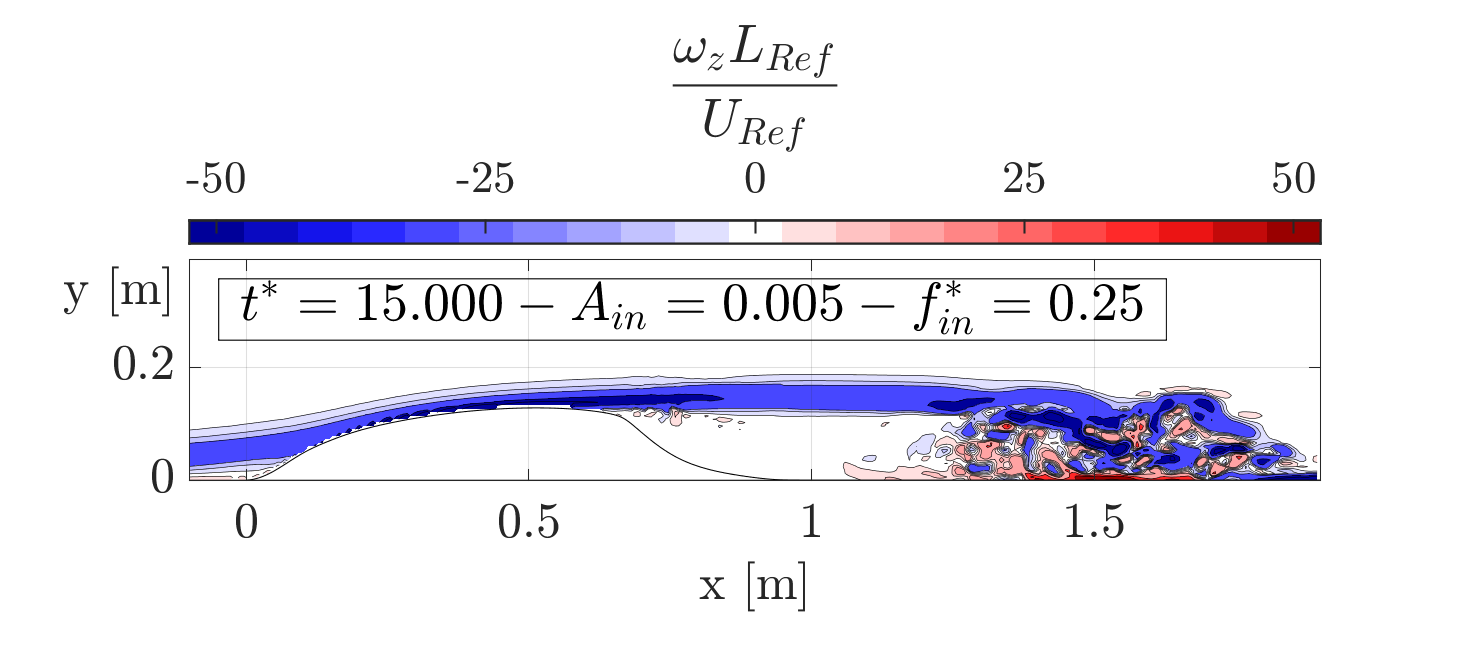} \\
\includegraphics[width=.48\textwidth,trim={0.35in 0.4in 0.75in 1.7in},clip] {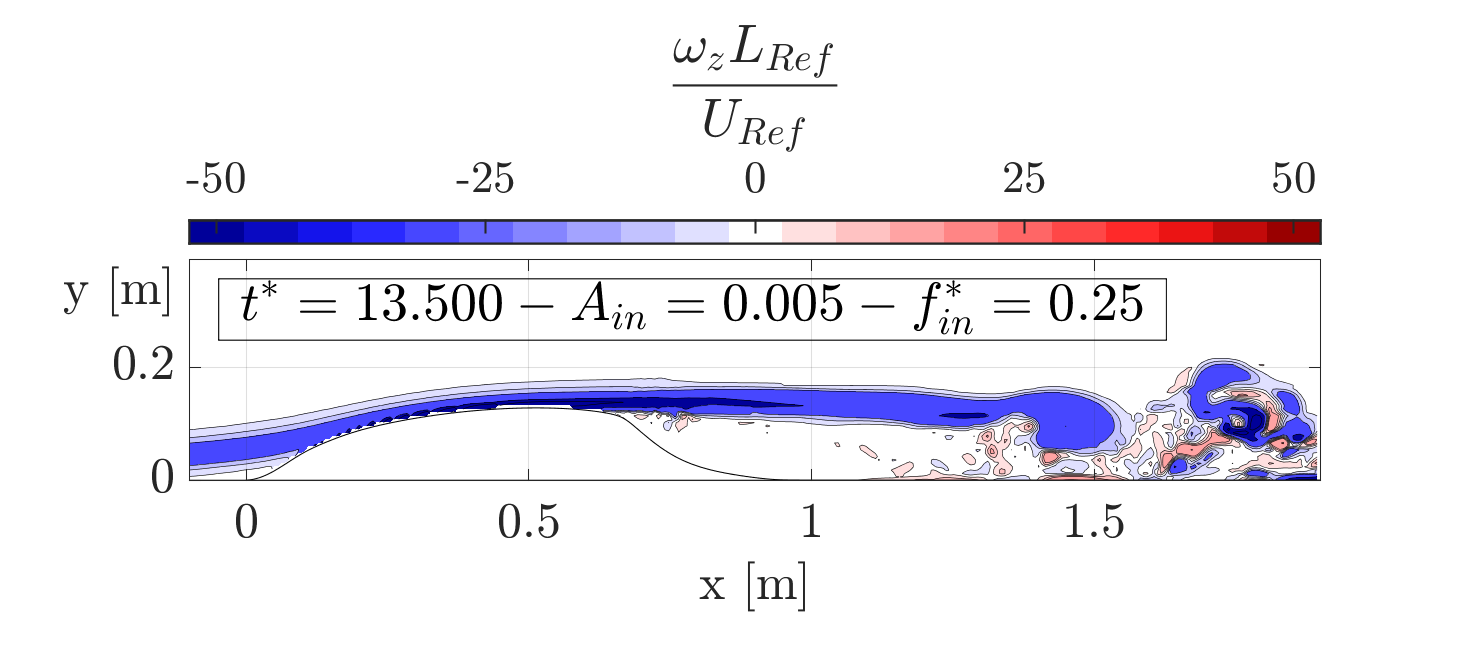} &
\includegraphics[width=.435\textwidth,trim={1.2in 0.4in 0.75in 1.7in},clip] {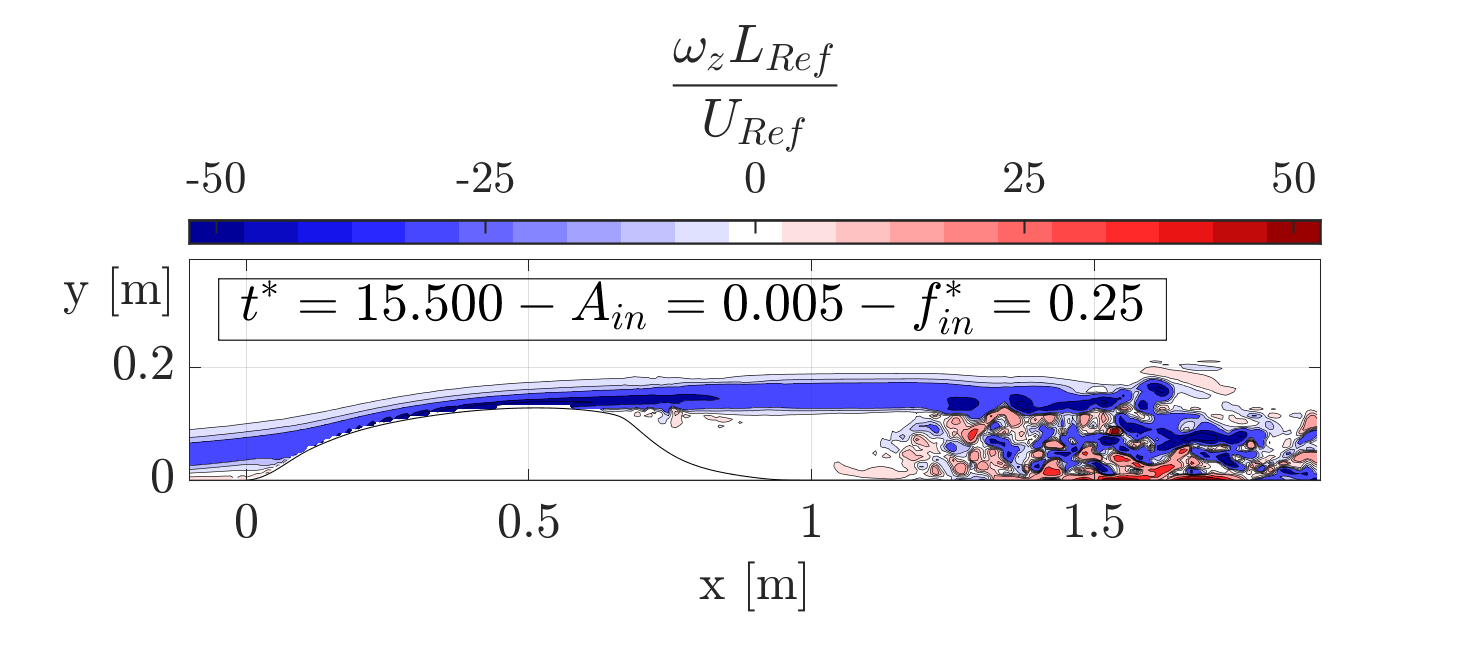} \\ \\
\end{tabular}}
\caption{Instantaneous spanwise vorticity for the NASA hump. $A_{in}=0.005$ and $f_{in}^{*}=0.25$. 
}
\label{fig:insta_vort_A0005_f025_NASA}
\end{figure}

\begin{figure}[h!]
\centering{
\includegraphics[width=.5\textwidth,trim={0.0in 2.3in 0in 0in},clip] {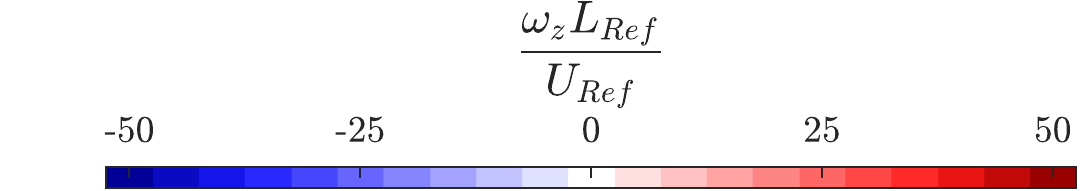}
\begin{tabular}{cc}
\includegraphics[width=.48\textwidth,trim={0.35in 1.1in 0.75in 1.7in},clip] {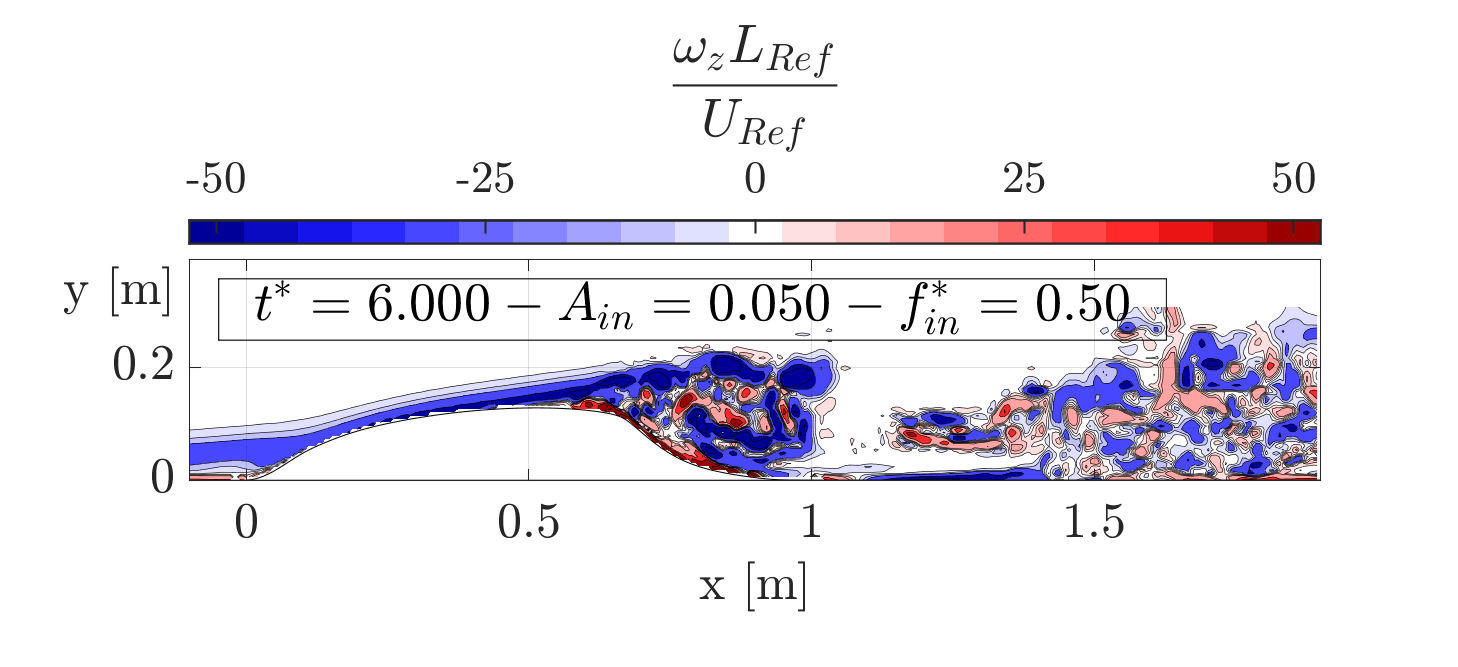} &
\includegraphics[width=.435\textwidth,trim={1.2in 1.1in 0.75in 1.7in},clip] {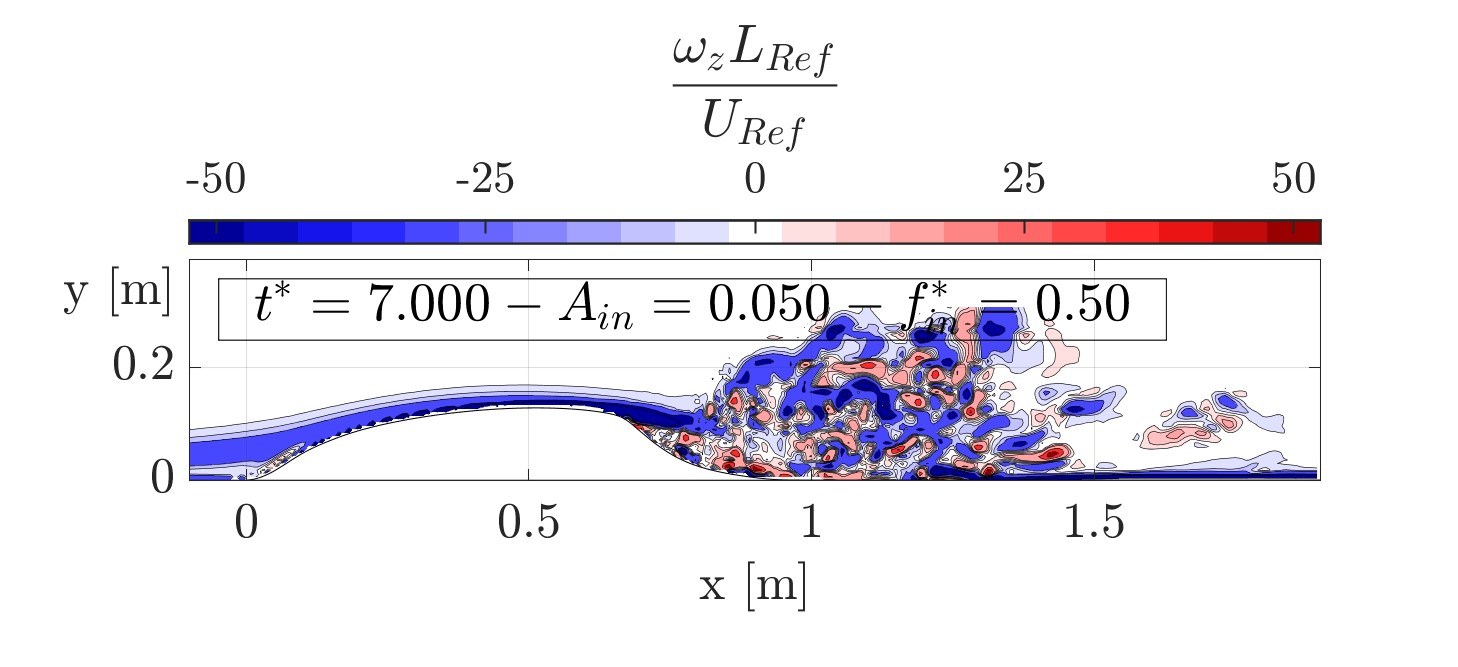} \\
\includegraphics[width=.48\textwidth,trim={0.35in 1.1in 0.75in 1.7in},clip] {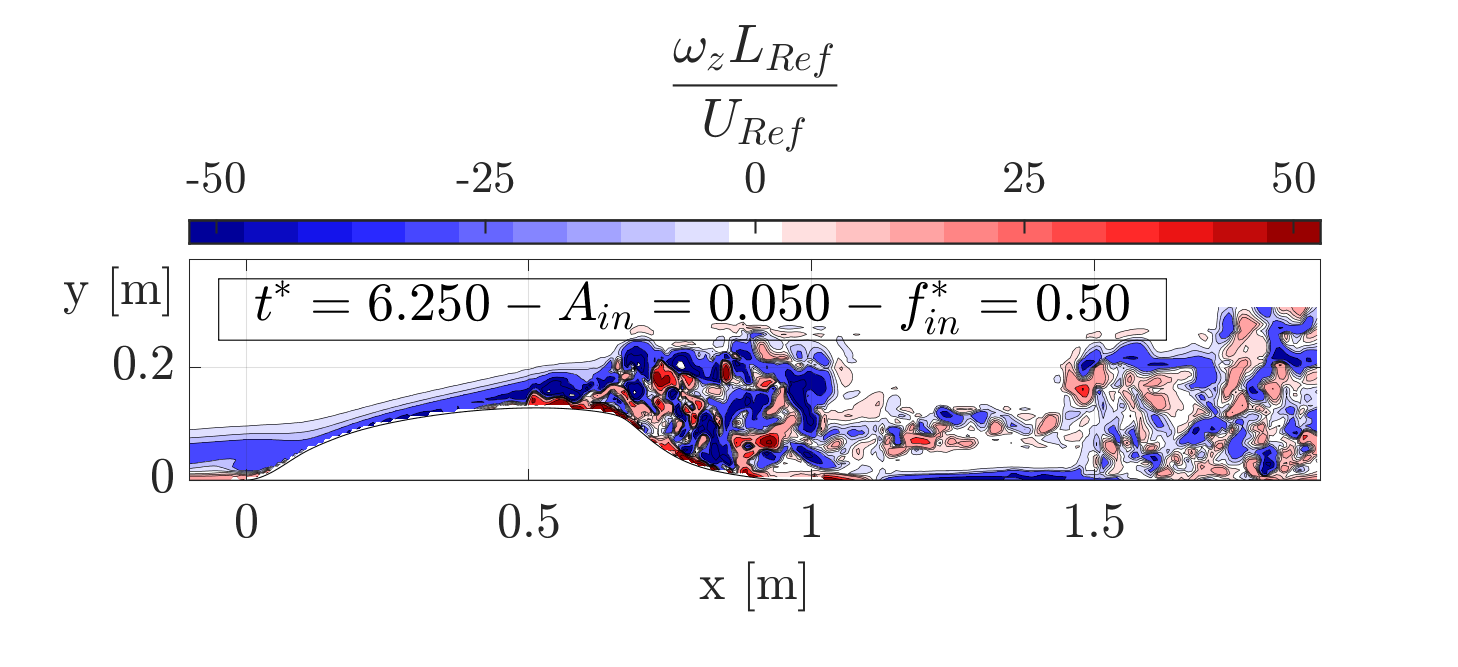} &
\includegraphics[width=.435\textwidth,trim={1.2in 1.1in 0.75in 1.7in},clip] {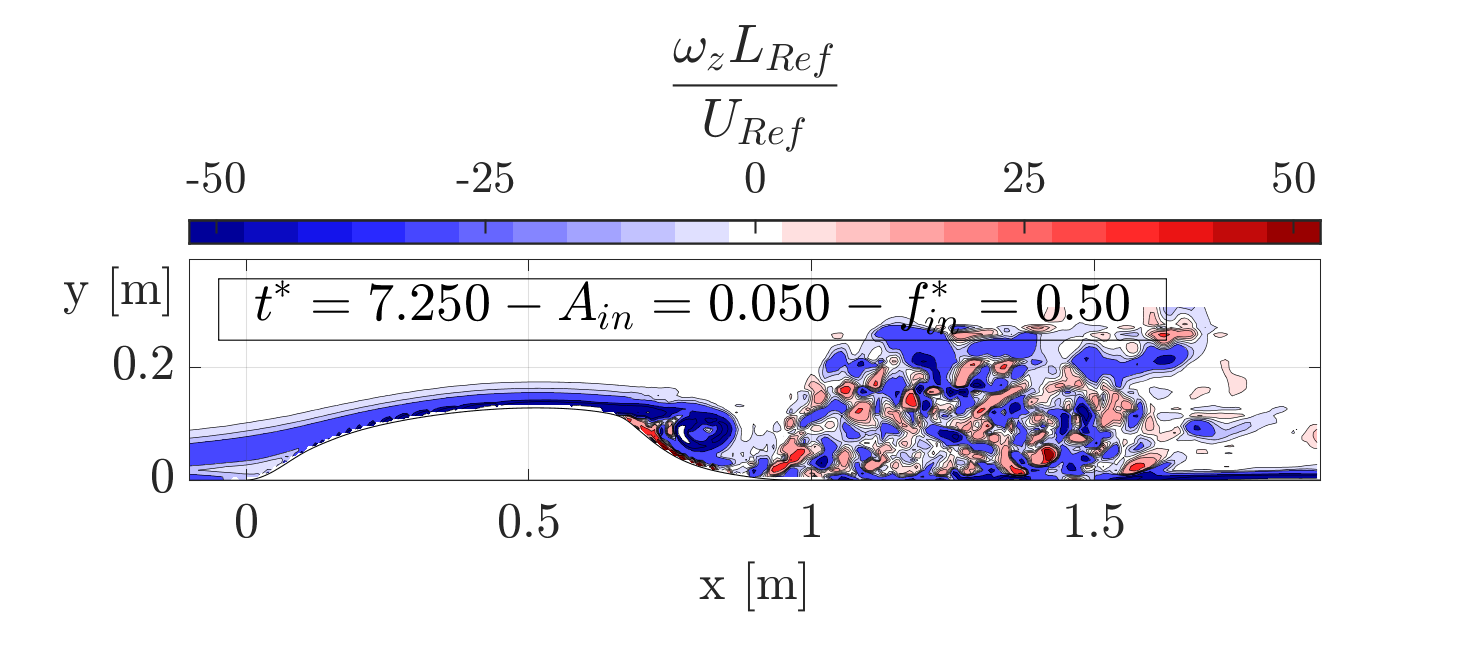} \\
\includegraphics[width=.48\textwidth,trim={0.35in 1.1in 0.75in 1.7in},clip] {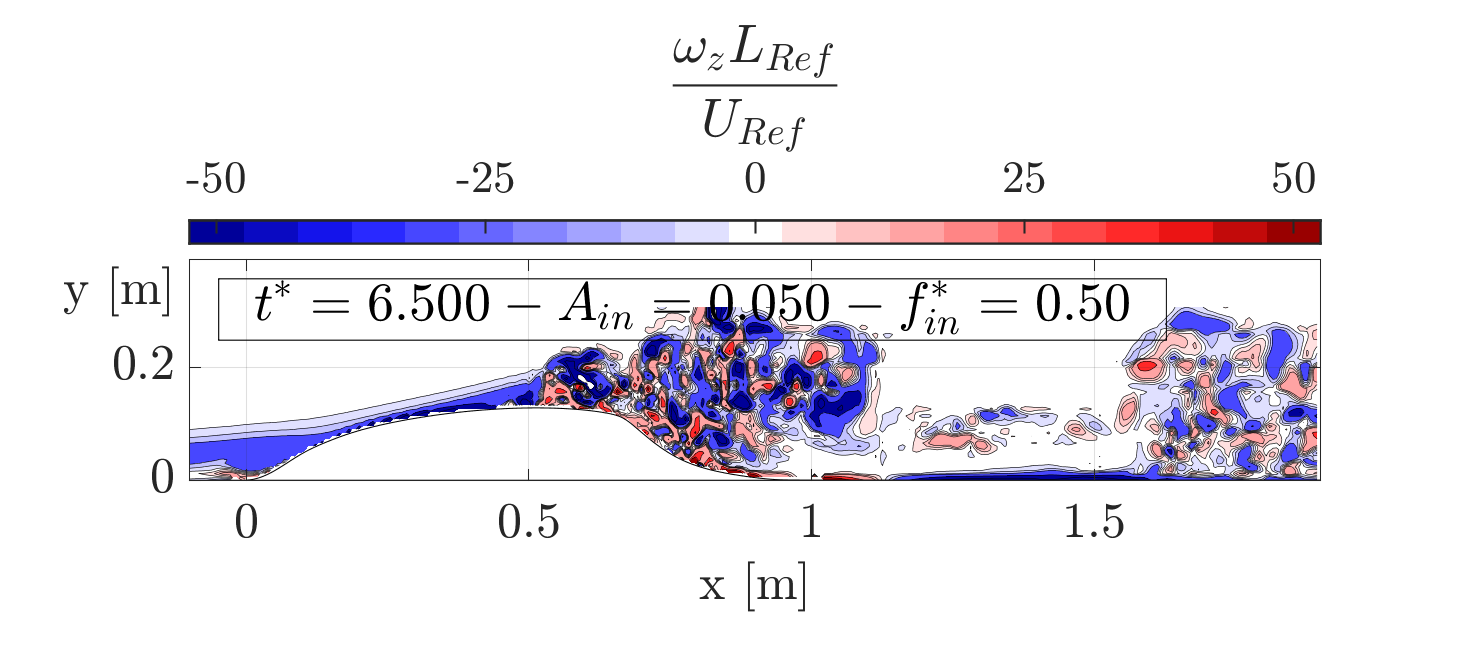} &
\includegraphics[width=.435\textwidth,trim={1.2in 1.1in 0.75in 1.7in},clip] {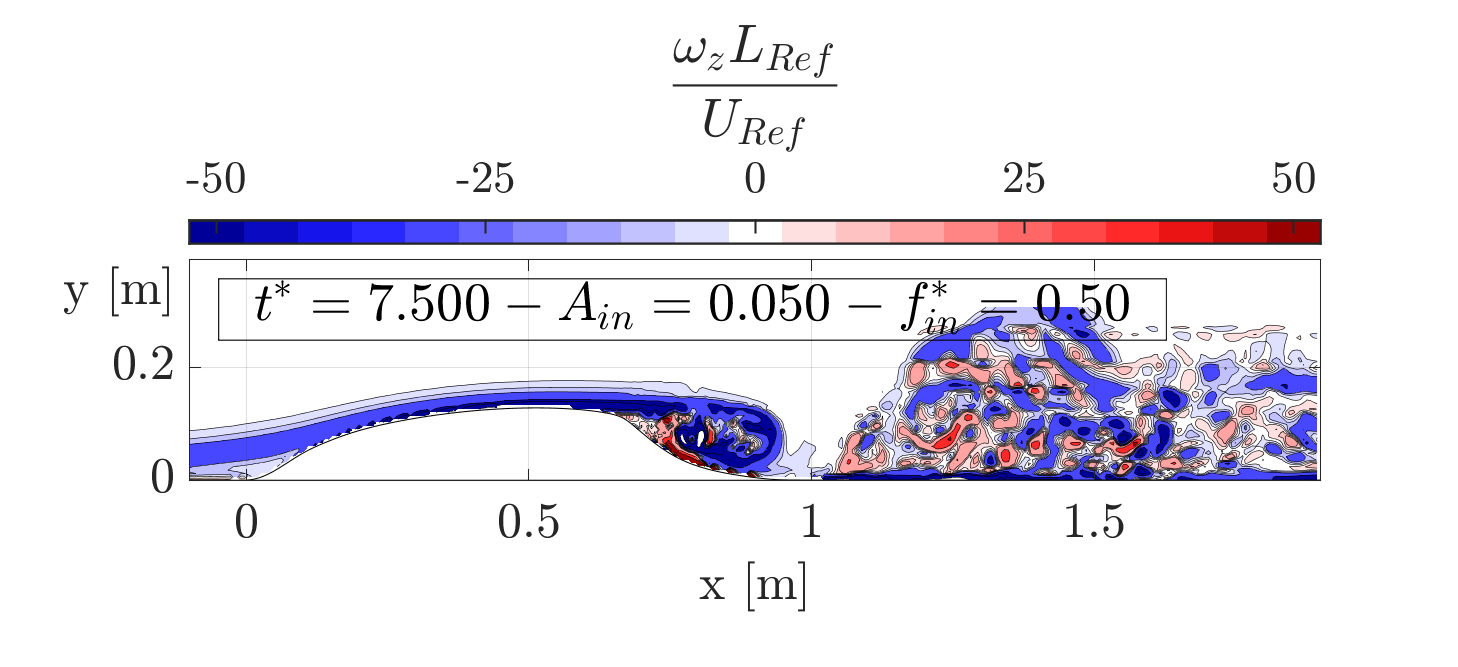} \\
\includegraphics[width=.48\textwidth,trim={0.35in 0.4in 0.75in 1.7in},clip] {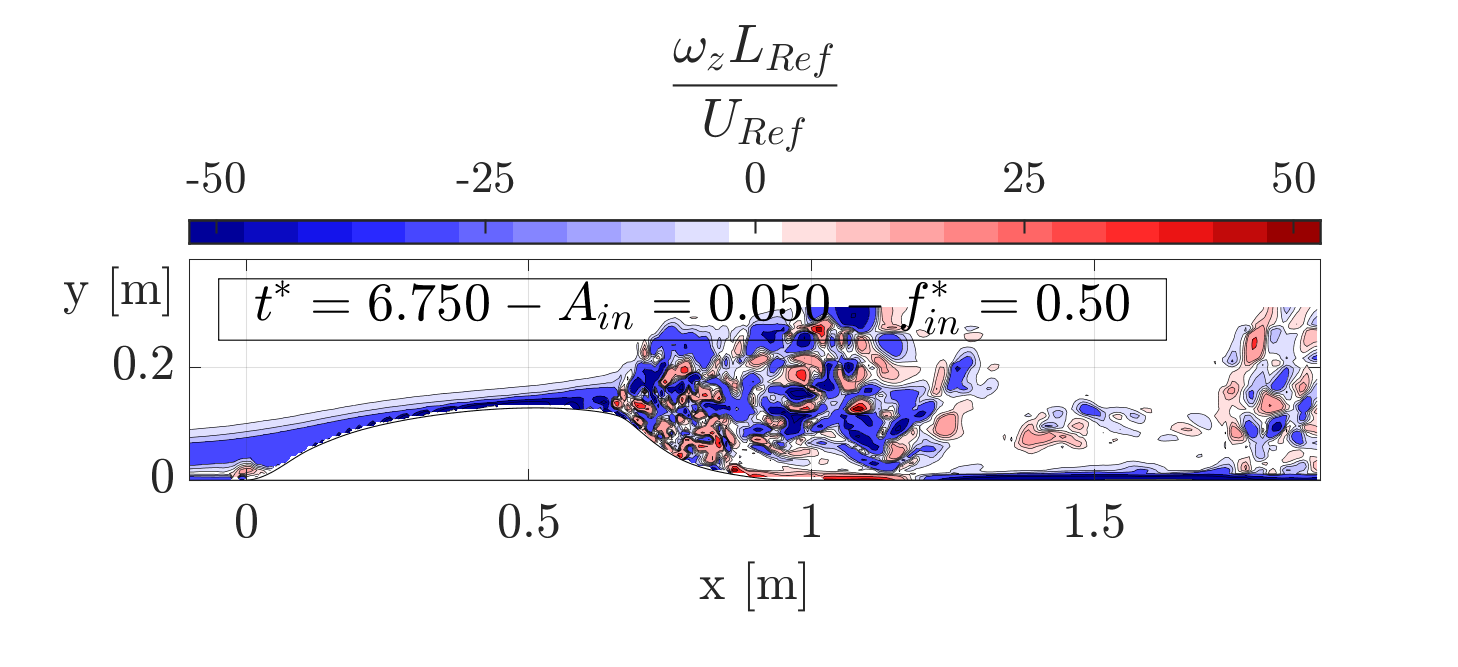} &
\includegraphics[width=.435\textwidth,trim={1.2in 0.4in 0.75in 1.7in},clip] {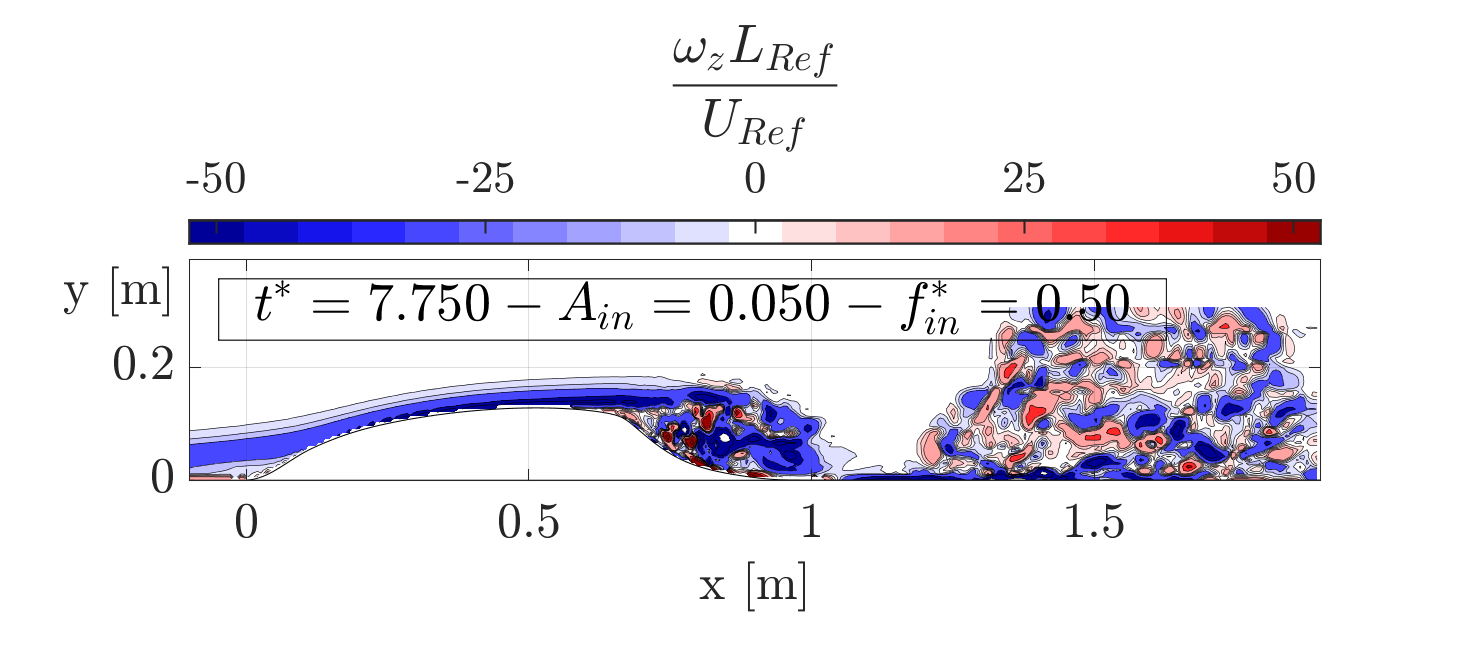} \\
 \\
\end{tabular}}
\caption{Instantaneous spanwise vorticity for the NASA hump. $A_{in}=0.05$ and $f_{in}^{*}=0.5$. 
}
\label{fig:insta_vort_A005_f05_NASA}
\end{figure}

Table \ref{tab:simulations_details_NASA_hump} summarizes the cases simulated, showing the harmonic inflow parameters, the mean and oscillation amplitude of the streamwise velocity at a reference point, $(x,y,z)=(0,0.9,0.25)$ m, the time-averaged length of the separated flow region and its reduction with respect to the steady inflow.  The instantaneous spanwise vorticity for the harmonic inflow cases ($A_{in}=0.005$, $f^*_{in}=0.25$) and ($A_{in}=0.05$, $f^*_{in}=0.5$) are presented in Figs. \ref{fig:insta_vort_A0005_f025_NASA} and \ref{fig:insta_vort_A005_f05_NASA}, respectively. The first case recovers scenario (i), dominated by KH vortices, while the second one shows the phased-locked formation and release of the vortex cluster, characteristic of scenario (iii).
The classification of the cases for the NASA hump geometry is given in Table \ref{tab:K_downstream_NASA_Hump} along with the minimum value of the acceleration parameter components computed at the wall-normal coordinate $y=2.778 h$. This coordinate is the same used for the computation of $K$ in section \ref{sec:Results}. 

\begin{table}[h]
\caption{Summary of the cases simulated for the NASA hump, including the definition of the inlet pressure condition and the characterization of time-averaged and phase-dependent values of the reference streamwise velocity and the length of the recirculation region $L_s$.  }
\label{tab:simulations_details_NASA_hump}
\begin{ruledtabular}
\begin{center}
\def~{\hphantom{0}}
\begin{tabular}{ccccccc} 
$t^*_{data}$ & $A_{in}$ & $f_{in}^*$ & $\bar{u}^*_{@ Ref}$ & $\Delta {u}^*_{@ Ref}$ & $L_s$ [m]   & $\Delta{L_s}$ [m] \\ [1pt] \hline

27 & -  & -  & 1.0958     & -       &  1.2460  &  - \\
 & & & & & & \\
 16 & 0.005 & 0.25 &  1.0974  & 0.0339  &  1.2284  &  -0.0176 \\
 16 & 0.005 & 0.5  &  1.0967  & 0.0348  &  1.0012  &  -0.2448 \\
 16 & 0.025 & 0.25 &  1.0953  & 0.1699  &  0.8610  &  -0.3850 \\
 12 & 0.025 & 0.5  &  1.0932  & 0.1744  &  0.4531  &  -0.7929 \\
 12 & 0.05 &  0.5 &  1.0887  & 0.3498  &  0.4079  &  -0.8381\\
\end{tabular}
\end{center}
\end{ruledtabular}
\end{table}

\begin{table*}[h!]
\caption{Classification of the cases for the NASA hump geometry. Minimum values of the time-averaged and phase-averaged streamwise acceleration parameter and reduction of the time-averaged length of the separated flow region. The first row corresponds to the steady inflow case}
\label{tab:K_downstream_NASA_Hump}
\begin{ruledtabular}
\begin{center}
\def~{\hphantom{0}}
\begin{tabular}{cclcccccc} 
$A_{in}$ & $f_{in}^*$ & $A_{in} f_{in}^*$ & ${\overline{K}}_{min}$  &  $  <K> _{min} $  & $<K_{\partial u / \partial x}>_{min} $ & $<{K}_{\partial u / \partial t}>_{min}$  & $\Delta L_s / L_{s,steady}$ & Scenario \\  
& & & $\times 10^{-6}$& $\times 10^{-6}$& $\times 10^{-6}$& $\times 10^{-6}$ & & \\ [1pt] \hline
& & & & & & & & \\
- & - & - & -6.05 & -6.05 & -6.05 & - & - & (i) \\
 \\
 0.005 & 0.25 & 0.00125  & -6.78 &  -15.24  &  -15.28 &    -0.18 & -0.0141  & (i) \\
0.005 & 0.5 & 0.0025   & -8.51 &  -19.61  &  -19.85  &  -0.27   &  -0.1965 & (ii) \\
0.025 & 0.25 & 0.00625 & -9.15 &  -20.53  &  -20.40  &  -0.28   & -0.3090  & (ii) \\
0.025 & 0.5 & 0.0125   & -10.02 &  -24.48  &  -25.29 &  -3.66   &  -0.6364  & (ii) \\
0.05 & 0.5 & 0.025     & -16.70 &  -42.77  &  -43.74  &  -8.33   & -0.6726  & (iii) \\

\end{tabular}
\end{center}
\end{ruledtabular}
\end{table*}

\newpage
\vspace*{\fill}
\newpage

%

\end{document}